\documentclass[preprint,journal]{vgtc}       

\ifpdf
  \pdfoutput=1\relax                   
  \usepackage{graphicx}                
  \DeclareGraphicsExtensions{.pdf,.png,.jpg,.jpeg} 
\else
  \usepackage{graphicx}                
  \DeclareGraphicsExtensions{.eps}     
\fi%

\graphicspath{{figures/}{pictures/}{images/}{./}} 

\usepackage{microtype}                 
\PassOptionsToPackage{warn}{textcomp}  
\usepackage{textcomp}                  
\usepackage{mathptmx}                  
\usepackage{times}                     
\usepackage{cite}                      
\usepackage{tabu}                      
\usepackage{booktabs}                  
\usepackage{float}
\usepackage{soul}

\usepackage{amsfonts,amsmath,amssymb,amsthm,amsopn}
\usepackage{enumitem}

\setlist{nolistsep,leftmargin=*}
\newcommand{\field}[1]{\mathbb{#1}}
\newcommand{\R}{\field{R}}

\newcommand{\domain}{\mathcal{M}}
\newcommand{\range}{\R}

\newcommand{\persistenceDiagram}{\mathcal{D}}
\newcommand{\Index}{\mathcal{I}}

\newcommand{\reebGraph}{\mathcal{R}}

\newcommand{\vect}[1]{\boldsymbol{\mathbf{#1}}}
\newcommand{\bspline}{\Psi}

\usepackage{xcolor}

\newcommand{\response}[1]{{\color{black}#1}}

\newcommand{\grammar}[1]{{\color{black}#1}}
\newcommand{\stgrammar}[1]{{\color{black}\st{}}}

\onlineid{1081}

\vgtccategory{Research}
\vgtcpapertype{Empirical Study}

\title{\response{The Effect of Data Transformations on Scalar Field Topological Analysis of High-Order FEM Solutions}}

\author{Ashok Jallepalli, Joshua A.\,Levine, and Robert M.\,Kirby}
\authorfooter{
\item
 Ashok Jallepalli and Robert M.\,Kirby are with SCI Institute, University of Utah. E-mail: \{ashokj,kirby\}@sci.utah.edu.
\item
 Joshua A.\,Levine is with Department of Computer Science, University of Arizona. E-mail: josh@email.arizona.edu.
}

\shortauthortitle{Biv \MakeLowercase{\textit{et al.}}: The Effect of Data Transformation Methodologies on the Topological Analysis of Scalar Fields from High-Order FEM Solutions}

\abstract{
High-order finite element methods (HO-FEM) are gaining popularity in the simulation community due to their success in solving complex flow dynamics.
There is an increasing need to analyze the data produced as output by these simulations.  
Simultaneously, topological analysis tools are emerging as powerful methods for investigating simulation data. 
However, most of the current approaches to topological analysis have had limited application to HO-FEM simulation data for two reasons.
First, the current topological tools are designed for linear data (polynomial degree one), but the polynomial degree of the data output by these simulations is typically higher (routinely up to polynomial degree six). 
Second, the simulation data and derived quantities of the simulation data have discontinuities at element boundaries, and these discontinuities do not match the input requirements for the topological tools. 
One solution to both issues is to transform the high-order data to achieve low-order, continuous inputs for topological analysis.
Nevertheless, there has been little work evaluating the possible transformation choices and their downstream effect on the topological analysis.
We perform an empirical study to evaluate two commonly used data transformation methodologies along with the recently introduced L-SIAC filter for processing high-order simulation data.
Our results show diverse behaviors are possible. We offer some guidance about how best to consider a pipeline of topological analysis of HO-FEM simulations with the currently available implementations of topological analysis.
}


\CCScatlist{ 
 \CCScat{K.6.1}{Management of Computing and Information Systems}%
{Project and People Management}{Life Cycle};
 \CCScat{K.7.m}{The Computing Profession}{Miscellaneous}{Ethics}
}

\teaser{
	  
		\centering
		\subcaptionbox{Sampled vorticity.  \label{fig-bvort-Seg2-dGSamp}}{\includegraphics[width=0.35\linewidth]{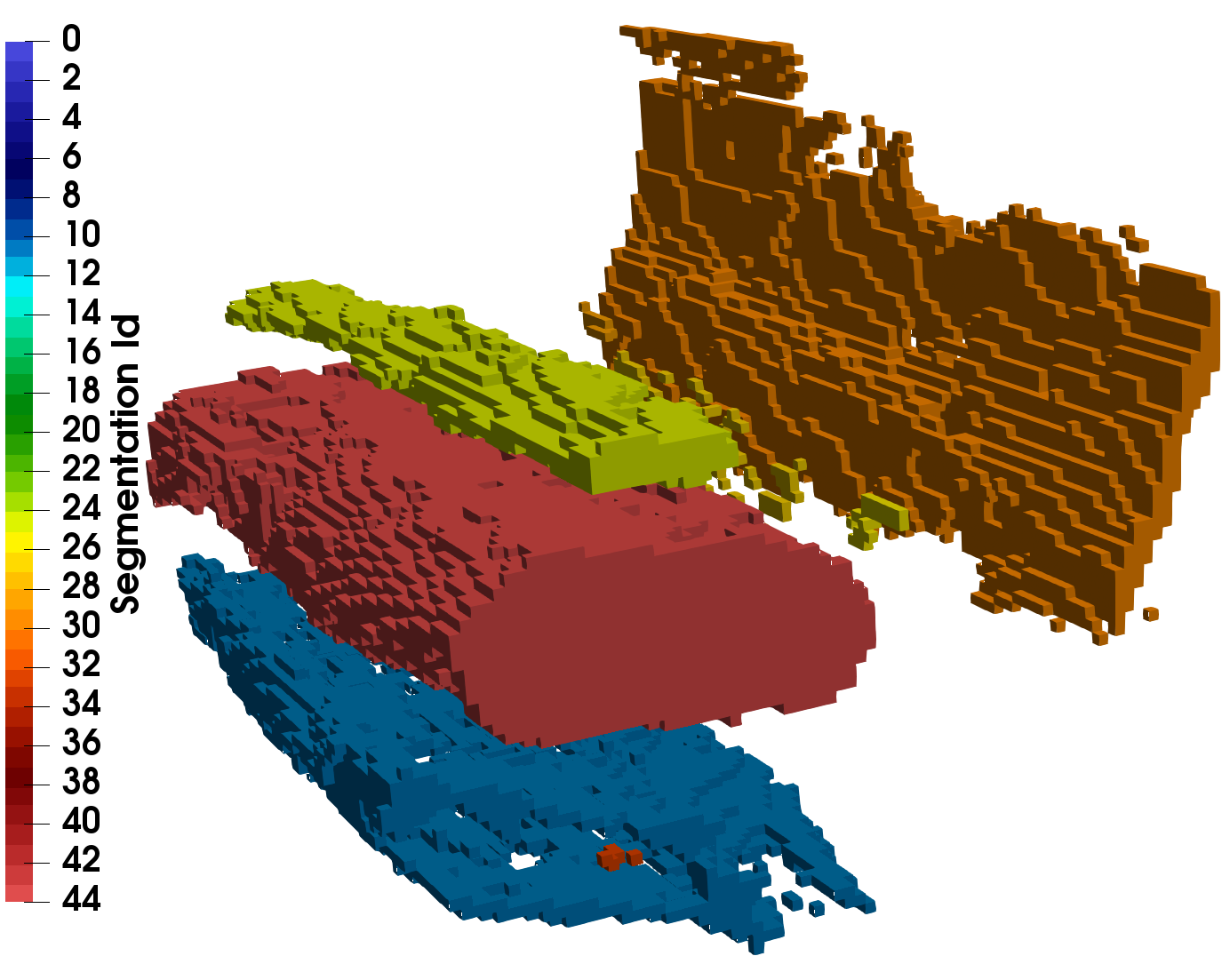}}%
		\subcaptionbox{Subdivided vorticity.  \label{fig-bvort-Seg2-Subdiv}}{\includegraphics[width=0.32\linewidth]{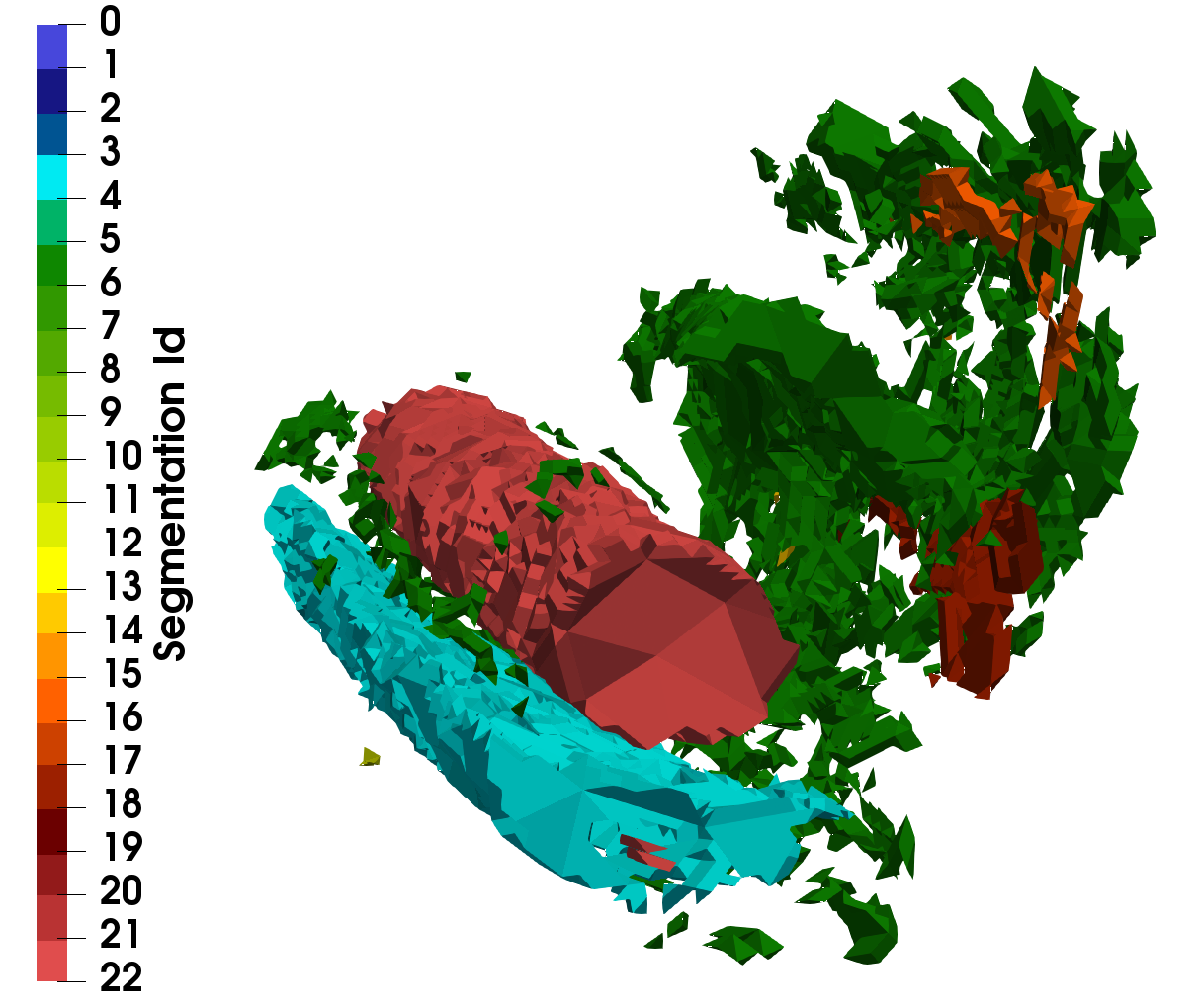}}%
		\subcaptionbox{L-SIAC vorticity. \label{fig-bvort-Seg2-LSIAC}}{\includegraphics[width=0.32\linewidth]{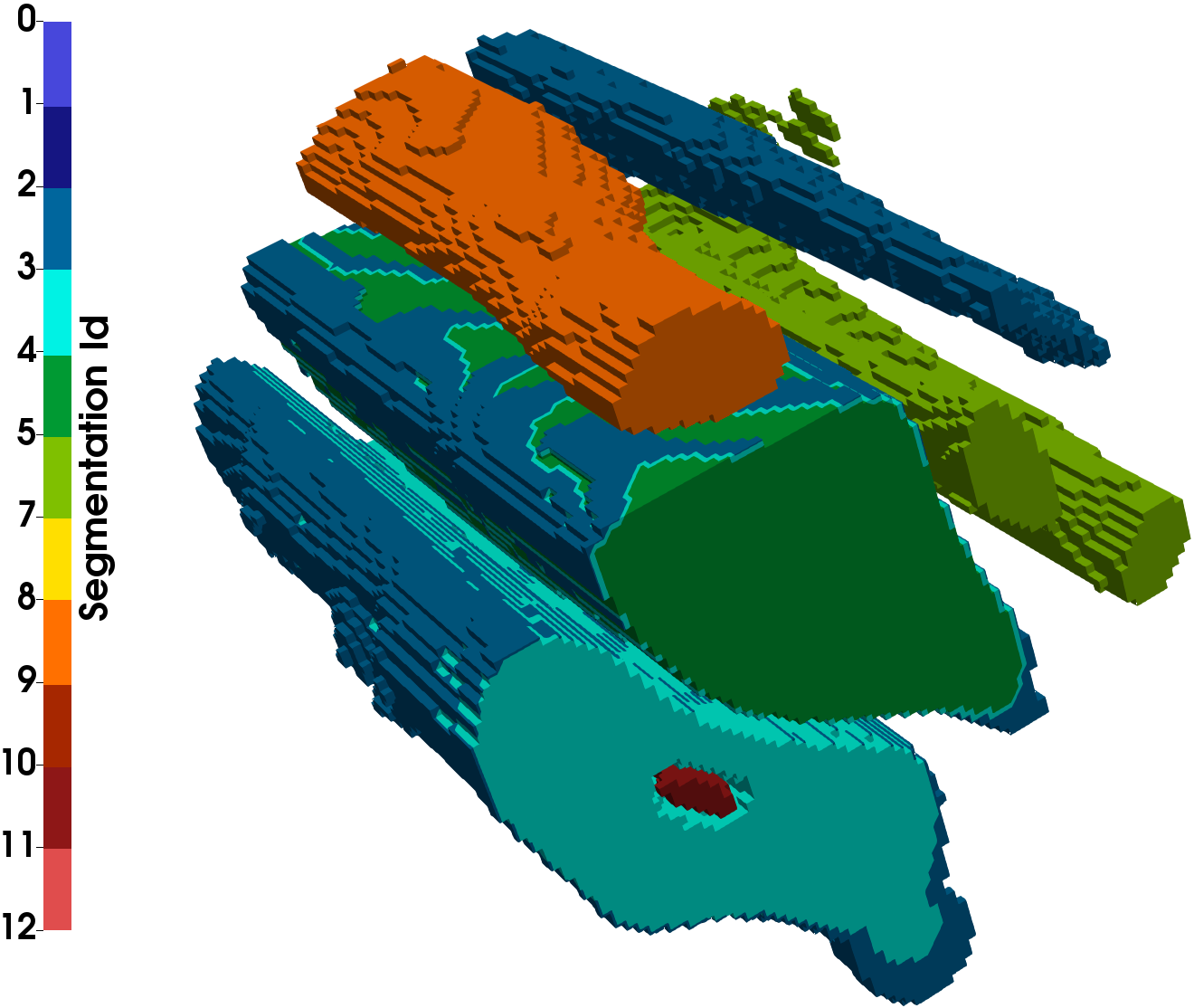}}
		\caption{Topological segmentation of counter-rotating vortex sampled using different methodologies discussed in the paper and by \grammar{filtering} the contour tree for segments \stgrammar{which}\grammar{that} resemble vortex-like structures.  The number, shape, and boundaries of the segments are different for the three techniques.\label{fig-bvort-Seg2-compare}}
			
}

\vgtcinsertpkg
\usepackage[labelfont=sf]{subcaption}
\begin{document}
	\setlength{\abovedisplayskip}{2pt}
	\setlength{\belowdisplayskip}{2pt}
	\setlength{\abovedisplayshortskip}{0pt}
	\setlength{\belowdisplayshortskip}{0pt}
	
	\maketitle
	
\section{Introduction}
\label{sec-intro}

The development of robust, efficient solvers that utilize high-order finite element methods (HO-FEM),
also sometimes classified as spectral/$hp$ element methods, is an area of considerable interest to the simulation community at
present. The use of higher order polynomial expansions within elements carries a number of benefits\grammar{,} as seen 
from two main perspectives. \grammar{1)} Numerically, these methods exhibit far lower levels of numerical dispersion and dissipation at
higher polynomial orders. This makes them a particularly well-suited approximation choice in areas such as computational fluid dynamics, where the
accurate time-advection of energetic structures such as vortices is a key concern (e.g., \cite{vincent2014,lombard-2016a}). 
\grammar{2)} Given current hardware trends, the most appealing property of these methods in recent years has been their computational performance.
Although the cost per degree of freedom in terms of algorithmic floating point operations (FLOPS) increases substantially with polynomial order, 
the use of higher order expansions leads to formulations of the underlying equations of state that involve dense, 
compact kernels for key finite element operators, such as inner products and derivatives. This is important from the perspective of modern
hardware, where increasingly the bottleneck in performance is memory bandwidth as opposed to the clock speed of processors. 
The underlying arithmetic intensity of the algorithm at hand (i.e., the number of floating-point operations performed
for each memory operation) is therefore key to attaining optimal
performance. This is where high-order methods \grammar{have} a significant advantage over
lower order methods (e.g., \cite{fehn-2018,Moxey2019}).

These trends indicate that there will continue to be an increasing need for visualization of high-order finite element solutions.  
Topological analysis has a rich history of providing methods to extract and visualize structural properties in simulation data.  
For example, one can use topology to study vortex breakdown patterns~\cite{tricoche2004visualization}, 
vortex merging~ \cite{al2012vortex}, and shedding patterns. 
These techniques are built with the assumption of continuity in the data, but as presented in \cite{jallepalli2017treatment}, one of the challenges associated with the visualization, and in this case the topological analysis, of HO-FEM fields (and their corresponding derived fields) is their lack of continuity at element interfaces.  
Fundamentally, topological analysis is based on extracting properties in a continuum: these properties are designed to be invariant to deformation but sensitive to topological events such as cutting or splitting.  At first glance, it seems an impossible task to analyze a discontinuous function using standard topological properties, but we note that even if the solution is represented with a discontinuous approximation, the simulated phenomena of interest \grammar{are} often assumed to be continuous (from the physical modeling perspective).

Besides the issues with discontinuous data, \grammar{a practical problem is that most} of the mathematics and implementations of topological analysis for scalar fields focus on piecewise linear interpolants.  There has been limited work in pushing the limits beyond piecewise linear \grammar{interpolants}, and the typical extensions consider only a specific interpolant~\cite{norgard2013robust,pascucci2004parallel,acharya2015parallel} as opposed to general methodologies for high-order interpolants~\cite{carr2009representing, nucha2017computing}.  \response{Chen et al. considered the effects of interpolants on Morse decompositions of vector fields, but did not explicitly use the high-order, discontinuous methods we target~\cite{Chen:MCG:2008}.}
This \grammar{lack of prior art} is justifiable, as piecewise linear representations have the advantage that they help to limit the space of combinatorial possibilities within a given mesh (e.g., in a piecewise linear element, there \grammar{may be only one} critical point).  The typical assumption is that any more complex functions can be represented by simply increasing mesh resolution.  Nevertheless, using increased mesh resolutions can require an infeasible amount of storage overhead, and as we show in our work, they may not always achieve an ideal characterization of the shape of topological features.  

To bridge the gap between HO-FEM simulations and topological methods, we \grammar{need to either} transform the data or redesign 
(and re-verify) the topological techniques. 
In this work, we consider the first option.
Specifically, we explore three main methodologies for converting the HO-FEM data to match the input requirements of topological tools.
We then conduct an empirical study to analyze the effect of these methodologies on downstream topological analyses.
Two of the methodologies we consider are sampling the data on a regular grid (avoiding discontinuities) and subdividing the elements on the simulation mesh (avoiding discontinuities and judiciously producing low-order elements). These methodologies are commonly used due to their ease of implementation and lower computational cost, but they have the disadvantage of creating undesirable artifacts~\cite{jallepalli2017treatment}. 
We use these methodologies to provide baselines to discuss the interplay between the sampling parameter and its effect on the intensity of the discontinuities and interpolation artifacts. We also provide insight into selecting the sampling parameter to help reduce the artifacts.  Simultaneously, we consider using topological analysis to filter out the artifacts as well as identifying what significant features may be retained. 
The third methodology we consider is the Line-SIAC (L-SIAC) filter, which is computationally expensive, but has been successful in identifying a more extended range of features compared to the other two techniques.

In this paper, we demonstrate that through the use of the L-SIAC filtering methodology, we can transform HO-FEM data so as to take full 
advantage of the suite of techniques and intuitions developed within the topological analysis field.  Furthermore, we provide a set of 
insights into how these tools -- L-SIAC and topological analysis -- interact, which helps show how they might be used together in the future.

\response{Specifically, our contributions are
	\begin{itemize}
		\item An empirical study to evaluate data transformations of high-order scalar fields that enable the use of existing topological analysis tools;
		\item Experimentation using three different data transformation methods and two different datasets; and
		\item Evaluation of the downstream effects in topological structures including persistence diagrams and curves, the position and number of critical points, and the segmentation based on the contour tree.
	\end{itemize}
	These contributions are the first step towards a better understanding of existing issues between data interpolant and topological feature extraction. Such a study may help motivate the development of new topological techniques that can adapt to HO-FEM data without the need to first transform it.
}
\section{Previous Work}
\label{sec-prework}
\response{In this section, we review the main concepts of topological analysis in regard to scalar fields for linear and high-order polynomial data. The current challenges are to extend these techniques to HO-FEM data and to introduce the data transformation methodologies. These methodologies transform the data into continuous piecewise linear data to enable topological analysis.}


\subsection{Topological Analysis}

We review the main concepts and structures of topological analysis that we employ in our empirical study.  For a more complete introduction to the mathematics of computational topology, see Edelsbrunner and Harer~\cite{edelsbrunner09}.  For details of their usage, Heine et al.~provide a recent survey of topology-based methods in visualization and analysis~\cite{heine16}. 

Topological analysis provides a collection of tools to extract features from  data that characterize its structure.  Moreover, it also provides mechanisms to rank and filter these features, thus offering an analyst the ability to summarize a scalar field at multiple scales.  While this general framework is applicable to a variety of data modalities, a key area of focus in the visualization community is the analysis of scalar fields. 
From scalar field data, one can \grammar{visualize the} structure through collections of important feature points (i.e.,~critical points~\cite{banchoff70}), graph structures that encode level sets (i.e.,~Reeb graphs~\cite{pascucci07, biasotti08, tierny_vis09} and contour trees~\cite{carr2003computing}), and segmentations of gradient flow behavior (i.e.,~Morse-Smale complexes~\cite{gyulassy_vis08, Defl15}).  
The structures have seen direct usage in visualization of scalar fields, for example in selecting isosurfaces~\cite{vanKreveld97}, topologically-guided simplification~\cite{tierny_vis12}, feature tracking~\cite{sohn06}, transfer function design~\cite{weber07}, isosurface simplification~\cite{CARR201042}, and  similarity estimation~\cite{thomas14}. 
Even though topological analysis is a purely mathematical framework, a key reason for its success has been the mapping of these mathematical abstractions to application-specific features, as demonstrated by its use across a wide variety of domains\grammar{,} including astrophysics~\cite{shivashankar2016felix, sousbie11}, battery design~\cite{gyulassy_vis15}, combustion~\cite{bremer2010analyzing, Day09, gyulassy_ev14}, chemistry~\cite{cazals03, chemistry_vis14}, porous media~\cite{gyulassy07}, turbulence~\cite{laney_vis06}, and vortex extraction~\cite{kasten_tvcg11}.

In this paper, we focus on level set topology as our main tool of interest.  
Let $f: \domain \rightarrow \range$ be a continuous scalar field, defined on a manifold $\domain$ that is \grammar{usually} a subset of $\R^2$ or $\R^3$.
Given a scalar value $i \in \range$, called the \emph{isovalue}, we can study the structure of a scalar field through its \emph{level sets}, $f^{-1}(i) = \{ p \in \mathcal{M} ~ | ~ f(p) = i \}$.  Level sets encode the subset of $\domain$ that is the preimage of the value $i$.  They are also called isosurfaces, and they are the generalization of contour lines on a topographic map. \grammar{Although} individual level sets are often descriptive, the relationships between level sets when $i$ changes describe important relationships.

The points of $\domain$ where the gradient, $\nabla f$, is equal to zero are the \emph{critical points} of $f$~\cite{banchoff70}.  Critical points are particularly important as a building block for studying level sets, as they correspond to values where level sets undergo topological changes.
\grammar{In particular, critical points may be classified by their \emph{index} $\Index$, which equals $0$ for minima and $d$ for maxima points.  Critical points with indices between $1$ and $d-1$ are called saddles. } 

If we imagine sweeping through all possible isovalues $i$, extremal critical points (minima and maxima) are positions where level set components are created and destroyed, and saddles indicate where level set components merge and split.  During such a sweep, we can imbue critical points with a notion of scale.  To compute this notion, we apply Elder's rule~\cite{edelsbrunner09}, \grammar{which} pairs critical points such that each critical point appears in only one pair $(c_i, c_j)$ with $f(c_i) < f(c_j)$ and $\Index(c_i) = \Index(c_j) - 1$.  The height of the pair $p = f(c_j) - f(c_i)$\grammar{, }called the \emph{persistence}, encodes the \grammar{life span} of the level set created at $f(c_i)$ and destroyed at $f(c_j)$~\cite{cohen-steiner05, edelsbrunner02}.  Persistence introduces a measure that is frequently used to distinguish topological ``noise''---features that \grammar{live only} at small scales---from more significant, \grammar{highly persistent} features.  More formally, the field of persistent homology studies the evolution of the homology groups characterized by critical points.

A topological abstraction known as the \emph{persistence diagram} offers a view of the distribution of critical points of $f$.  The persistence diagram $\persistenceDiagram(f)$ embeds each pair $(c_i, c_j)$ in the plane such that its horizontal coordinate equals $f(c_i)$, and the vertical coordinate\grammar{s} of $c_i$ and $c_j$ are $f(c_i)$ and $f(c_j)$.   In $\persistenceDiagram(f)$, there is a vertical bar for the pair, and the height of these vertical bars is the persistence.  We can also encode the distribution of critical points as a simple curve known as the \emph{persistence curve}, which shows how many pairs of critical points in $f$ possess a given persistence threshold or lower. Persistence diagrams and persistence curves offer \grammar{high-level} building blocks to see the range of scales at which features exist as well as to guide tools such as topological simplification (Figure~\ref{fig-solution-ttk} shows an example of both). 

Building further on critical points and persistence, the \emph{Reeb graph} is a topological abstraction that segments $\domain$ into regions where the connectivity of $f^{-1}(i)$ does not change, clustering points if they belong to the same connected component of the level set.  Let $f^{-1}(f(p))_p$ be the connected component of $f^{-1}(f(p))$ containing $p$. The Reeb graph $\reebGraph(f)$ is a one-dimensional simplicial complex defined as the quotient space $\reebGraph(f) = \domain / \sim$ by the equivalence relation $p_1 \sim p_2$, which holds if $p_2 \in f^{-1}(f(p_1))_{p_1}$. 
The contour tree (the loop-free variant of $\reebGraph(f)$) is often preferred in the context of visualization as it is efficient to compute~\cite{carr2003computing, tarasov98} when we can guarantee that $\domain$ is simply connected.  

Reeb graphs and contour trees have been well studied by the visualization community, offering a number of approaches for their computation in two- and three-dimensional settings~\cite{biasotti08, cole03, doraiswamy12,  doraiswamy13, gueunet_ldav16, parsa12, pascucci07, patane08, tierny_vis09}.  In our work, we construct segmentations of the input domain based on the contour tree, and rely on the fact that these segmentations correspond to regions of interest in the data (specifically, we consider vortex identification).  We rely on persistence to help us identify which portions of this segmentation are significant by helping a user navigate the space of interesting topological features.

\paragraph{Topological Analysis for Higher Order Data} 

Thus far, topological analysis has been utilized for higher order data in only limited settings because the recovery of topological structures for \grammar{low-order} representations has presented significant computational challenges. Most techniques assume piecewise linear interpolants or utilize Forman's discrete Morse theory~\cite{forman98}; in both cases the combinatorial complexity of topological structures stay\grammar{s} manageable.  Notable exceptions consider quadratic~\cite{dillard2009topology}, bilinear~\cite{norgard2013robust}, and piecewise trilinear interpolants~\cite{pascucci2004parallel,acharya2015parallel}.  Carr and Snoeyink propose an abstract framework for interpolants of arbitrary order~\cite{carr2009representing}.  In general\grammar{,} though, implementing a complete topological analysis is challenging for higher order interpolants \grammar{because }\response{finding critical points involves determining roots of equations of degree greater than five which is analytically intractable}.  A notable exception is the work of Nucha et al.~that considers contour tree computations for 2D piecewise polynomial functions~\cite{nucha2017computing}. \grammar{This} work advances the state-of-the-art, \grammar{but} it still \grammar{offers only} one facet of the analysis (contour trees). \grammar{Most} importantly, it does not consider cases where higher order discontinuous functions are used, such as HO-FEM derived fields and the discontinuous Galerkin method that we consider here.  Therefore, we see our work as offering a stopgap to answer the question of how best \grammar{to} make use of the widely available low\grammar{-}order tools as approximations for complex, high\grammar{-}order data that suffer from the problems mentioned above.  In such a setting, we cannot expect a full, verifiably correct topological extraction~\cite{etiene2011topology}, so instead we focus on characterizing the behavior of these approximations and their effects on downstream analysis.

\subsection{Data Transformation Methodologies}
We refer to the mapping from discontinuous high-order polynomial data to continuous piecewise linear data as our \emph{data transformation 
	methodology}. Due to their computational simplicity, two simple methods are commonly used as data transformations methods. 
The first \grammar{method} is to resample the original (unstructured) data onto a regular grid. \response{This resampling can induce minor persistent features}. The second \grammar{method is to subdivide the original simulation mesh hierarchically and then evaluate it}. In both cases, this sampling choice
can be shown mathematically to act as a filter that ``removes'' discontinuities from the data. Although increasing the sampling rate improves the accuracy of the filter, the effect of discontinuities in the original data also becomes more pronounced in the resultant field as \grammar{we continue} to add
more sampling points. Therefore, we expect that it \grammar{will be} challenging to pick an ideal sampling rate to balance these issues.

Recently, SIAC filters \grammar{have gained} popularity for post-processing HO-FEM due to their ability to increase smoothness at element boundaries (i.e.,~remove discontinuities) while maintaining the order of accuracy, and in many cases \grammar{they have been} shown to improve the accuracy of the solution.
Extending the work done by Bramble and Schatz \cite{bramble1977higher}, Cockburn et al.~\cite{cockburn2000post, cockburn2003enhanced} introduced SIAC filters for increasing the accuracy of discontinuous Galerkin methods with uniform spacing. 
Later Mirzaee et al.~\cite{mirzaee2012efficient, mirzaee_unstructured2013, mirzaee2011smoothness, mirzaee2014smoothness} introduced techniques to apply the SIAC filter at arbitrary points, thus extending its application to unstructured meshes. 
In \cite{ryan2015one,van2011position,nguyen2016nonuniform}, variations of the SIAC filter, called \grammar{one-sided filters}, are introduced to deal with boundaries and mathematical discontinuities in the solution (e.g., shocks in supersonic compressible flows). To improve accuracy on hexagonal meshes,  Mirzargar et al. \cite{mirzargar2017hexagonal} proposed hexagonal SIAC filters. Li et al. \cite{li2016smoothness} discussed effective ways to calculate the derivative quantities using  SIAC and one-sided SIAC filters.

SIAC filters are \grammar{also frequently} used to create visualization data.  Steffen et al.~used SIAC filters for improving the streamline integration through the discontinuous field~\cite{steffen2008investigation}. Walfisch et al.~used 1D SIAC filters on 2D discontinuous data to create continuous streamlines~\cite{WRKH09}. Later, Docampo-S\'anchez et al.~proved that the Line-SIAC filter (L-SIAC)\grammar{,} when applied to 2D and 3D simulation data\grammar{,} has all the properties \grammar{of} the SIAC filter~\cite{Sanchez2016multi}, but is more computationally efficient than the traditional tensor-product-based SIAC filter in 
multiple dimensions.  Jallepalli et al. compared the commonly used data transformation methodologies to the L-SIAC filter and showed that the L-SIAC filter is a better tool \grammar{for} creating continuous visualization data~\cite{jallepalli2017treatment}.  This work, in part, motivates us to study its effect on downstream analysis with topological tools.   Recently, Jallepalli and Kirby introduced algorithms that significantly improve the computational 
speed of L-SIAC filter when used to create data for visualization~\cite{jallepalli2019Efficient}.

\subsubsection{L-SIAC}
\label{subsubsec:lsiac}

A Line-SIAC filter (L-SIAC) is defined as a 1D SIAC filter rotated at an angle ($\theta$), scaled with characteristic length ($H$), and \grammar{convolved} 2D or 3D field data. Thus, the L-SIAC filter can be defined as
\begin{align*}
u^*(\vect{x}) &=  \int\limits_{-\infty}^{\infty}{ K_{H}(t) * u_h(\vect{x}- \Gamma( t)) dt } \\
\Gamma(t) &= (t\cos(\theta),t\sin(\theta)), \theta \text{ is constant}
\end{align*}
\noindent where $u^*$ is the postprocessed solution,  $u_h$ is the dG solution of degree $k$, $H$ is the characteristic length defined as $h(cos(\theta)+sin(\theta))$,  h is the uniform element size, and the SIAC kernel is defined along $\Gamma(t)$ by the parameter $t$ as 
\begin{align*}
K^{2k+1,k+1}_{H}(t) &= \sum\limits_{\gamma=-k}^{k} c_{\gamma}\bspline^{k+1}_{H}\big(t-\gamma\big) = \frac{1}{H}\sum\limits_{\gamma=-k}^{k} c_{\gamma}\bspline^{k+1}\big(\frac{t}{H}-\gamma\big) ,  \\
K^{2k+1,k+1}_{H}(t) &= \frac{1}{H}K^{2k+1,k+1}\big(\frac{t}{H}\big) \text{.} 
\end{align*}
For nonuniform meshes used in practical applications, dynamically adapting the characteristic length \cite{jallepalli2018adaptive} based on the neighborhood of the filter produced a more accurate solution. The B-splines used in the postprocess are well studied and can be computed using the recurrence relation.
\begin{align*}
\bspline^1  &= X_{[-1/2,1/2]},\\
\bspline^{k+1} &= \frac{1}{k} \bigg( \Big(t+\frac{k+1}{2}\Big)\bspline^{k}(t+\frac{1}{2}) +  \Big(\frac{k+1}{2}-t\Big)\bspline^{k}(t-\frac{1}{2})\bigg), k\geq 1 \text{.}
\end{align*}
The coefficients of the kernel, $c_{\lambda}$, can be found by using the property that the kernel must not destroy the accuracy of the approximation. More specifically, the coefficients reproduce polynomials of degree $2k$ by convolution. When using a symmetric B-spline kernel, we can solve the coefficients and store them for reuse.  All algorithms described  in this paper assume that a symmetric filter can be applied at the location where the L-SIAC filter is enforced. 

The result of applying the L-SIAC filter to either a continuous or discontinuous Galerkin field $u_h(\vect{x})$ is an updated field $u^*(\vect{x})$ that is of higher degree within each element {\em and} \grammar{that} has higher levels of continuity between elements (continuity as high as $k-1$ when using a $k^{th}$ degree filter). \grammar{We} can now sample this more accurate and more continuous representation of the data for input to topological analysis tools.

\section{Methods}
\label{sec-Methods}

For our empirical study, we use an analysis pipeline that transforms higher order input data to linear data, which is then analyzed through the lens of level set topology.  Our pipeline has multiple parameters that we discuss in this section.  We experiment with three different transformation filters in this analysis pipeline, each of which produce\grammar{s} linear representations of scalar fields.  We refer to these methods as ``sampled'' (for data sampled on a grid), ``subdivided'' (for data where we subdivide the input mesh), and ``L-SIAC'' (for data transformed using the L-SIAC filter).  For each method, we manually select parameters that help to provide the best possible transformation. 

Next, we employ the Topology ToolKit (TTK)~\cite{TiernyFLGM18} to extract topological features of our filtered data using ParaView~\cite{paraview}.  TTK requires piecewise linear representations of the input data, encoded on a simplicial mesh, and it can accept inputs as both unstructured input meshes \grammar{and} implicit triangulations of data sampled on regular grids.  As mesh resolution is intimately related to analysis quality, we experiment with multiple possible resolutions.

Finally, our analysis pipeline involves computing an overview of features with TTK, in particular extracting critical points and computing their persistence, as well as visualizing persistence diagrams and persistence curves.  We use this information to guide a segmentation of the data domain based on using the contour forests method for fast extract\grammar{ion} of contour trees~\cite{gueunet_ldav16}.  We consider multiple scales of features, based on a manual analysis of the persistence diagram to set persistence thresholds.  To restrict our analysis to features at a particular scale, we perform topological simplification \response{on the transformed data (i.e., continuous piecewise linear data)}, relying on the method of Tierny and Pascucci~\cite{tierny_vis12} to simplify the underlying scalar field \response{and reduce noise }while preserving topological features that are above a user-specified persistence threshold.

\begin{figure}[!ht]
	\centering
	\subcaptionbox{Original function \label{fig-solution}}{\includegraphics[width=0.32\linewidth]{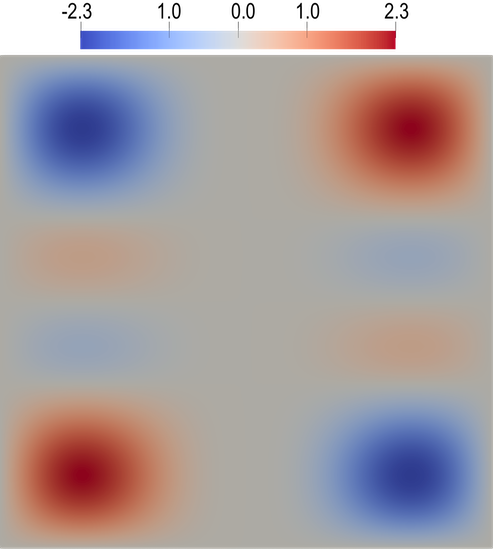}}%
	\hfill
	\subcaptionbox{Simulation mesh \label{fig-simMesh}}{\includegraphics[width=0.32\linewidth]{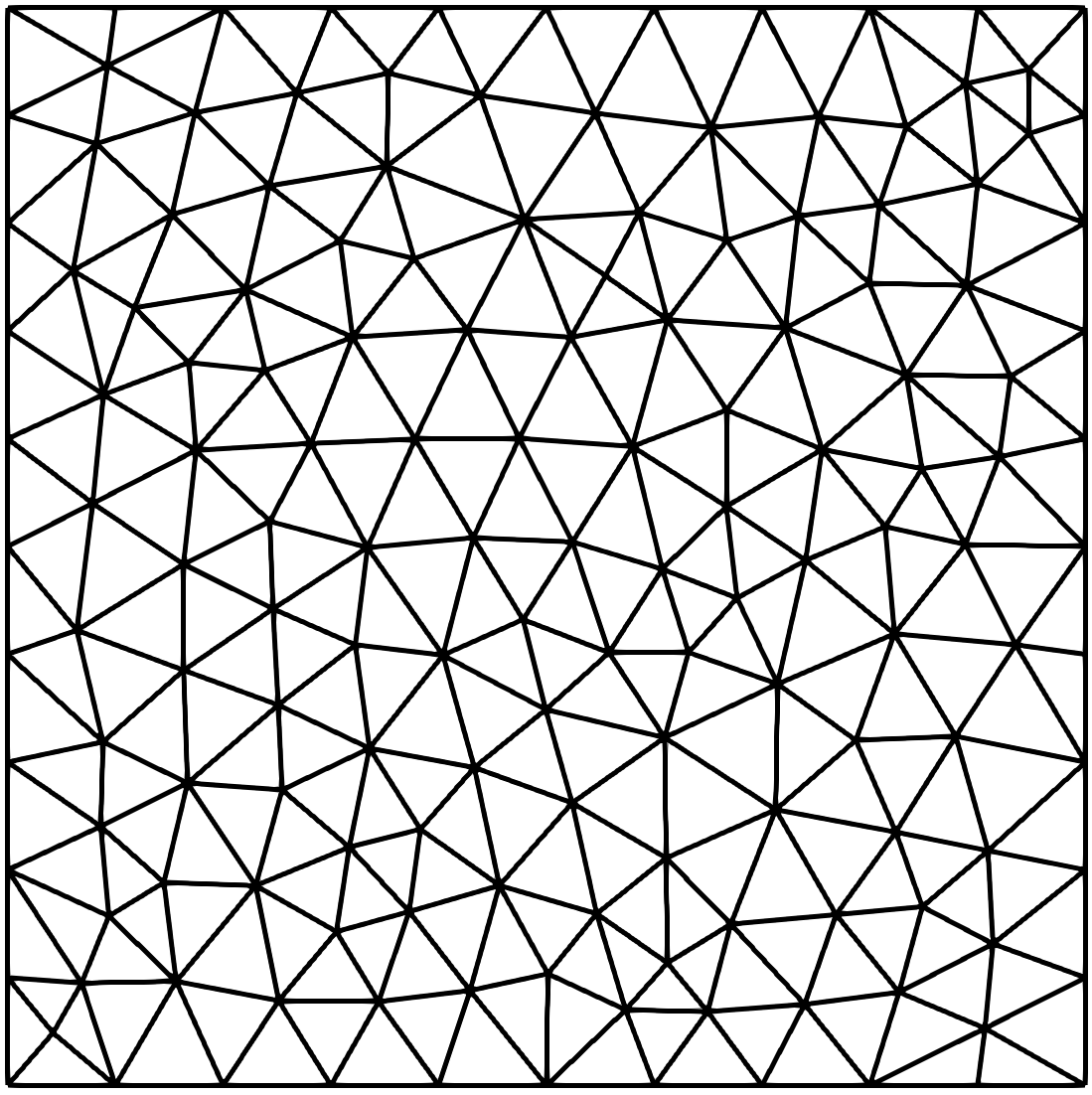}}%
	\hfill
	\subcaptionbox{Projected data \label{fig-projData}}{\includegraphics[width=0.32\linewidth]{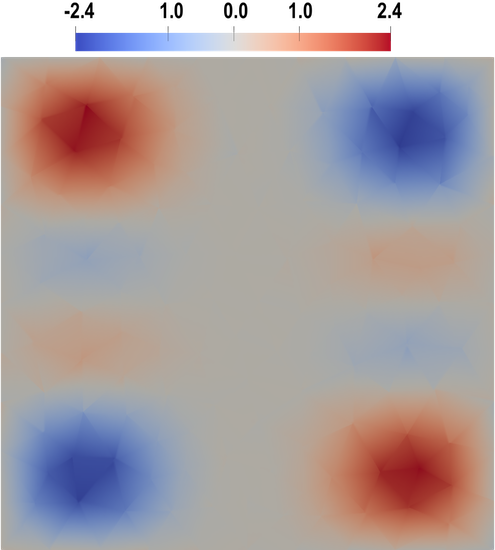}}%
	\vspace{-0.5em}
	\caption{The function $f(x,y)$, simulation mesh, and the projected data.} 
	\label{fig-solution-mesh}
	\vspace{-1em}
\end{figure}

\begin{figure}[!ht]
	\centering
	\subcaptionbox{Persistence curve \label{fig-solution-PC}}{\includegraphics[width=0.32\linewidth]{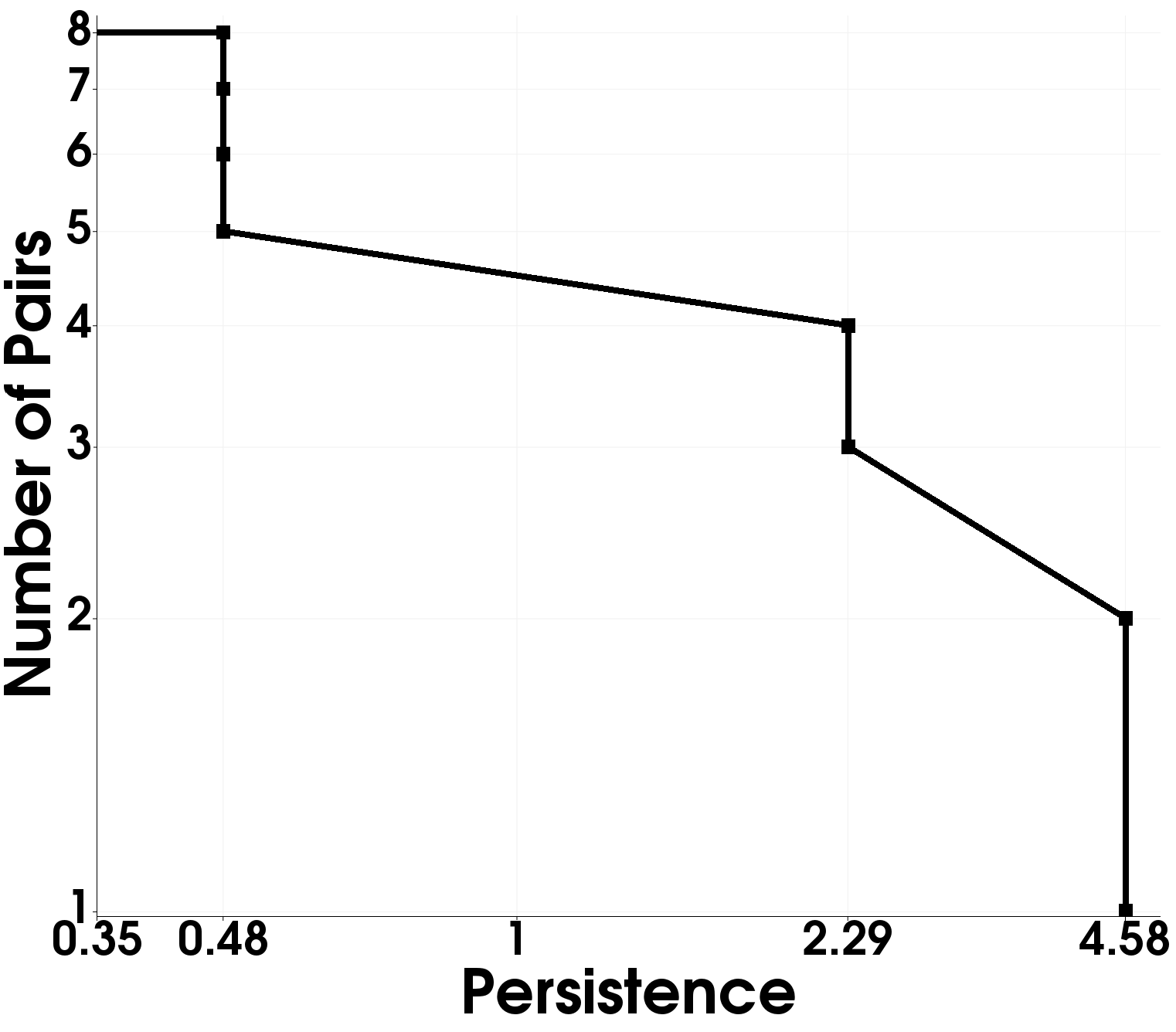}}%
	\hfill
	\subcaptionbox{Persistence diagram }{\includegraphics[width=0.32\linewidth]{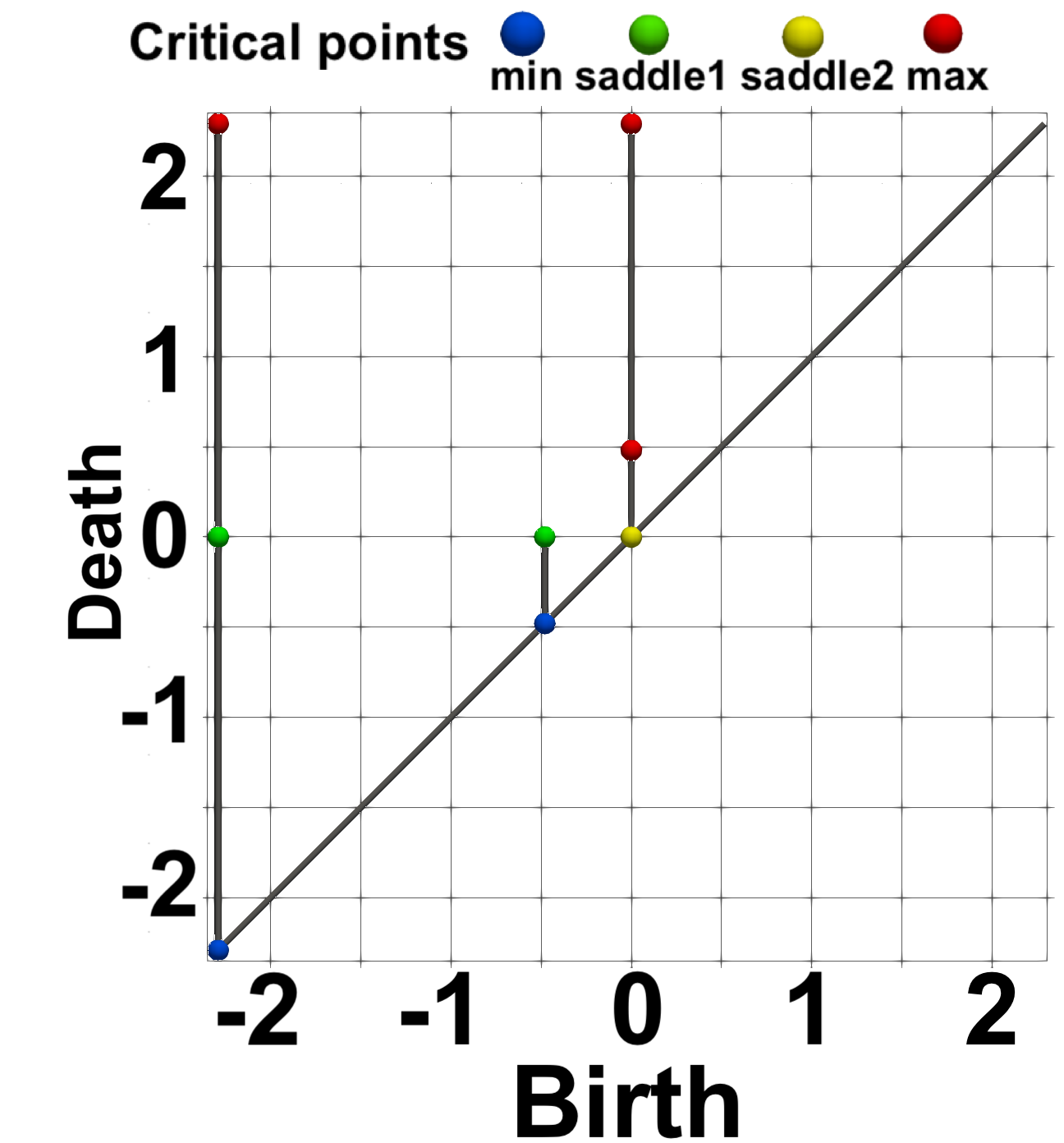}}%
	\hfill
	\subcaptionbox{Segmentation}{\includegraphics[width=0.32\linewidth]{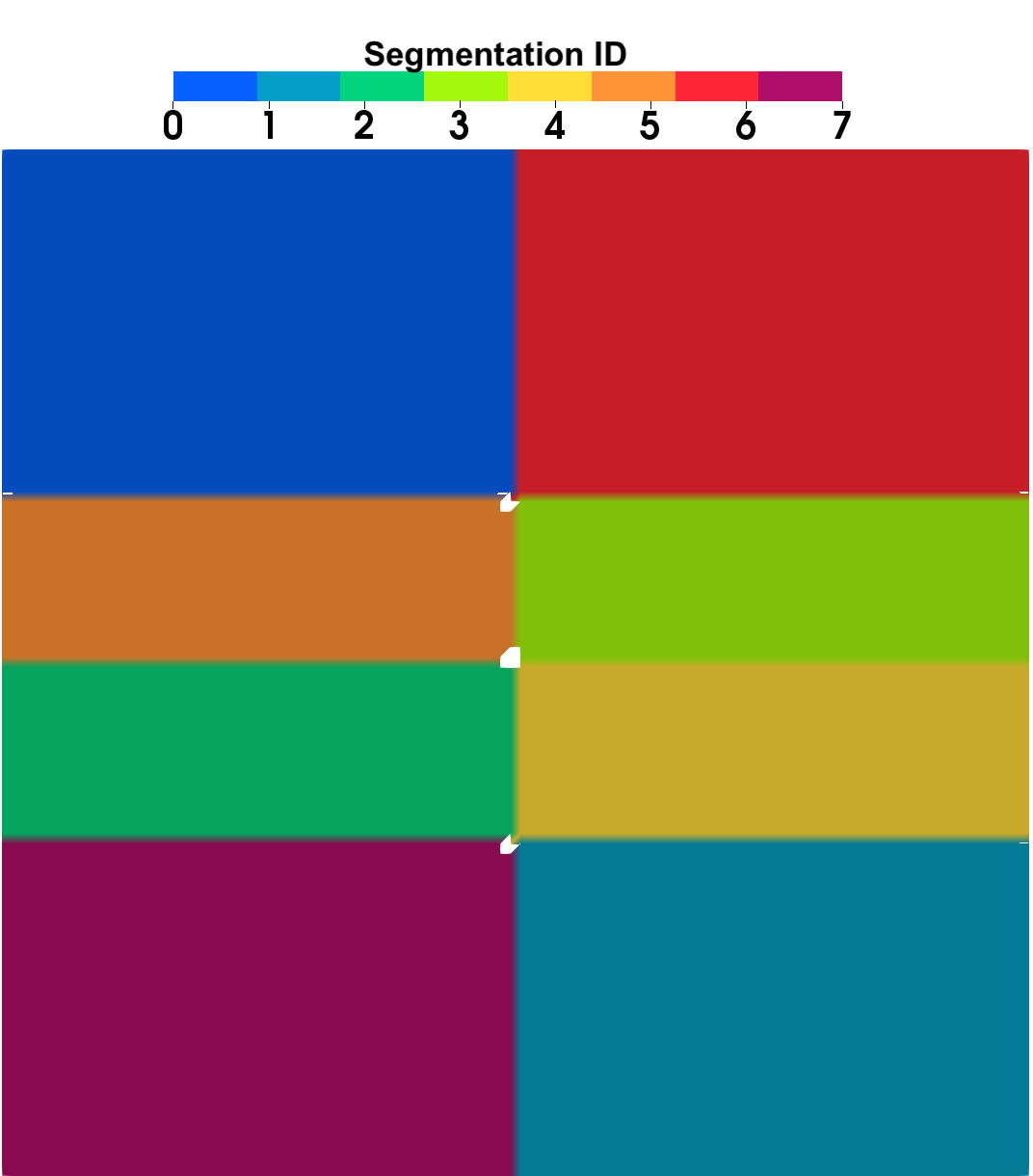}}
	\vspace{-0.5em}
	\caption{The persistence curve, the persistence diagram, and the segmentation of the original function $f(x,y)$ (Figure~\ref{fig-solution}).}
	\label{fig-solution-ttk}
	\vspace{-1em}
\end{figure}

As an example to describe our analysis pipeline, we consider a simple, 2D function
\[f(x,y) = (sin(2\pi x) + \frac{1}{2}sin(4\pi x) )(sin(2\pi y) + sin(4\pi y))\] which is asymmetric in the $x$ and $y$ directions.  It also contains topological features that persist at two different sets of scales. The first set of features with maximum persistence  (two maxima and two minima) \grammar{is} located near the corners of the domain, and the second set of features with lower persistence \grammar{is} located between the first set of features along the $x$-axis as shown in Figure~\ref{fig-solution}. 
To create a ground truth comparison, we sample $f(x,y)$ at a sufficiently high resolution to capture its topological features using TTK, as shown in its persistence diagram, curve, and segmentation using the contour tree in Figure \ref{fig-solution-ttk}. 
This decomposes the domain into mainly 8 regions, \grammar{each} associated with \grammar{a} leaf of the contour tree for the 8 extrema. \grammar{The} small flat region near the center where the level sets merge at saddles accounts for the 3 small white holes (since we exclude segments associated with interior arcs).

To create a higher order representation of $f(x,y)$, we project it on an unstructured triangular element mesh (Figure \ref{fig-simMesh}) over the domain $[0,1]\times[0,1]$ with degree $2$ polynomial expansions on each element (hereafter referred to as the simulation mesh).  
With this simple example, we can mimic the elementwise discontinuities \grammar{we get} when computing derived fields from high-order FEM data. In the following subsections, the projected data (shown in Figure \ref{fig-projData}) is transformed using our three different filtering methodologies to create \grammar{low-order} data suitable for input to TTK.

\subsection{Transformation by Sampling to a Grid\\(Method: Sampled)}
\label{subsec-methods-grid}

Our first transformation methodology is to sample the simulation data on an equispaced grid and then implicitly triangulate this grid using TTK.  In this technique, sampling the projected data acts as a crude filter with only one free parameter: the sampling rate itself.  This transformation thus applies no special considerations for the discontinuous piecewise polynomial data; it simply overlays a continuous piecewise linear mesh and interpolates the higher order data using the nearest element.   

\begin{figure}[!ht]
	\centering
	\includegraphics[width=\linewidth]{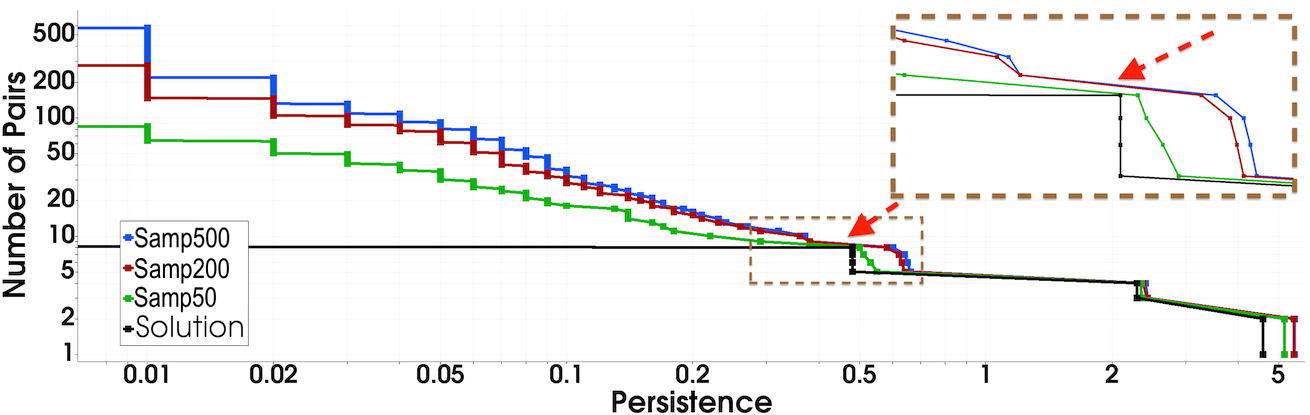}
	\vspace{-0.5em}
	\caption{Persistence curves of the sampled data (at three resolutions) as compared to the original function solution.\label{fig-sampGrid-PC}}
	\vspace{-1em}
\end{figure}

Figure~\ref{fig-sampGrid-PC} shows the persistence curves for the projected data sampled using different resolutions ($50\times 50$, $200\times 200$, and $500\times 500$ referred to as Samp50, Samp200, and Samp500, respectively). 
\response{The average time taken to compute vorticity at each location on a machine with a $2.4$ GHz (Intel CPU E7-4870) processor is $25$ microseconds.}
We were surprised to discover that an increase in the sampling rate increased the number of low persistence pairs, which is counterintuitive to the idea that by simply using a higher resolution mesh we can approximate data better.  This \grammar{increase of low-persistence pairs} is a direct byproduct of hav\grammar{ing} additional resolution (which allows more critical points to exist), \grammar{and} this added resolution near the discontinuities creates low-persistence perturbation (resulting in the staircase artifacts on the left side of Figure \ref{fig-sampGrid-PC}). Fortunately, using the persistence curve of the actual solution as a guide, we \grammar{can} classify all the persistence pairs to the right of the red arrow in Figure~\ref{fig-sampGrid-PC} as significant features in the solution. 

\begin{figure}[!ht]
	\centering
	\subcaptionbox{Persistence Diagram: Samp50 \label{fig-sampGrid-PD-R50}}{\includegraphics[width=0.32\linewidth]{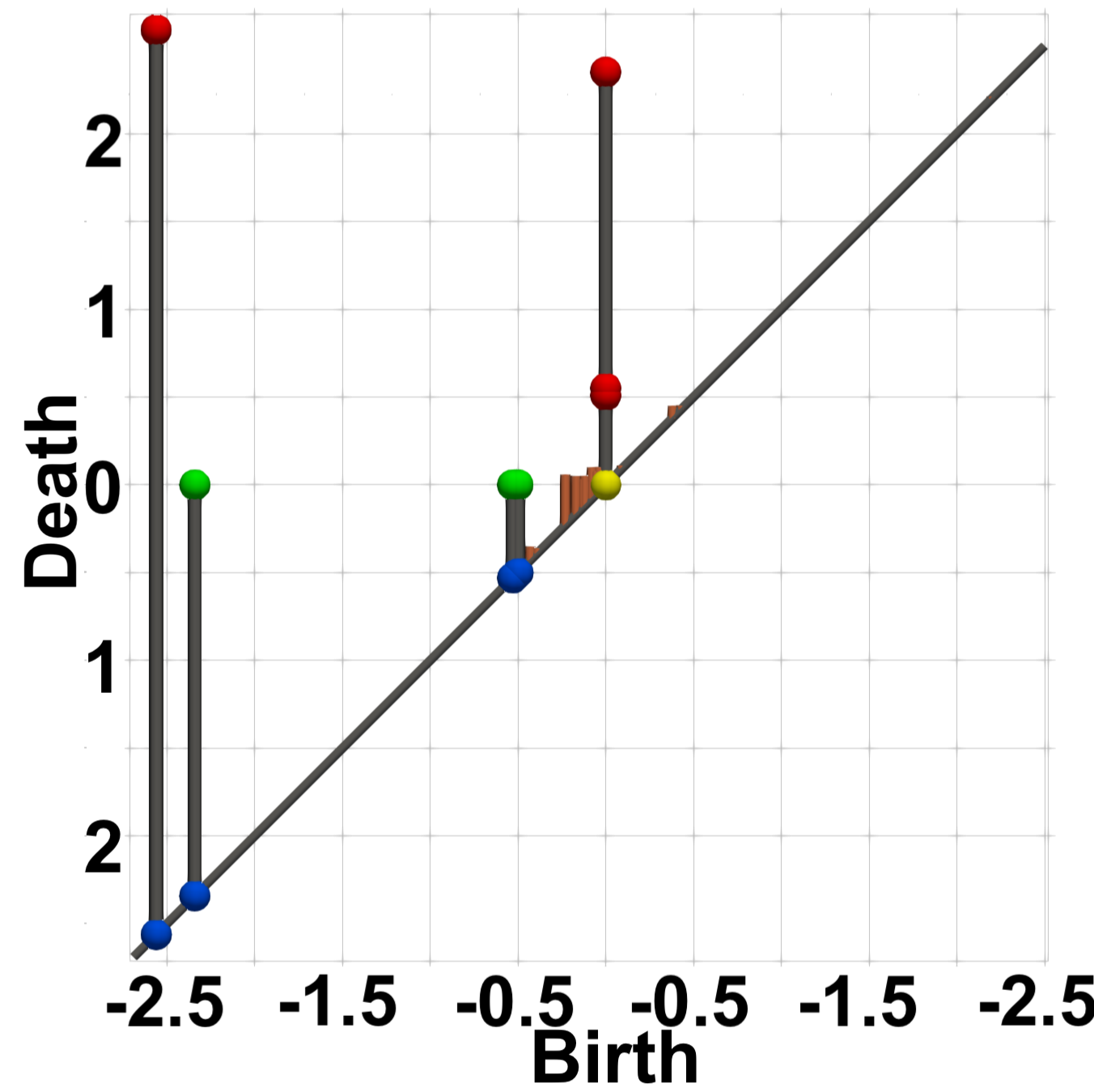}}
	\hfill
	\subcaptionbox{Persistence Diagram: Samp200 \label{fig-sampGrid-PD-R200}}{\includegraphics[width=0.32\linewidth]{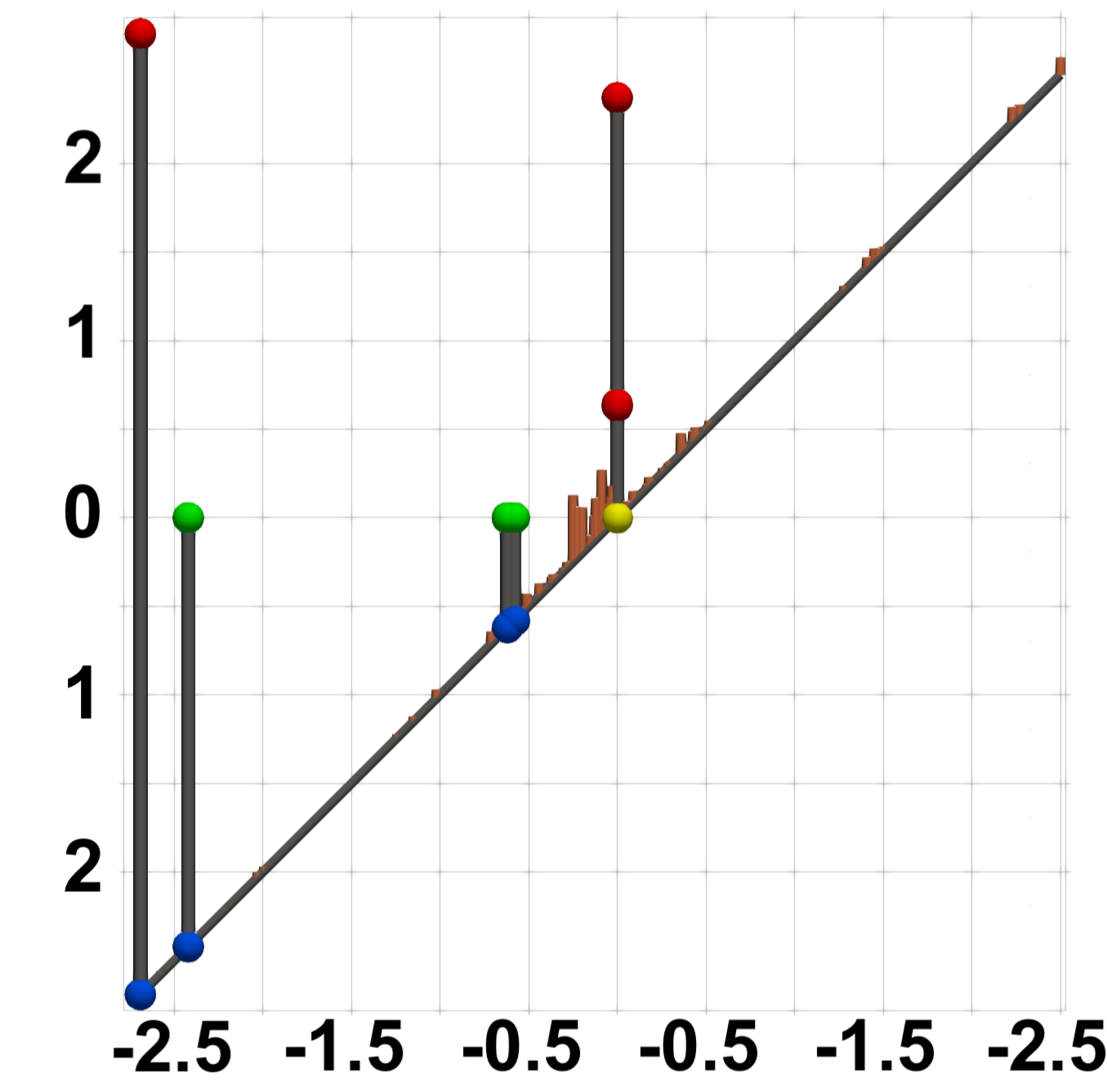}}
	\hfill
	\subcaptionbox{Persistence Diagram: Samp500 \label{fig-sampGrid-PD-R500}}{\includegraphics[width=0.32\linewidth]{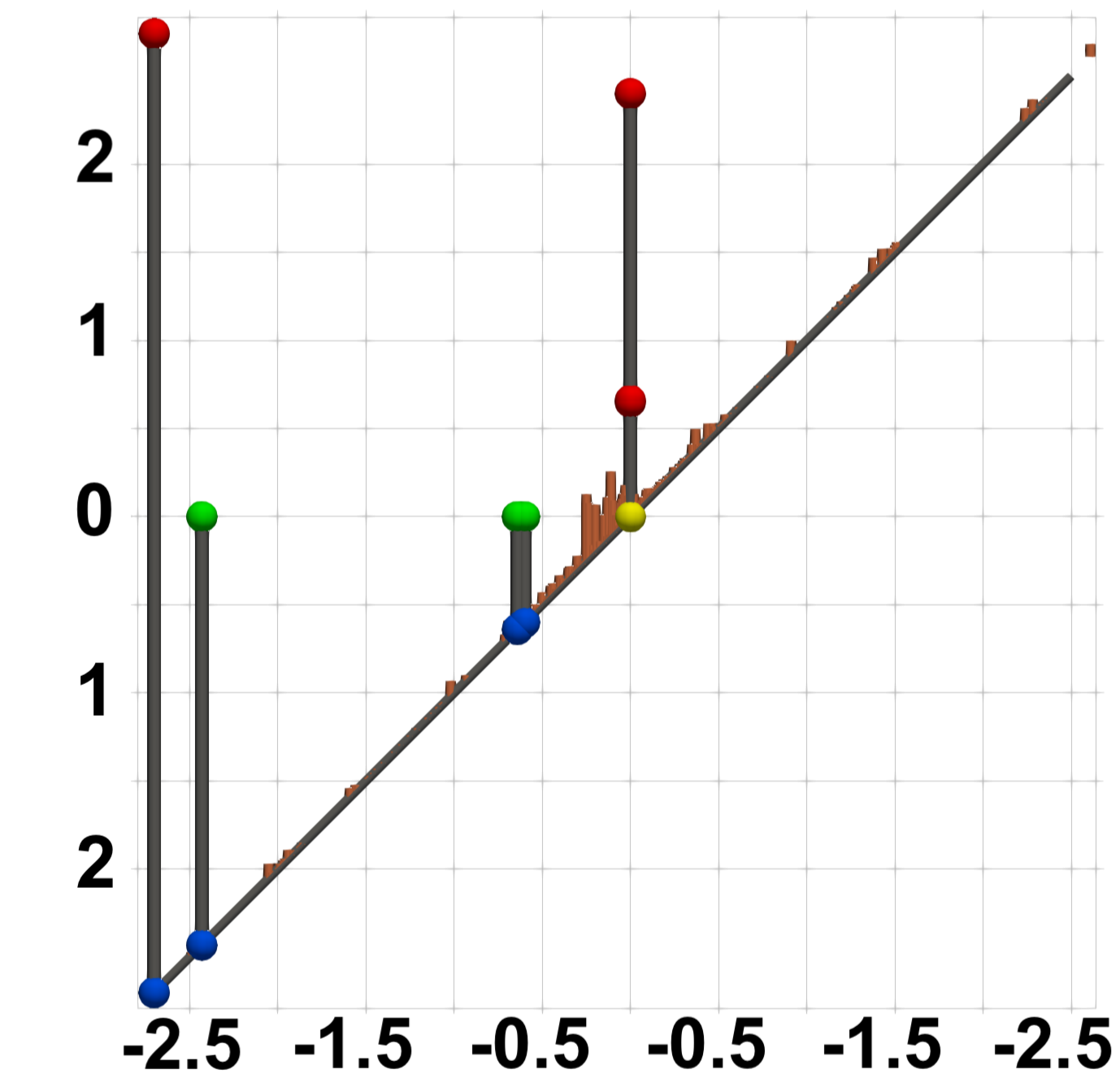}}
	
	\subcaptionbox{Segmentation: Samp50 \label{fig-sampGrid-Seg-R50}}{\includegraphics[width=0.32\linewidth]{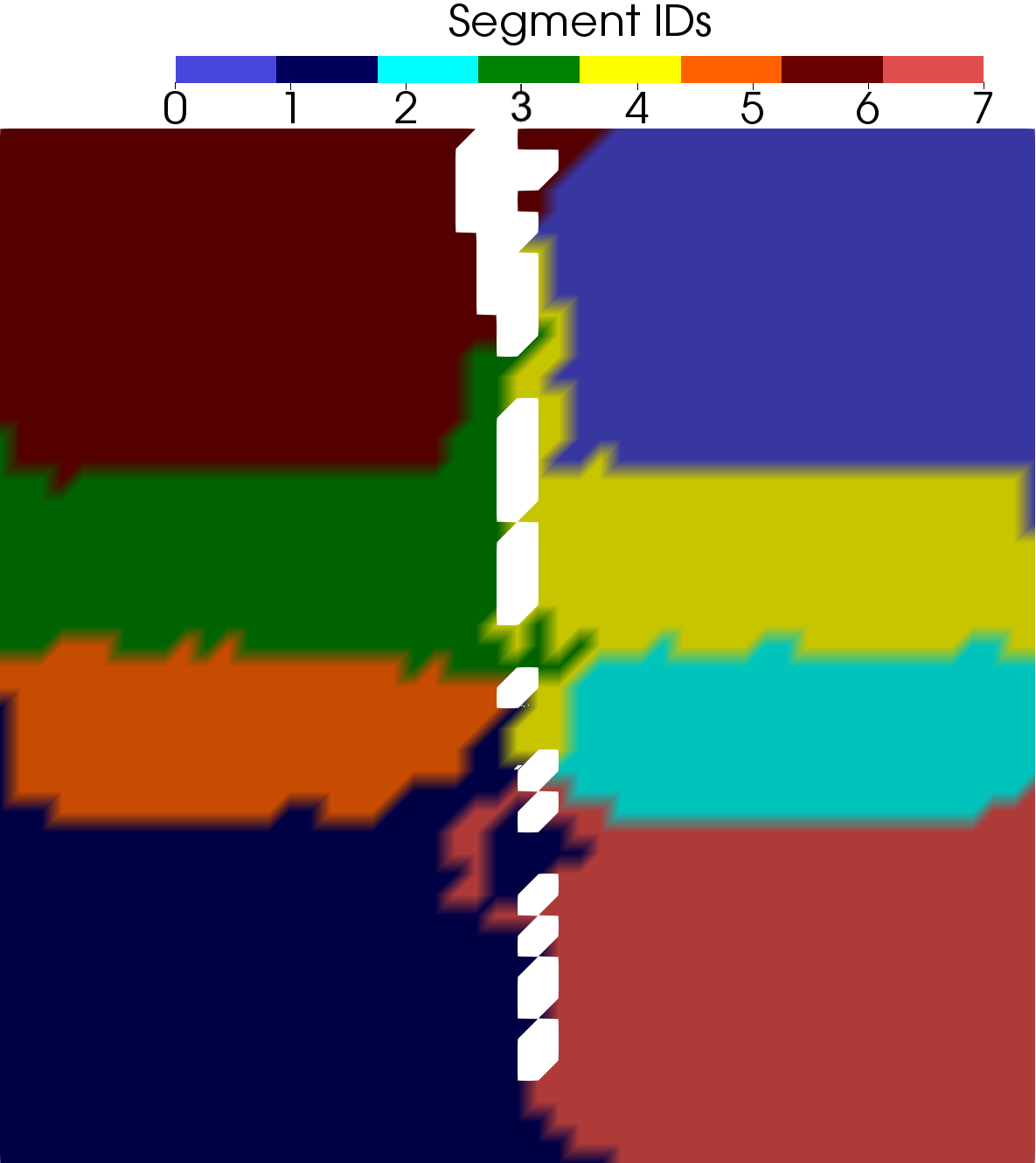}}
	\hfill
	\subcaptionbox{Segmentation: Samp200 \label{fig-sampGrid-Seg-R200}}{\includegraphics[width=0.32\linewidth]{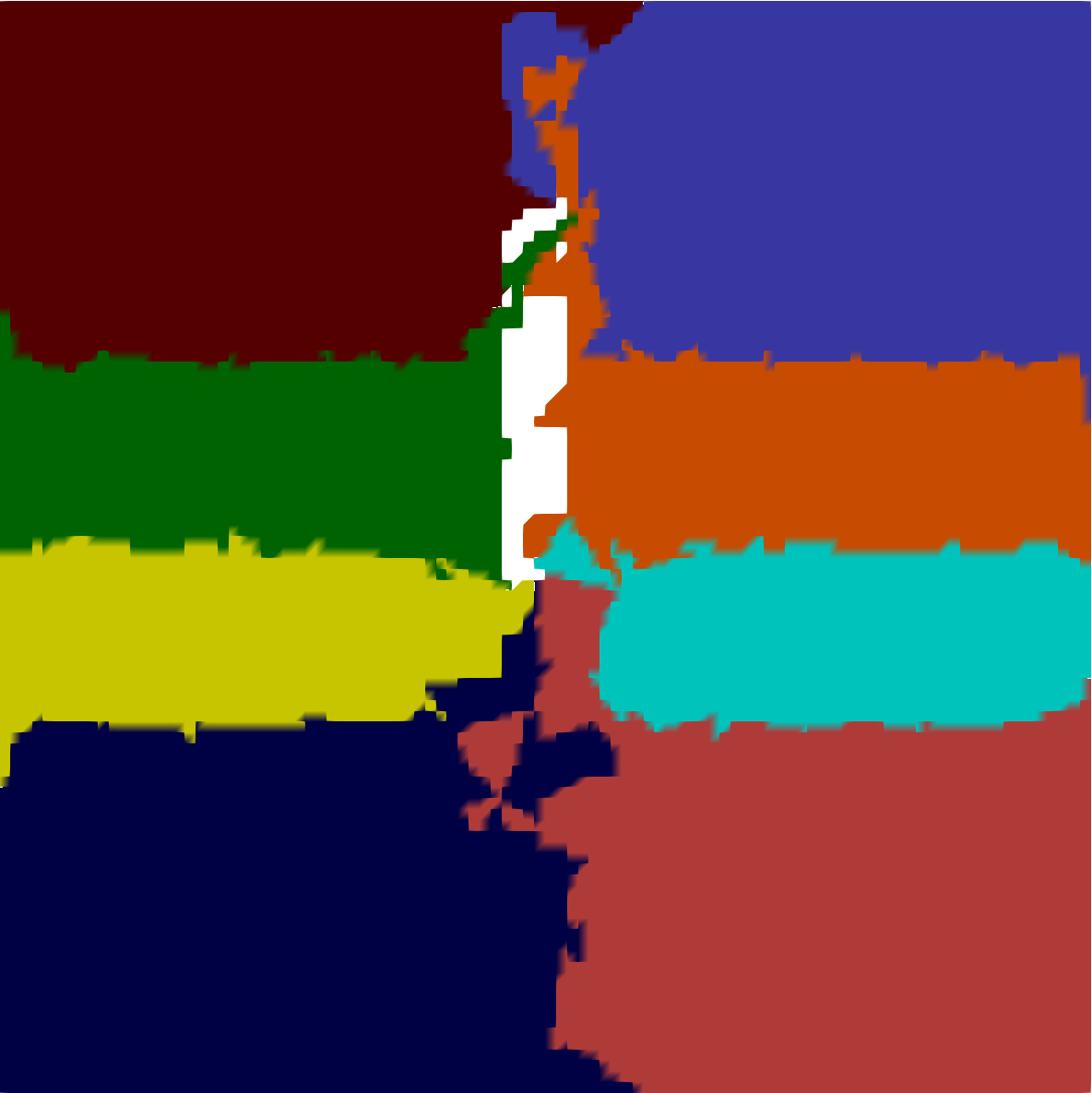}}
	\hfill
	\subcaptionbox{Segmentation: Samp500 \label{fig-sampGrid-Seg-R500}}{\includegraphics[width=0.32\linewidth]{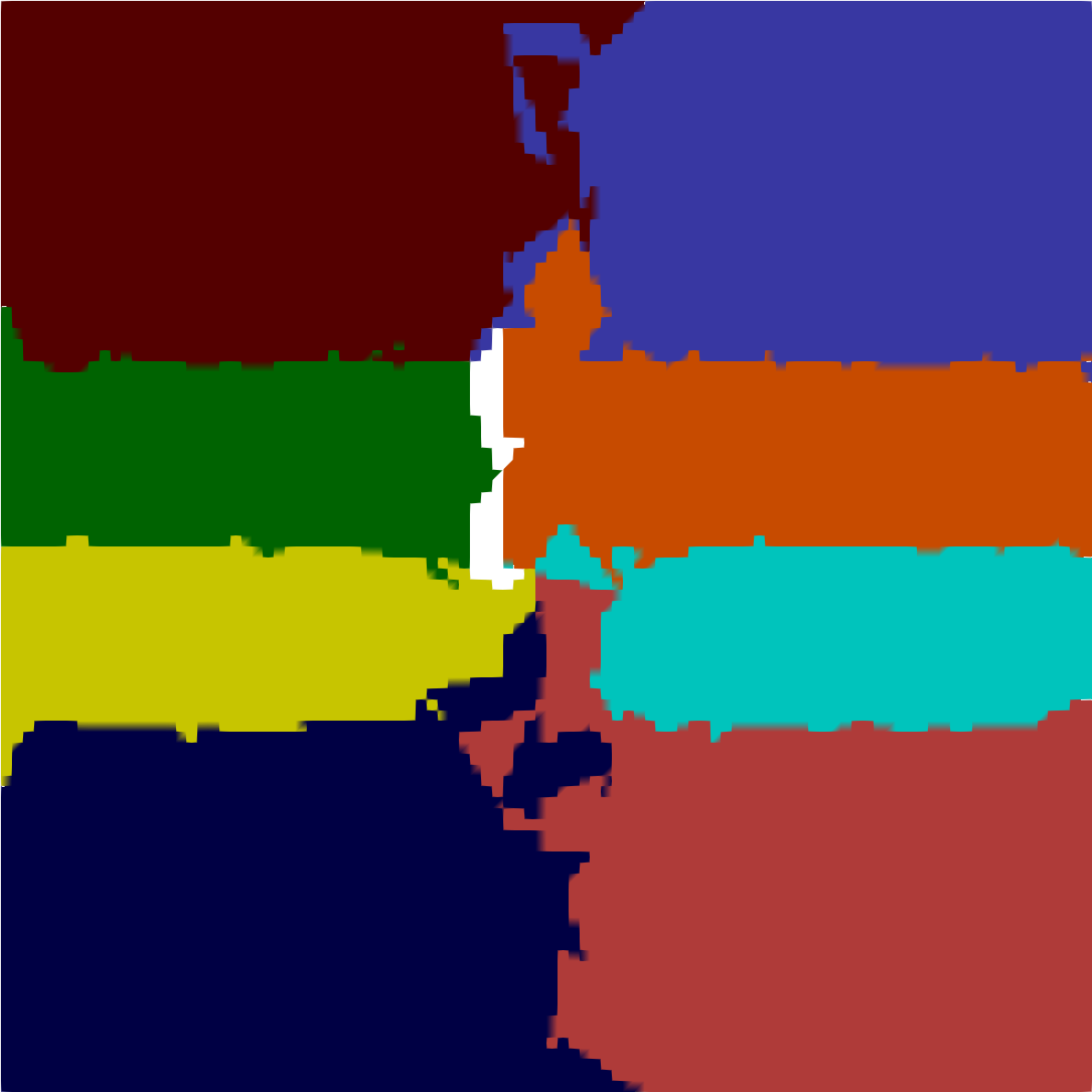}}
	\vspace{-0.5em}
	\caption{Visualizations of the persistence diagrams ((a), (b), and (c)) and segments ((d), (e), and (f)) of the sampled data at different resolutions
		\label{fig-sampGrid-2}}
	\vspace{-1em}
\end{figure}

Typically, one searches for plateau regions in persistence curves to find stable ranges for topological simplification.  For example, in the actual solution (Figure \ref{fig-solution-PC}), there is a stable region of $8$ critical pairs to the left of $0.48$, and another stable region between $0.48$ and $2.29$.  In the area isolated by the dotted box in Figure~\ref{fig-sampGrid-PC}, we see that the sampled data also \grammar{have} similar plateaus, but \grammar{the plateaus} persist through different ranges of persistence.  For the highest resolution, this gap is slightly wider than it is for the less sampling rates, suggesting it may be easier to identify the stable region. \grammar{Notably, however}, between $8$ and $4$ pairs, all three sampled data do not drop instantaneously, but rather take four small steps.  This \grammar{change in stable regions} suggests that\grammar{,} in the sample data, we have slightly perturbed the values of the minima and maxima that define these features.  These observations are also highlighted in the persistence diagrams for the sampled data in Figures \ref{fig-sampGrid-PD-R50}, \ref{fig-sampGrid-PD-R200}, and \ref{fig-sampGrid-PD-R500}. The persistence pairs below $0.4$ threshold (visualized using orange bars) increase with an increase in sampling resolution.

To segment the sampled data, we first simplify the data using the persistence threshold $0.4$ units (slightly left of the red arrow in Figure \ref{fig-sampGrid-PC}) and then \grammar{segment the data} using a contour tree. Visualizations for the segmented data are shown in Figures \ref{fig-sampGrid-Seg-R50}, \ref{fig-sampGrid-Seg-R200}, and \ref{fig-sampGrid-Seg-R500}. We observe that the boundaries of the segmentation are affected by the element boundaries of the simulation mesh, in particular in the flat region in the center vertical strip.  This effect is more pronounced in highly sampled data (Samp500). 

Hence, while sampling the projected data for topological analysis, it is not always the case that \grammar{the }highest resolution is preferred.  For our experiments where ground truth is not known, we need to take care to pick a sampling resolution sufficiently high to capture the underlying features, but not too high lest  we might start picking up the artifacts of discontinuities in the simulation data as features.

\subsection{Transformation by Subdividing the Input Mesh\\ (Method: Subdivided)}
\label{subsec-methods-subdiv}

In practical applications, simulation meshes are usually designed to capture the variation in the simulation output, and often utilize non-uniform resolutions and unstructured alignments to better approximate interesting behaviors. To take advantage of the properties of the simulation mesh, in this technique we subdivide each element into an equal number of smaller elements 
to capture the underlying high-order polynomial data.  On the shared boundaries between elements (e.g., the edge that bounds multiple triangles, or a vertex that is adjacent to multiple simplices), if there is a discontinuity in the higher order data\grammar{,} we must \grammar{choose which} value we use for our piecewise linear approximation.  In such cases, we choose to average between different values at all adjacent elements to create continuous piecewise linear data.  This technique of subdividing the elements into smaller elements is a filter that aims to both \grammar{resolve} the discontinuities by averaging and sample sufficiently to represent the piecewise polynomial data as continuous, piecewise linear data. 

\begin{figure}[!ht]
	\centering
	\includegraphics[width=\linewidth]{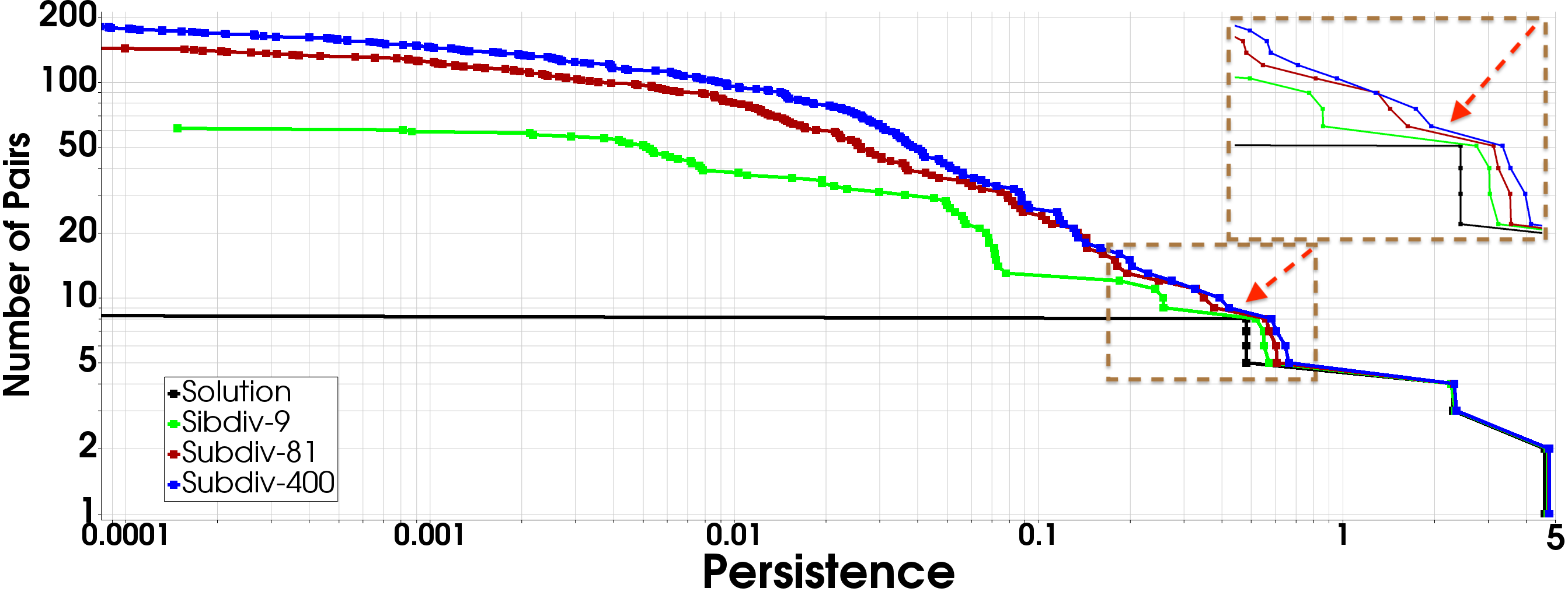}
	\vspace{-0.5em}
	\caption{Persistence curves of the subdivided data (at three resolutions) compared to the original function solution.\label{fig-subdiv-PC}}
	\vspace{-1em}
\end{figure}

To capture the projected data on the simulation mesh, which consists of triangular elements of polynomial degree two,
we subdivide each element into $9$, $81$, and $400$ triangular elements and refer to them as Subdiv-9, Subdiv-81, and Subdiv-400, respectively. \response{The average time taken to compute vorticity at each location on a machine with a $2.4$ GHz (Intel CPU E7-4870) processor is $0.804$ microseconds.}
Typically, each element is subdivided based on the polynomial degree ($k$) and the dimension ($d$) of the element. \grammar{Each element} subdivides into $(k+1)^d$ elements.
Therefore, we consider Subdiv-9 (as $(2+1)^2 = 9$) to be \grammar{the minimum amount required for} subdivision.  We also investigate Subdiv-81 and Subdiv-400 to see how further refinement affects the result.  
In all cases, we evaluate the values at new vertices internal to the elements and average the boundaries when necessary.

\begin{figure}[!ht]
	\centering
	\subcaptionbox{Persistence Diagram: Subdiv-9 \label{fig-subdiv-PD-R9}}{\includegraphics[width=0.32\linewidth]{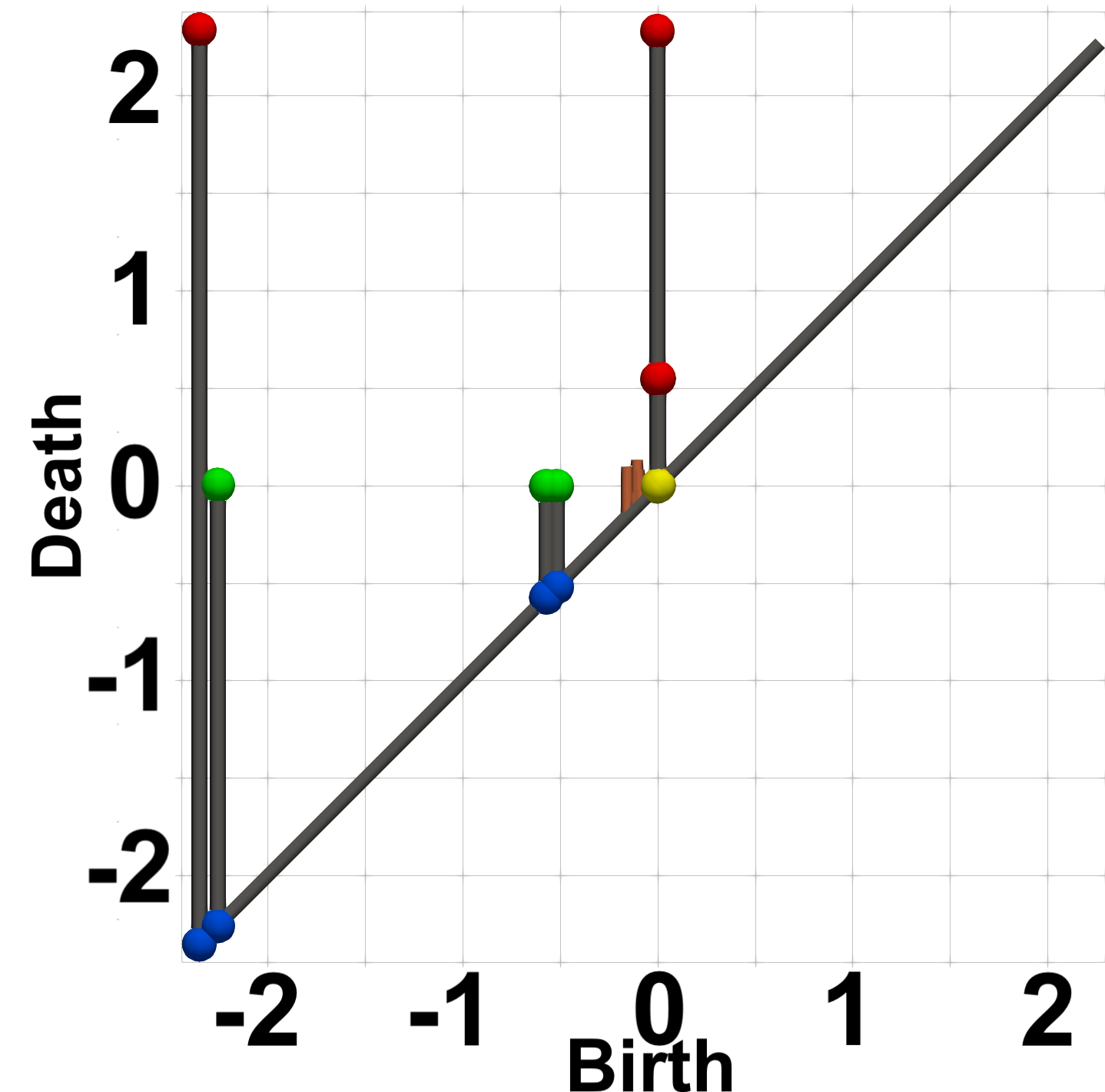}}
	\hfill
	\subcaptionbox{Persistence Diagram: Subdiv-81 \label{fig-subdiv-PD-R81}}{\includegraphics[width=0.32\linewidth]{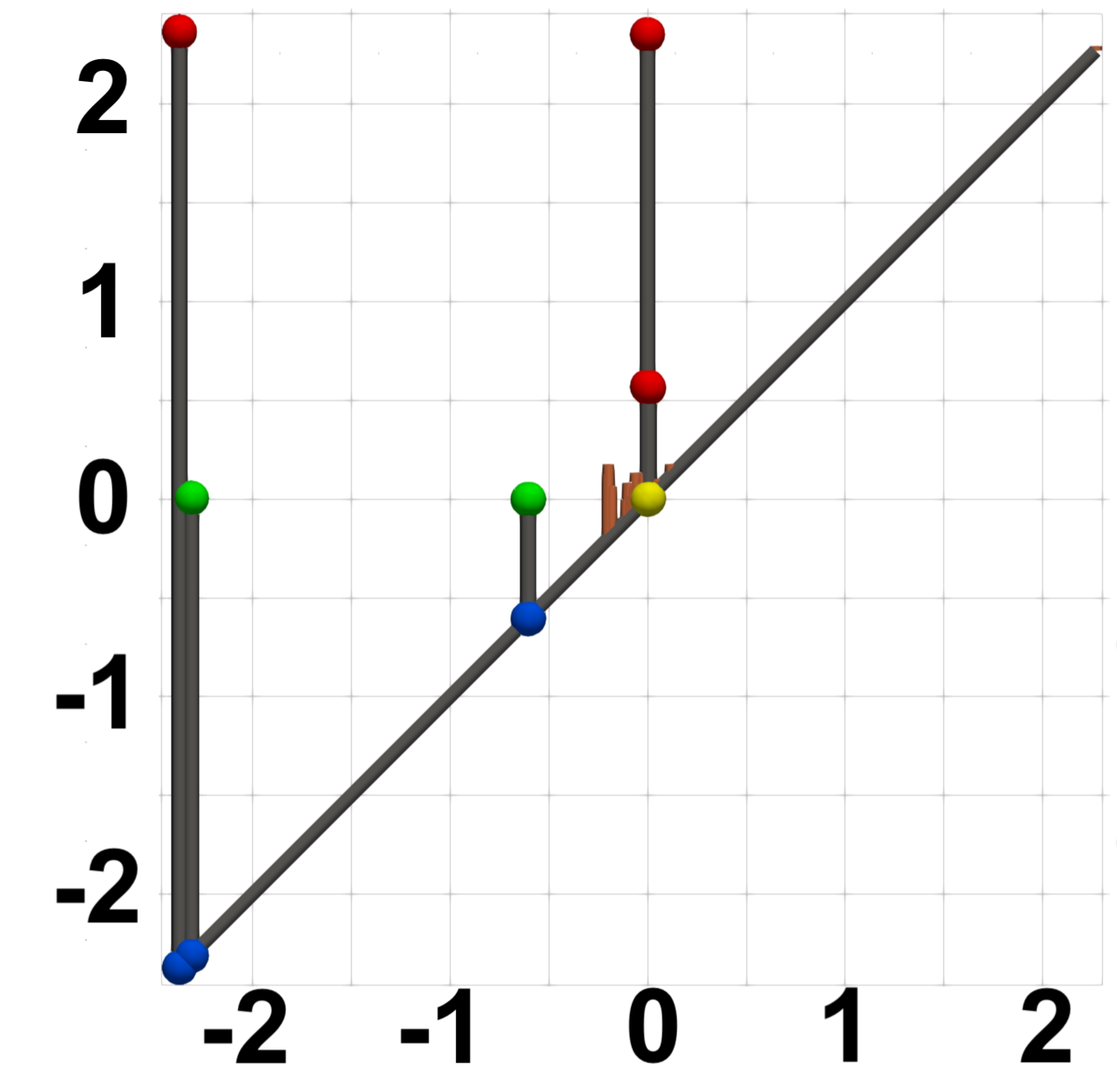}}
	\hfill
	\subcaptionbox{Persistence Diagram: Subdiv-400 \label{fig-subdiv-PD-R400}}{\includegraphics[width=0.32\linewidth]{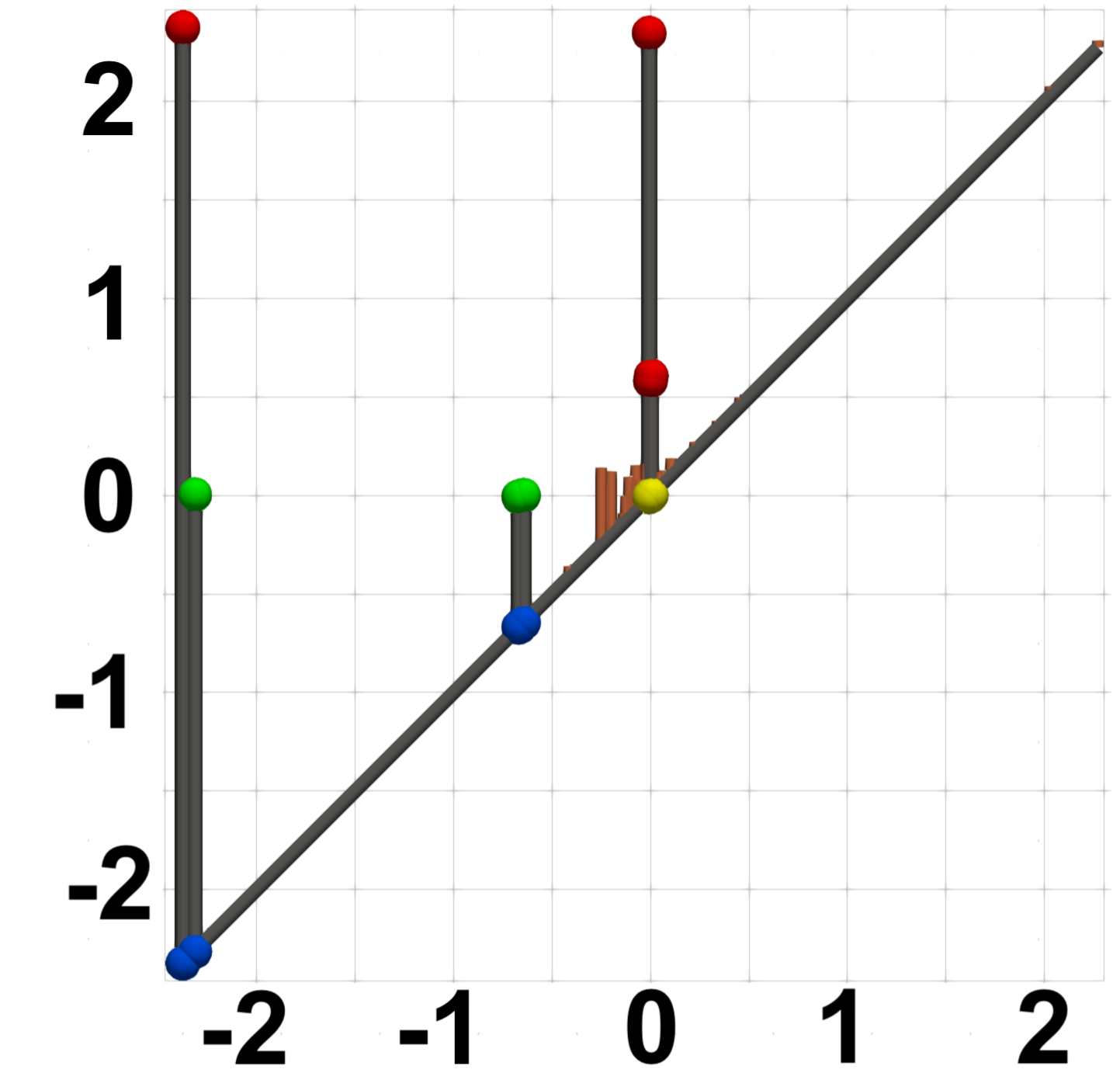}}
	
	\subcaptionbox{Segmentation: Subdiv-9\label{fig-subdiv-Seg-R9}}{\includegraphics[width=0.32\linewidth]{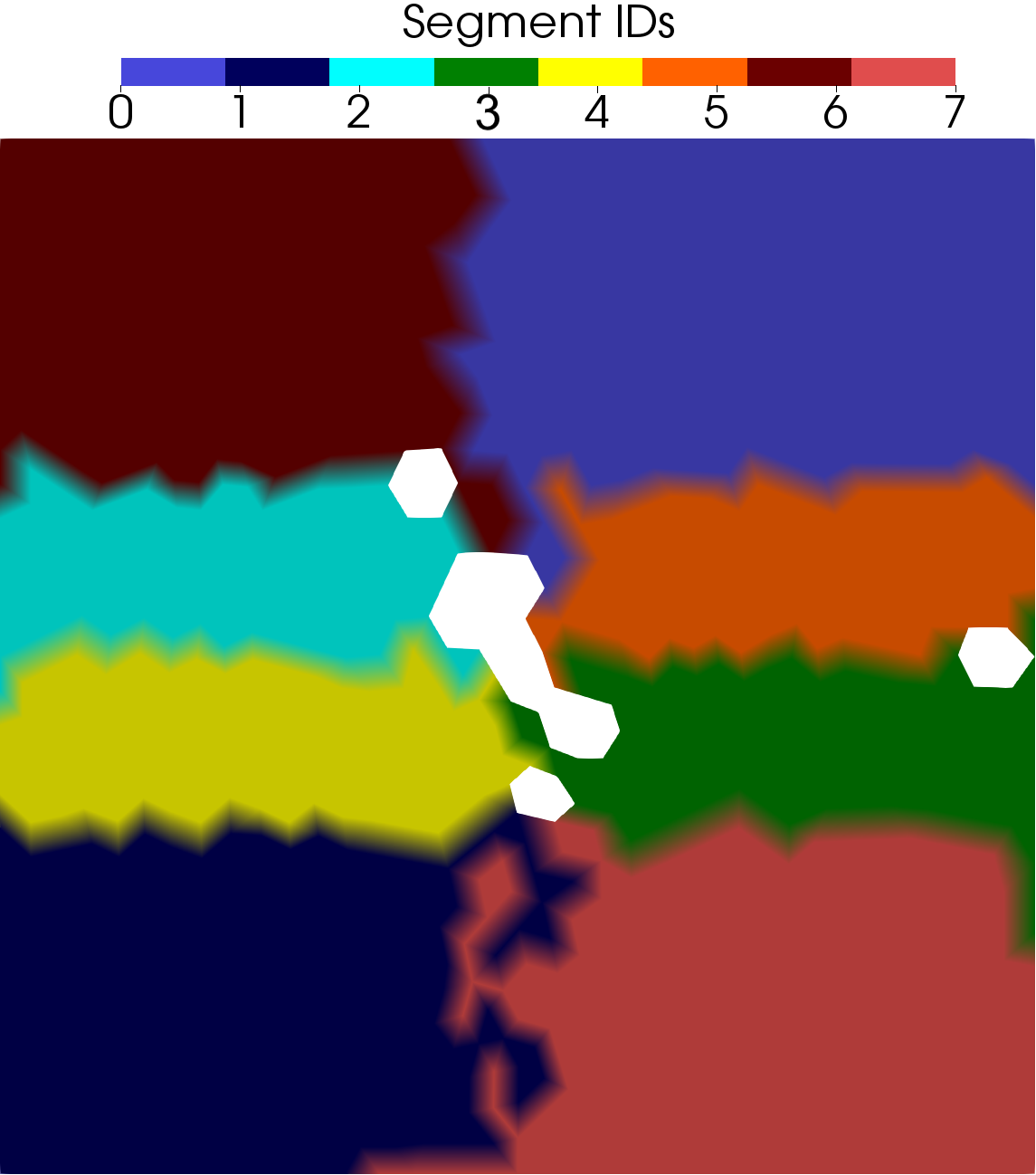}}
	\hfill
	\subcaptionbox{Segmentation: Subdiv-81\label{fig-subdiv-Seg-R81}}{\includegraphics[width=0.32\linewidth]{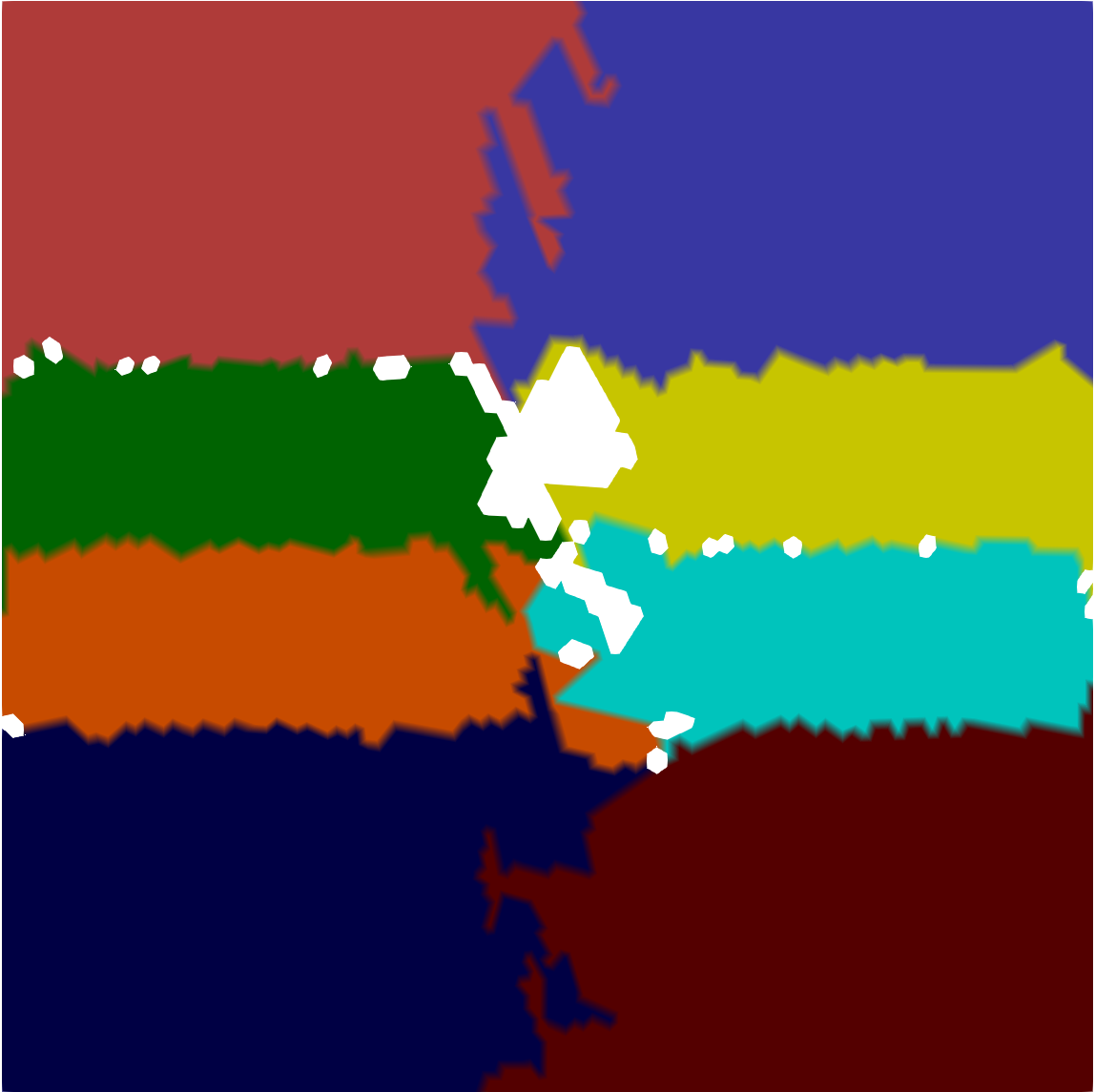}}
	\hfill
	\subcaptionbox{Segmentation: Subdiv-400\label{fig-subdiv-Seg-R400}}{\includegraphics[width=0.32\linewidth]{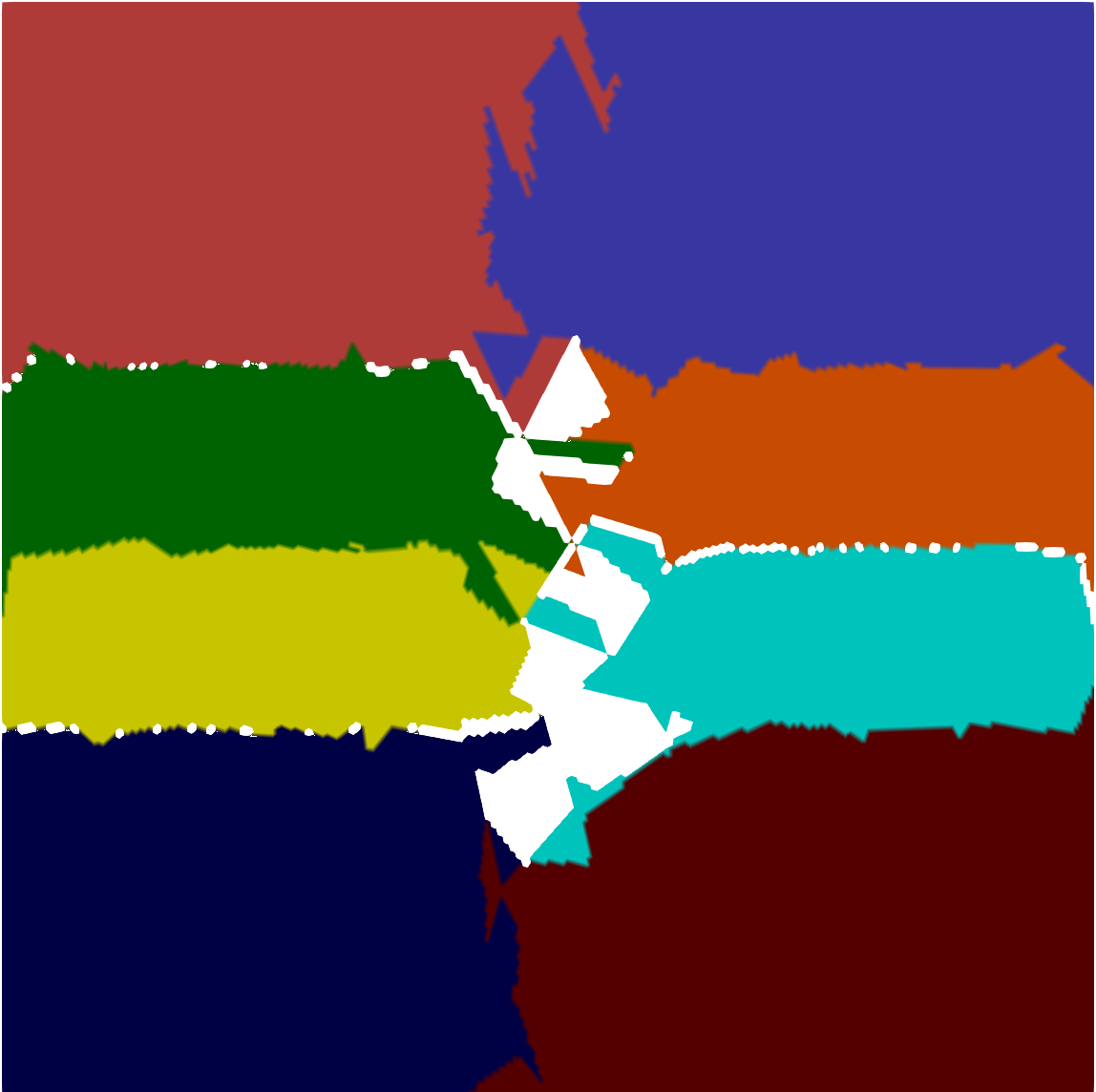}}
	\vspace{-0.5em}
	\caption{Visualizations of the persistence diagrams ((a), (b), and (c)) and segments ((d), (e), and (f)) of the subdivided data at different resolutions
		\label{fig-subdiv}}
	\vspace{-1em}
\end{figure}

Figure \ref{fig-subdiv-PC} shows the persistence curves for the projected data sampled on Subdiv-9, Subdiv-81, and Subdiv-400. Observe that an increase in the subdivisions again increases the initial number of persistence pairs, caused by the discontinuities being mapped to averages.  In this case, however, the behavior appears to more smoothly vary with persistence threshold and does not have the same ``staircase'' artifact.  Using the persistence curve of the actual solution as a guide, we examine the region associated with the first major topological change at the red arrow in Figure~\ref{fig-subdiv-PC}. 
We observe the gap between the persistence pairs due to the features and the discontinuities \grammar{gap decreases}, with an increase in subdivisions. 
Specifically, this plateau has a width of $0.26$ units for Subdiv-9, $0.18$ units for Subdiv-81, and $0.06$ units for Subdiv-400.
Similar observations are also highlighted in the persistence diagrams for the sampled data in Figures~\ref{fig-subdiv-PD-R9}, \ref{fig-subdiv-PD-R81}, and \ref{fig-subdiv-PD-R400}. The persistence pairs below $0.4$ threshold (visualized using orange bars) increase with an increase in sampling resolution.

To segment the sampled data, we first simplify the data using the persistence threshold $0.4$ units (red arrow in Figure \ref{fig-subdiv-PC}) and then segment using a contour tree. Visualizations for the segmented data are shown in Figures \ref{fig-subdiv-Seg-R9}, \ref{fig-subdiv-Seg-R81}, and \ref{fig-subdiv-Seg-R400}. We observe that the boundaries of the segmentation are affected by the element boundaries of the simulation mesh, and this effect is more pronounced \grammar{for the} mesh with \grammar{the most subdivisions} (Subdiv-400). 

For subdivided data as compared with sampled data, it turns out we have a similar challenge: we need to take care \grammar{in} specifying the output resolution as compared to the size of features we are interested in.  For our remaining experiments, which do not have ground truth, we use the resolution of $(k+1)^d$, which is the minimum resolution required to capture the underlying features.  We also remark that while averaging appears to provide a mathematical fix for discontinuities, it comes with the cost of potentially distorting the shapes of the segmentation.

\subsection{Transformation by Applying the L-SIAC Filter\\ (Method: L-SIAC)}
\label{subsec-methods-lsiac}

For our third transformation method, we use \grammar{the }L-SIAC filter to post-process the solution and define new values across the domain.  We can then sample the resulting output to a regular grid, using resolutions $h$, \grammar{which} are equivalent to the grid resolutions we used in the sampled data. For applying the L-SIAC filter at a point, we first choose the parameters: the characteristic length of the L-SIAC filter is adapted based on the size on the element~\cite{jallepalli2018adaptive}, the angle is chosen to be $0$ degrees, the order of B-splines equal to $k+1=3$ ($k$ is degree of the element), and the number of B-splines is $2k+1=5$. To apply the L-SIAC filter at a point in the mesh, we shift the filter to that point and then project the L-SIAC filter on\grammar{to} the mesh to find the overlapping elements. We then use Gaussian quadrature to integrate the L-SIAC kernel with the underlying mesh.
In this technique, the L-SIAC filter and sampling frequency work together to control the conversion process.

\begin{figure}[!ht]
	\centering
	\includegraphics[width=\linewidth]{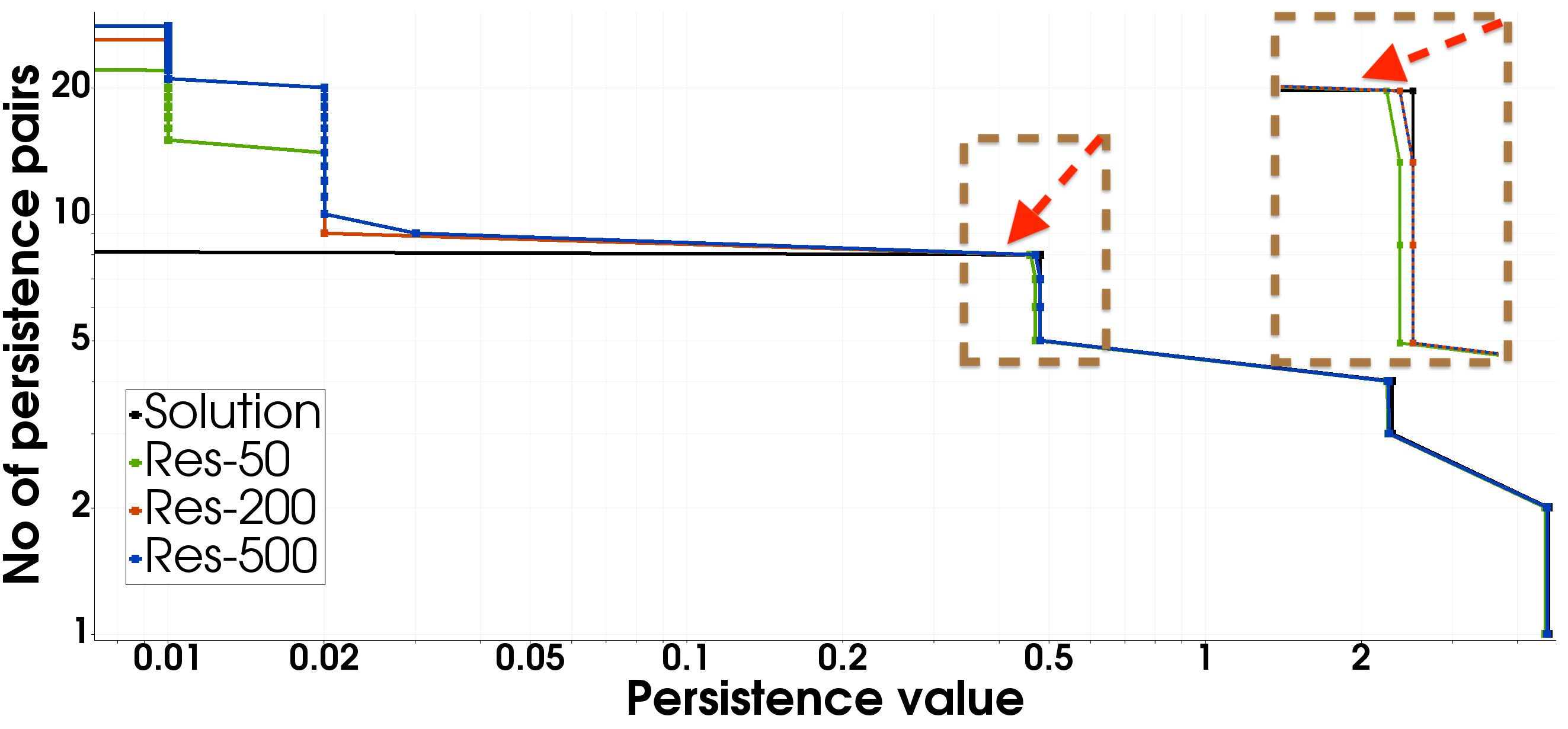}
	\vspace{-0.5em}
	\caption{Persistence curves of the L-SIAC data (at three resolutions) as compared to the original function solution. \label{fig-lsiac-PC}}
	\vspace{-1em}
\end{figure}

\begin{figure}[!ht]
	\centering	
	\subcaptionbox{Persistence Diagram: LSIAC-Samp50 \label{fig-lsiac-PD-R50}}{\includegraphics[width=0.32\linewidth]{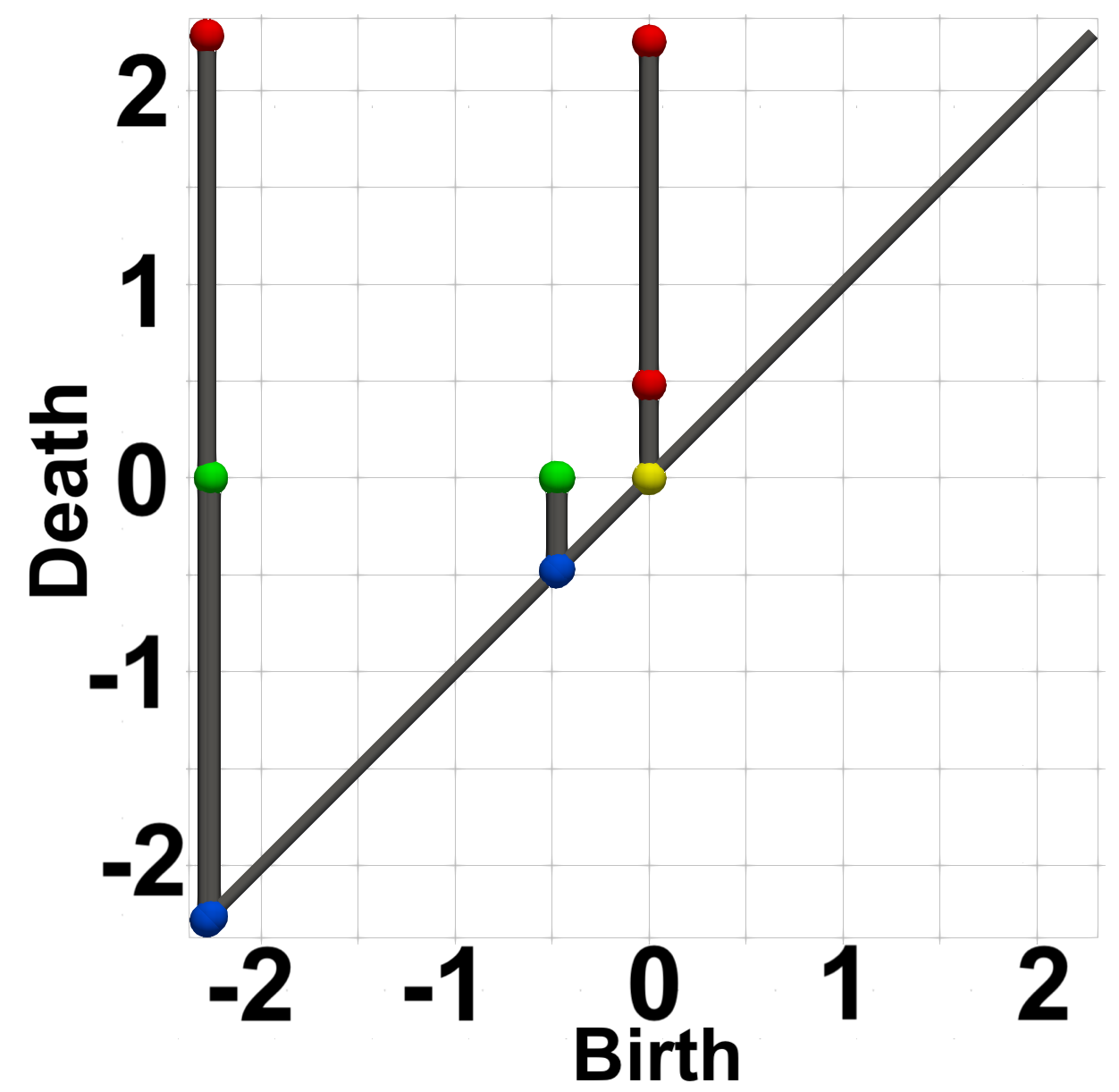}}
	\hfill
	\subcaptionbox{Persistence Diagram: LSIAC-Samp200 \label{fig-lsiac-PD-R200}}{\includegraphics[width=0.32\linewidth]{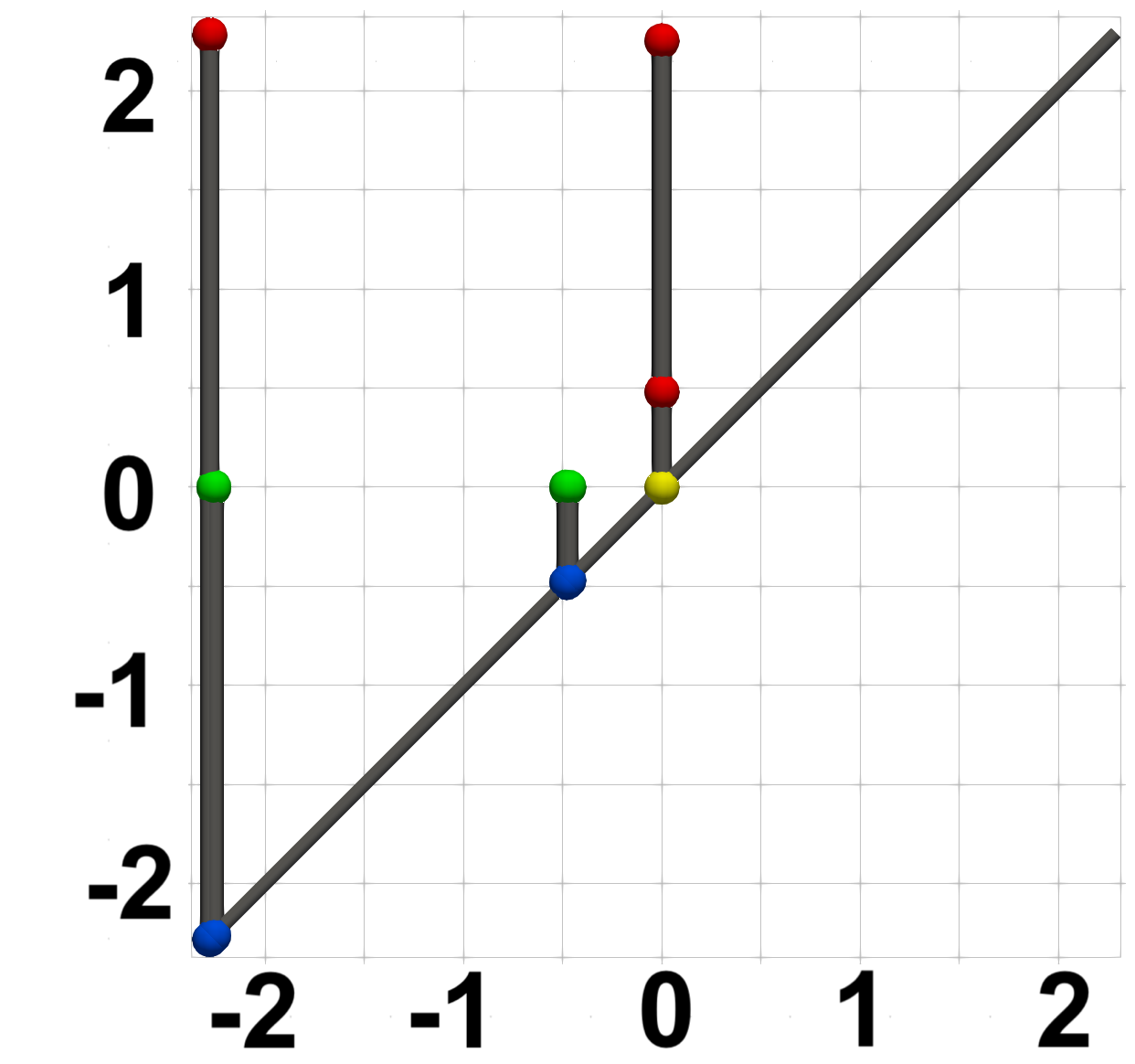}}
	\hfill
	\subcaptionbox{Persistence Diagram: LSIAC-Samp500 \label{fig-lsiac-PD-R500}}{\includegraphics[width=0.32\linewidth]{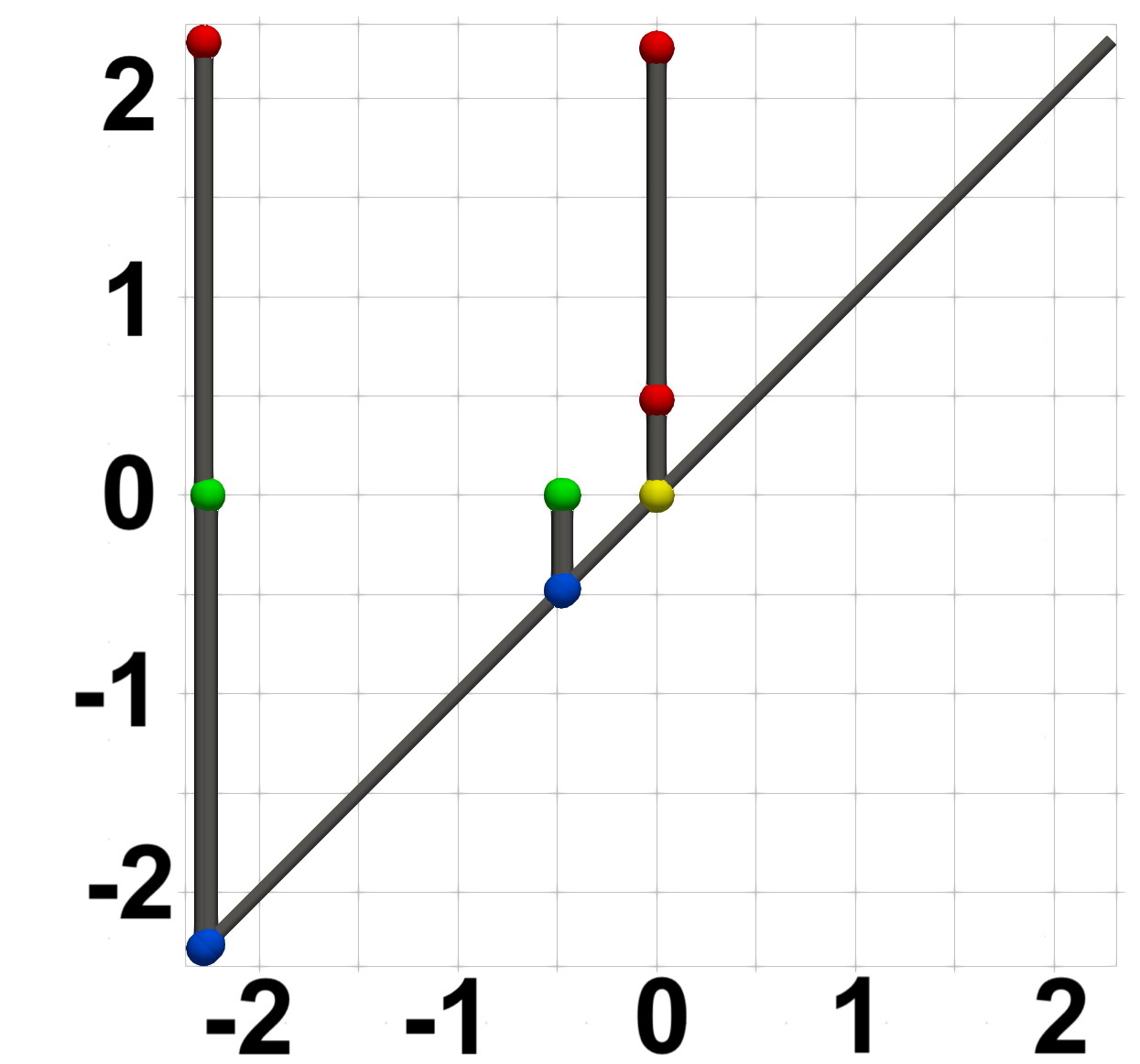}}
	
	\subcaptionbox{Segmentation: LSIAC-Samp50 \label{fig-lsiac-Seg-R50}}{\includegraphics[width=0.32\linewidth]{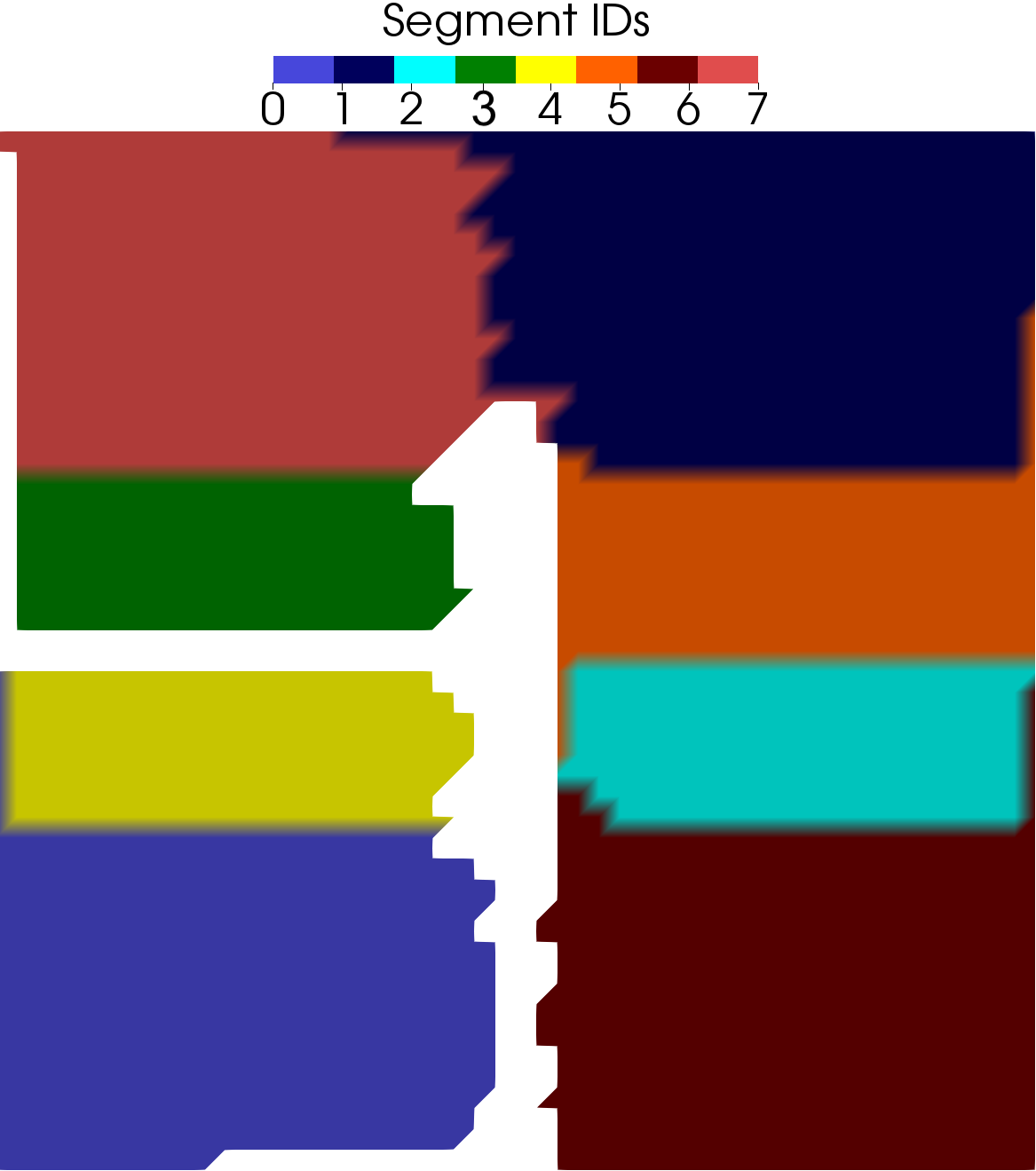}}
	\hfill
	\subcaptionbox{Segmentation: LSIAC-Samp200 \label{fig-lsiac-Seg-R200}}{\includegraphics[width=0.32\linewidth]{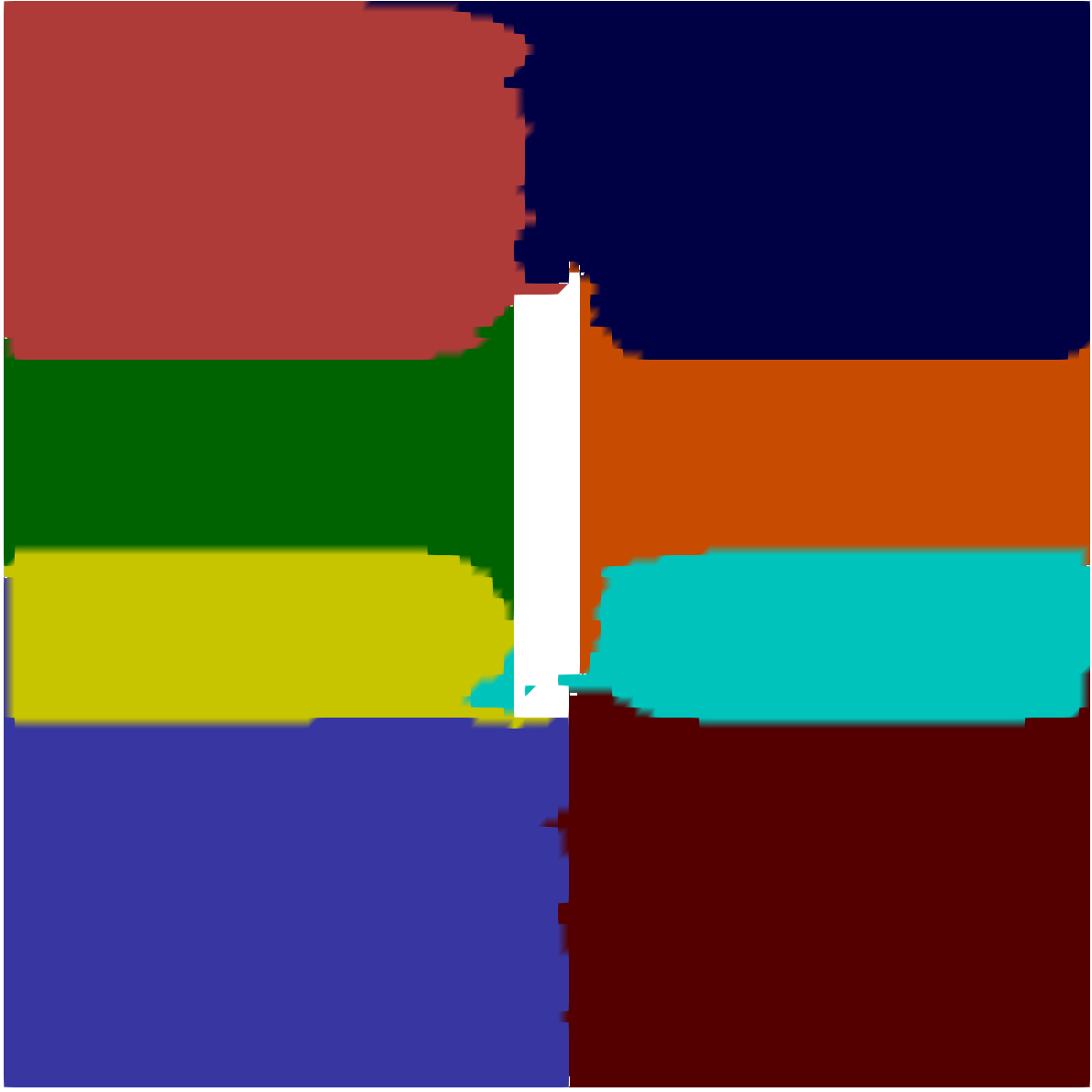}}
	\hfill
	\subcaptionbox{Segmentation: LSIAC-Samp500 \label{fig-lsiac-Seg-R500}}{\includegraphics[width=0.32\linewidth]{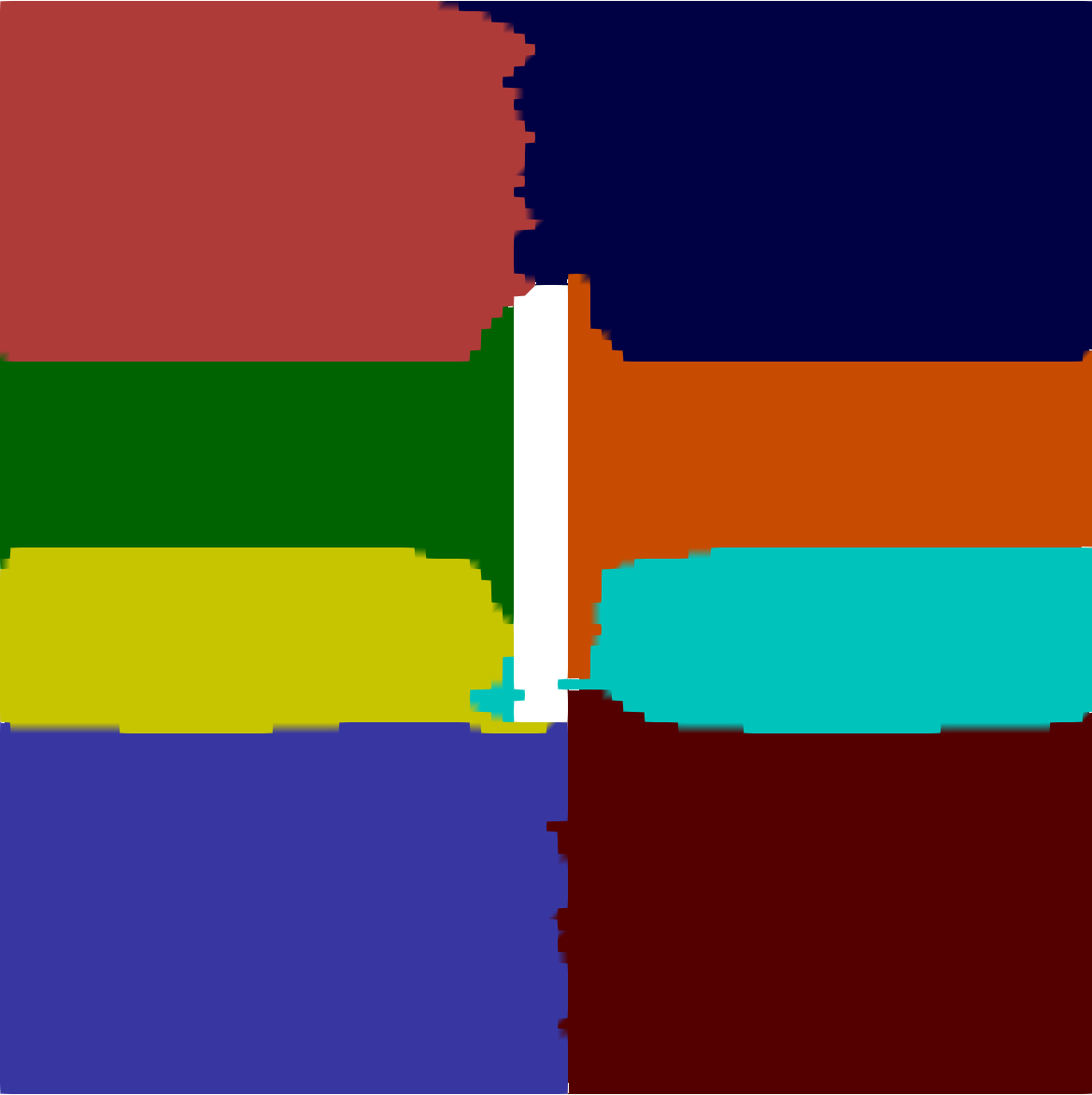}}
	\vspace{-0.5em}
	\caption{Visualizations of the persistence diagrams ((a), (b), and (c)) and segments ((d), (e), and (f)) of the L-SIAC data at different resolutions
		\label{fig-lsiac-2}}
	\vspace{-1em}
\end{figure}

Figure \ref{fig-lsiac-PC} shows the persistence curves for the projected data sampled using different resolutions ($50\times 50$, $200\times 200$, and $500\times 500$ referred to as LSIAC-Samp50, LSIAC-Samp200, and LSIAC-Samp500, respectively). \response{The average time taken to compute vorticity at each location on a machine with a $2.4$ GHz (Intel CPU E7-4870) processor is $334$ microseconds.}
\grammar{Although} L-SIAC mitigates most of the effect of discontinuities in the data, we still have some aberrant features that show up at low persistence values due to its error compared to the actual solution.  
However, we observe that these \grammar{features }disappear significantly early, creating a wider stable region that begins at far lower persistence thresholds.
Additionally, the sampling rate has a negligible influence on when this region starts and ends, and the behavior at larger persistence values almost matches the ground truth data.  These observations are also confirmed in the persistence diagrams in Figures \ref{fig-lsiac-PD-R50}, \ref{fig-lsiac-PD-R200}, and \ref{fig-lsiac-PD-R500}.

To segment the sampled data, we first simplify the data using the persistence threshold $0.4$ units (red arrow in Figure \ref{fig-lsiac-PC}) and then segment it using a contour tree. Note\grammar{, however,} we could use a much lower persistence threshold than before because L-SIAC appears to mitigate topological noise that was present in the sampled and subdivided data. Visualizations for the segmentation from the contour tree are shown in Figures \ref{fig-lsiac-Seg-R50}, \ref{fig-lsiac-Seg-R200}, and \ref{fig-lsiac-Seg-R500}. \grammar{Although we observe the same number of persistent features at all resolutions, they differ at coarse resolutions, at which the shapes} of these regions can vary.  In particular, we see more boundary artifacts for coarser data at the nearly flat region in the center of the domain.  Higher resolutions help to recover these boundaries more smoothly.  This \grammar{recovery }suggests that for more complex data, increasing the resolution of L-SIAC filter is helpful in terms of \grammar{both the} number and shape of features, albeit only as shown \grammar{for} this simple dataset.
\section{Results}
\label{sec-results}

In this section, we present both two\grammar{-}dimensional and three\grammar{-}dimensional results.  For our first example, we analyze the results of simulating
flow past a two\grammar{-}dimensional circular cylinder.  In our second example, we analyze a three\grammar{-}dimensional flow scenario that produces a collection
of co-rotating vortices.

\subsection{Flow Over a 2D Cylinder}
\label{subsec-flow2Dcylinder}
\begin{figure}[!ht]
	\centering
	\subcaptionbox{ Simulation mesh. \label{fig-2Dcyl-Intro-mesh}}{\includegraphics[width=0.47\linewidth]{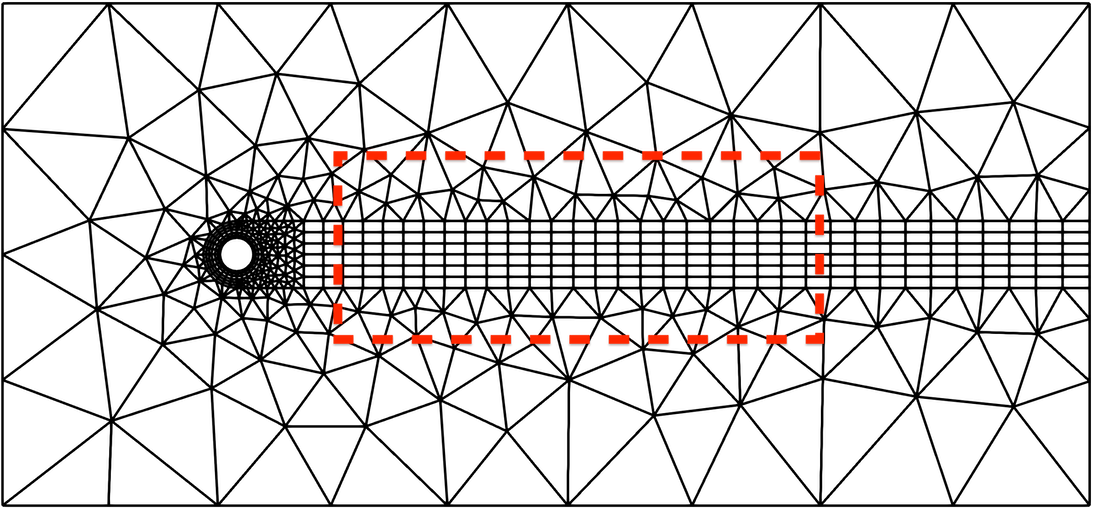}}%
	\hfill
	\subcaptionbox{Vorticity. \label{fig-2Dcyl-Intro-dG}}{\includegraphics[width=0.52\linewidth]{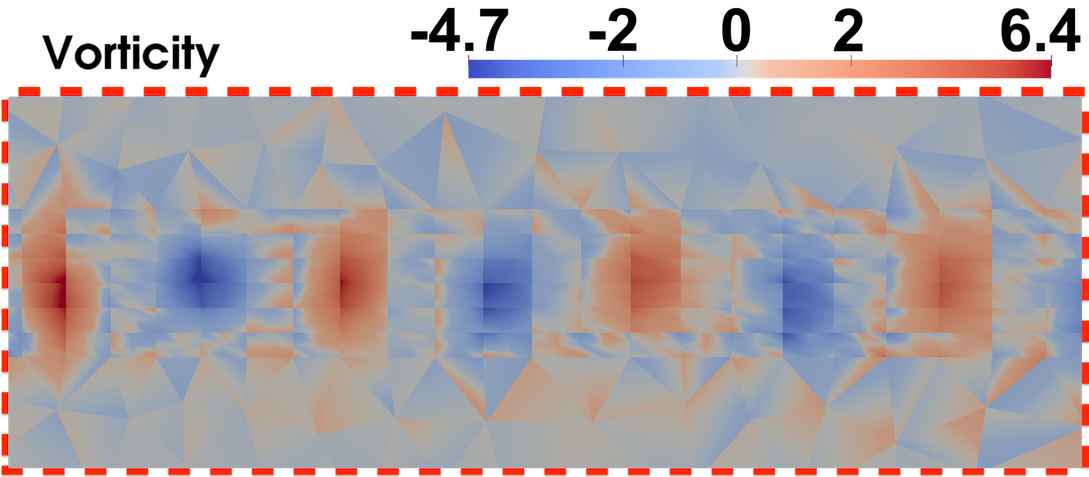}}%
	\vspace{-0.5em}
	\caption{(a) Simulation mesh of the flow over a 2D cylinder. (b) Vorticity calculated using element derivatives in the highlighted region of the simulation mesh.}
	\label{fig-2Dcyl-Intro}
	\vspace{-1em}
\end{figure}

The Nektar++\cite{nektar:15} solver suite, and in particular the incompressible Navier-Stokes solver, was used to generate the fluid 
flow results. Flow past a circular cylinder at the viscosity examined is a transient problem, but here we analyze only a single snapshot (in time) \grammar{topologically}. The mesh used for this simulation is shown in Figure~\ref{fig-2Dcyl-Intro-mesh}, which contains polynomial degree two ($P(2)$) elements: $500$ triangles and $330$ quadrilaterals found primarily in the wake region behind the circle.  A continuous Galerkin (FEM) methodology was used \grammar{that has} degree $2$ inside the element and has $C^0$ continuity at the element interfaces. The Reynolds number was set at $Re=500$, and the flow solver was run until shedding behind the cylinder generated consistently shaped vortices. 

The simulation data from the highlighted region of the simulation mesh (Figure \ref{fig-2Dcyl-Intro}) was chosen to be analyzed \grammar{topologically}.  We analyze the scalar field of vorticity.  
To calculate the vorticity, we need to apply a derivative on the $C^0$ vector field components $(u,v)$; thus creating discontinuity at the element boundaries as shown in Figure \ref{fig-2Dcyl-Intro-dG}. 
From this dataset, we use the methods described in Section~\ref{sec-Methods} to produce three different low-order datasets for topological analysis.  

To produce our ``sampled'' vorticity (Section \ref{subsec-methods-grid}), we sampled the simulation output on a regular grid of resolution $145\times80$ and calculated partial derivatives $(u_y,v_x)$ using a finite difference scheme.  We then used these derivatives to compute the vorticity defined as $\omega_z = v_x-u_y$.
To produce our ``subdivided'' vorticity (Section \ref{subsec-methods-subdiv}), we subdivided the simulation mesh such that each triangle was divided into $9$ triangles, and each quadrilateral was divided into $18$ triangles. 
The simulation data $(u,v)$ was sampled and stored at the vertices of the subdivided mesh. 
To calculate the partial derivatives, we first calculated the derivatives at the center of each triangle using the values at its vertices, and then the derivatives were interpolated back to the vertices using a weighted area of all the triangles connected to \grammar{each vertex}.
To produce \grammar{the} L-SIAC vorticity (Section \ref{subsec-methods-lsiac}), we use\grammar{d} the derivative L-SIAC filter ($D(K(3,6))$, where $D$ stands for derivative) as presented in \cite{jallepalli2017treatment}.  This filter calculates the partial derivatives $(u_y, v_x)$ at any point, and can be evaluated on a regular grid of resolution $145\times80$. \response{The average time taken to compute vorticity at each location on a machine with a $2.4$ GHz (Intel CPU E7-4870) processor is $86$, $0.496$, and $560$ microseconds for the sampled, subdivided, and L-SIAC vorticities, respectively.}

\begin{figure}[!ht]
	\begin{center}
		\includegraphics[width=1.0\linewidth]{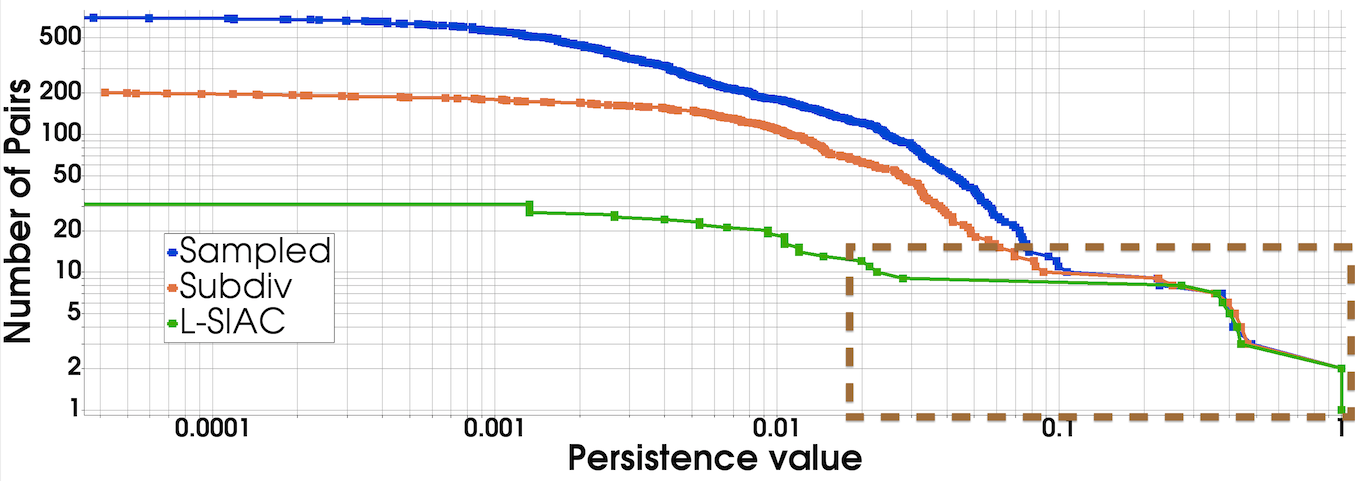}
	\end{center}
	\vspace{-0.5em}
	\caption{Persistence curves for the sampled, subdivided, and L-SIAC vorticity fields of the flow over a 2D cylinder.
		\label{fig-2Dcyl_PC_All}}
	\vspace{-1em}
\end{figure}

\begin{figure}[!ht]
	\begin{center}
		\includegraphics[width=1.0\linewidth]{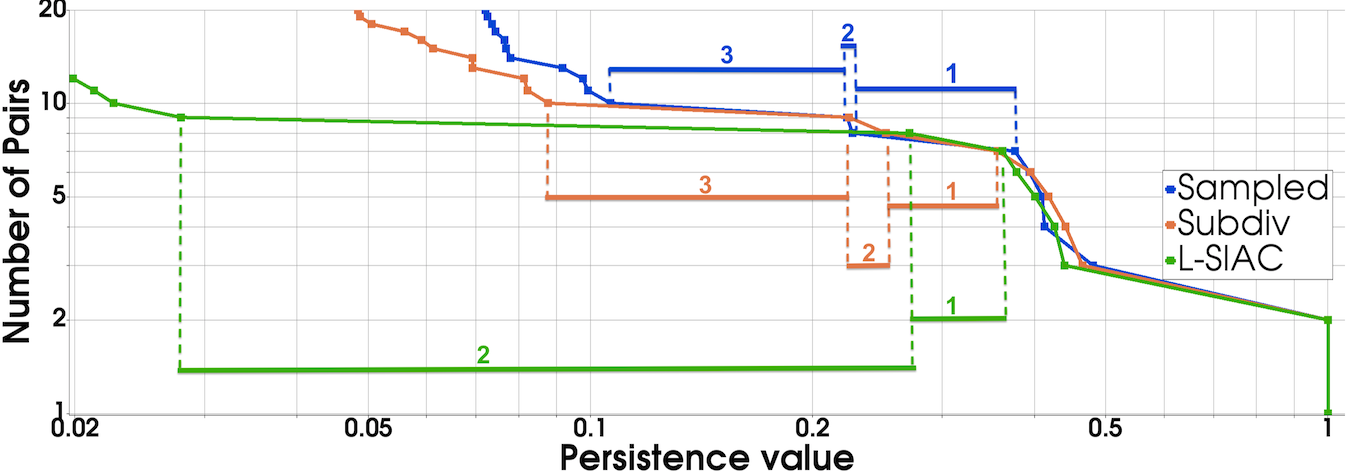}
	\end{center}
	\vspace{-0.5em}
	\caption{Zoom in on the region associated with Figure~\ref{fig-2Dcyl_PC_All}.  For each of datasets, we highlight two different stable regimes of critical features (numbered 1-3).
		\label{fig-2Dcyl_PC_zoom}}
	\vspace{-1em}
\end{figure}

Due to different approximation errors while calculating the vorticity, each scalar field has a different range.  
Specifically, the range of values is smallest for the L-SIAC ($[-3.85,3.69]$), and the subdivided vorticity ($[-4.36,4.67]$) has a smaller range than that of the sampled vorticity ($[-4.80,5.22]$).   
To enable comparison between the three datasets, we normalized their ranges to $[0,1]$. 
The persistence curves for the vorticity tests calculated above are shown in Figure \ref{fig-2Dcyl_PC_All}. 
We observe that the sampled vorticity and the subdivided vorticity have a large number of persistence pairs compared to the L-SIAC vorticity. 
The function of the persistence curve is to act as a guide to help choose the persistence threshold required to identify the features and simplify (remove) the noise. 
Key persistence thresholds are identified by detecting the plateau regions in the persistence curve. 
From the persistence curves in Figure \ref{fig-2Dcyl_PC_zoom}, we identified three regions of interest for the sampled and subdivided (indicated by $1$, $2$, and $3$), and two regions in case of L-SIAC (indicated by $1$ and $2$).
The regions indicated by $1$, $2$ and $3$ have $7$, $8$, and $9$ persistence pairs.
As it turns out, there are only 8 significant features in the dataset (determined by inspecting the data). 
The plateau corresponding to the $8$ features indicated by region $2$ is significantly shorter in the case of sampled and subdivided vorticity compared to that in the L-SIAC filter, and also shorter than its counterparts (regions indicated by $1$ and $3$).

\begin{figure}[!ht]
	\centering
	\subcaptionbox{Sampled vorticity.}{\includegraphics[width=0.33\linewidth]{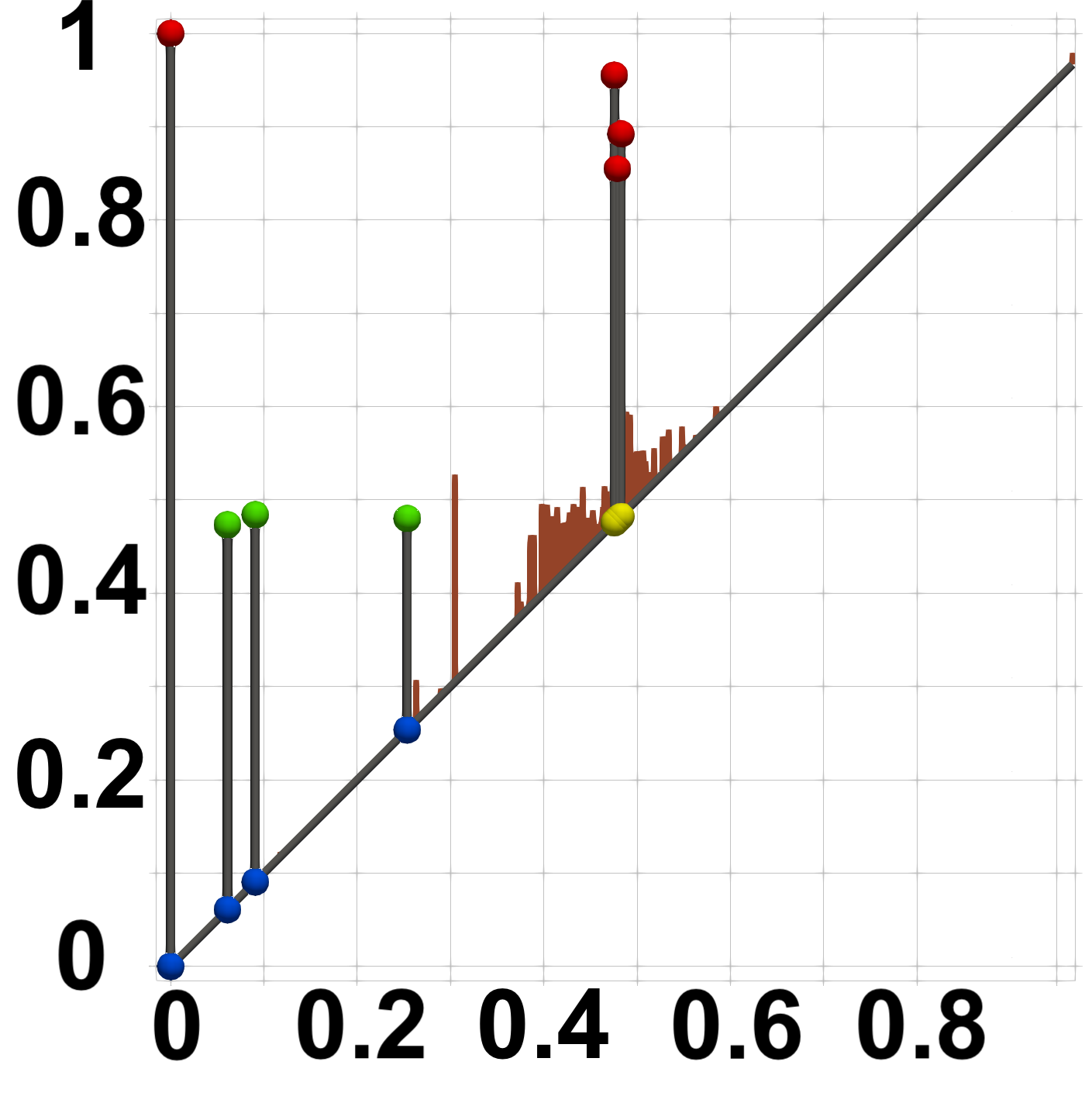}}%
	\hfill
	\subcaptionbox{Subdivided vorticity.}{\includegraphics[width=0.33\linewidth]{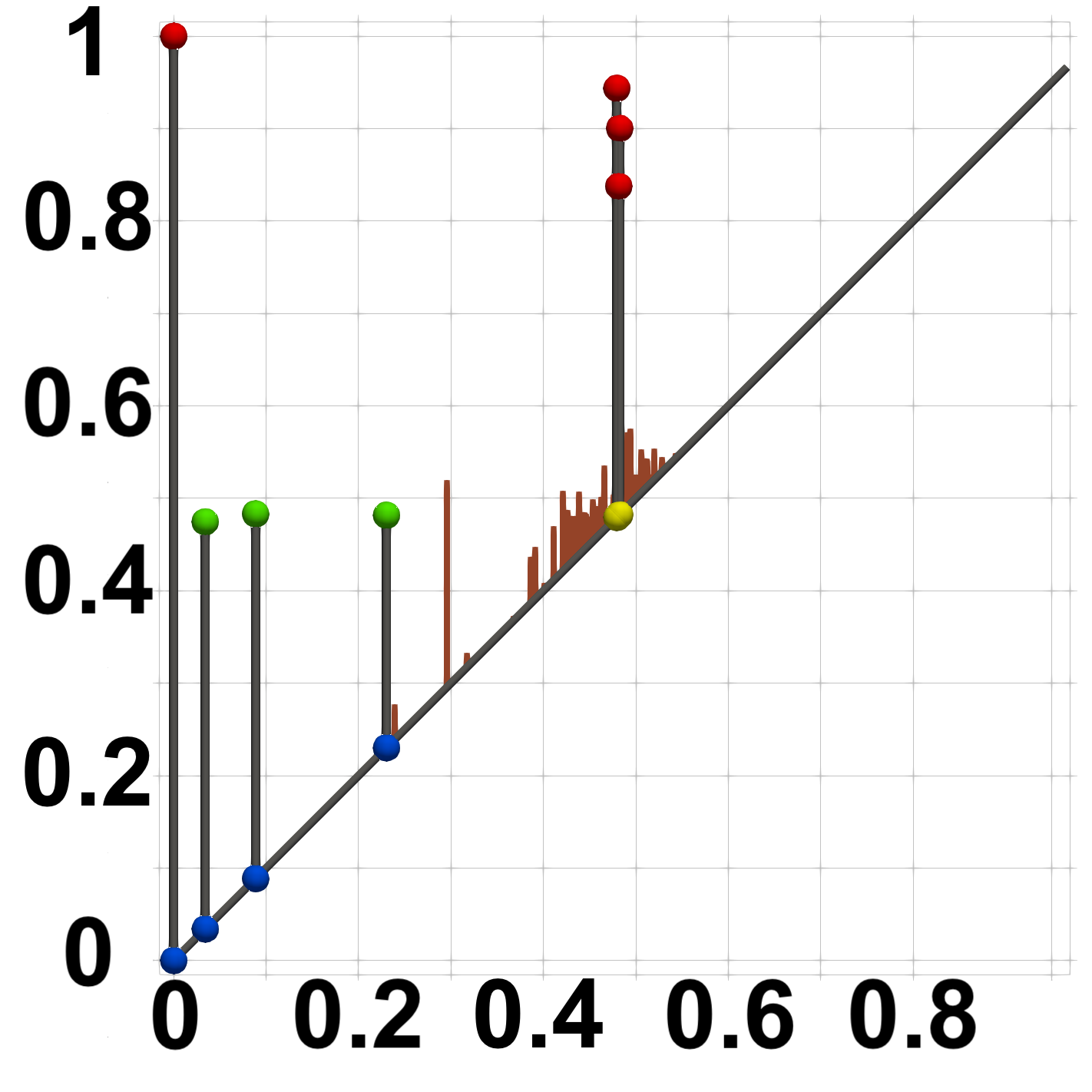}}%
	\hfill
	\subcaptionbox{L-SIAC vorticity. \label{fig-2Dcyl-PD-LSIAC}}{\includegraphics[width=0.33\linewidth]{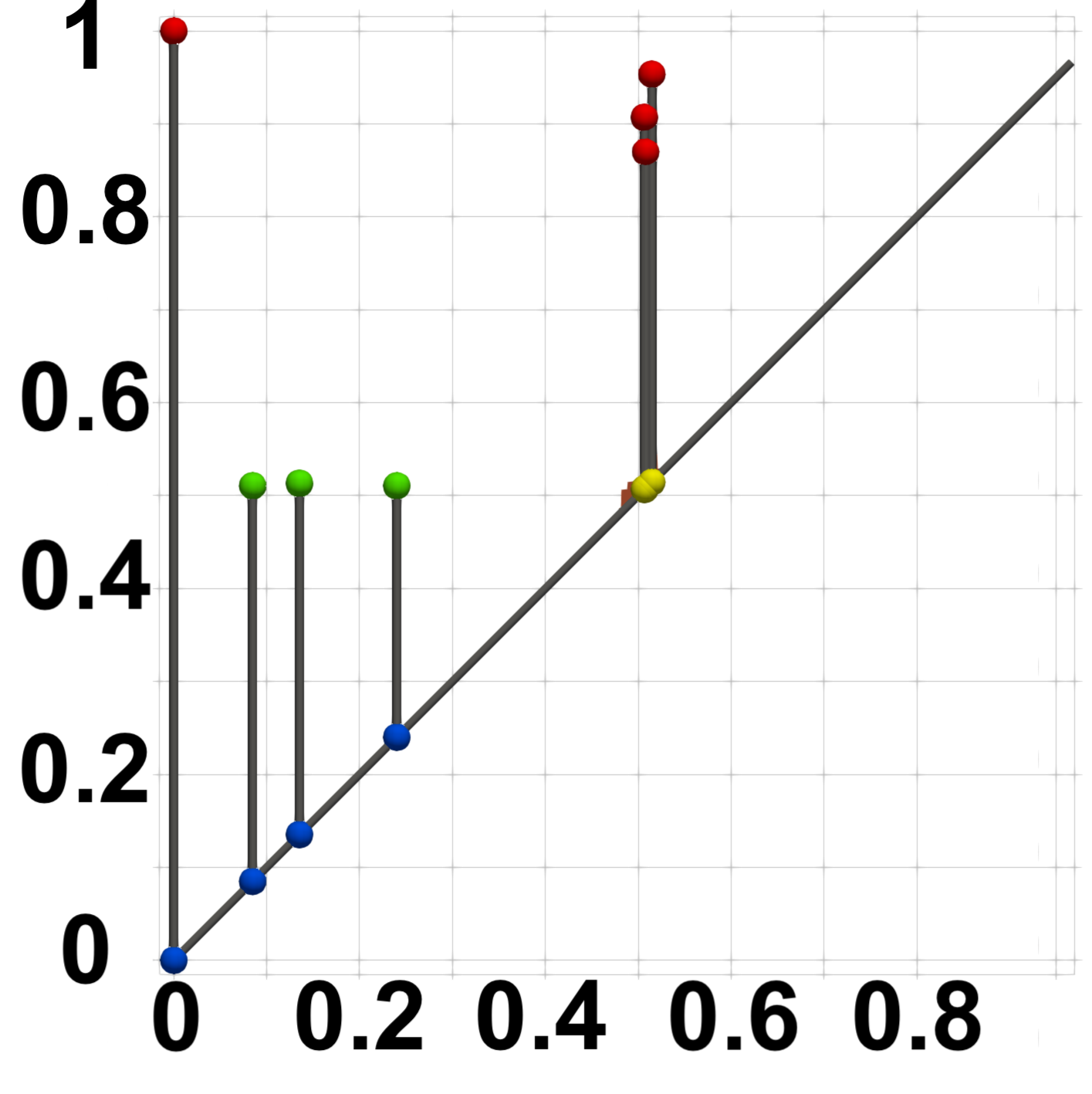}}
	\vspace{-0.5em}
	\caption{The persistence diagrams of the vorticity for the fluid flow past a circular cylinder described in Section \ref{subsec-flow2Dcylinder}. Thick gray lines are used to show the persistence pairs above the threshold $2.0$, and the lines in orange are used to indicate the persistence pairs below the threshold.}
	\label{fig-2Dcyl-PD-compare}
	\vspace{-1em}
\end{figure}

We pick a persistence value in the regions indicated by $2$ \grammar{in all} the datasets (i.e., $0.225$ for the sampled, $0.23$ for the subdivided, and $0.2$ in the case of the L-SIAC vorticity) to create the persistence diagrams shown in Figure \ref{fig-2Dcyl-PD-compare}. The persistence pairs below the threshold are shown by orange lines. Observe that in the case of subdivided and sample vorticity, there exist\grammar{s} a persistence pair (the tallest orange bar) \grammar{that} is very close to the threshold but does not exist in the case of the L-SIAC vorticity.  This feature corresponds to a boundary artifact that creates a spurious but high\grammar{-}persistence, feature that would be hard to separate without additional knowledge.

\begin{figure}[!ht]
	\centering
	\subcaptionbox{Sampled vorticity. \label{fig-2Dcyl-Seg-Samp}}{\includegraphics[width=0.48\linewidth,keepaspectratio]{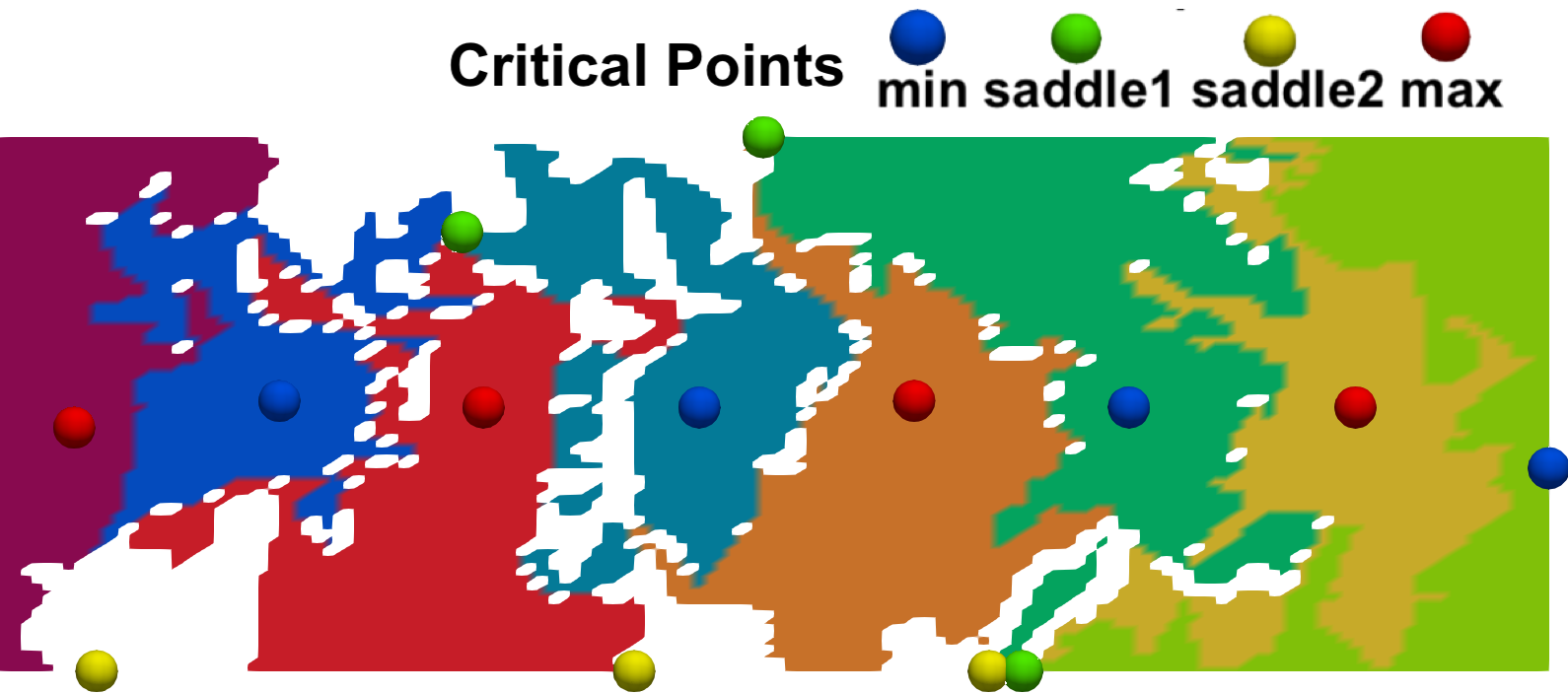}}
	\subcaptionbox{Subdivided vorticity. \label{fig-2Dcyl-Seg-SubDiv}}{\includegraphics[width=0.48\linewidth,keepaspectratio]{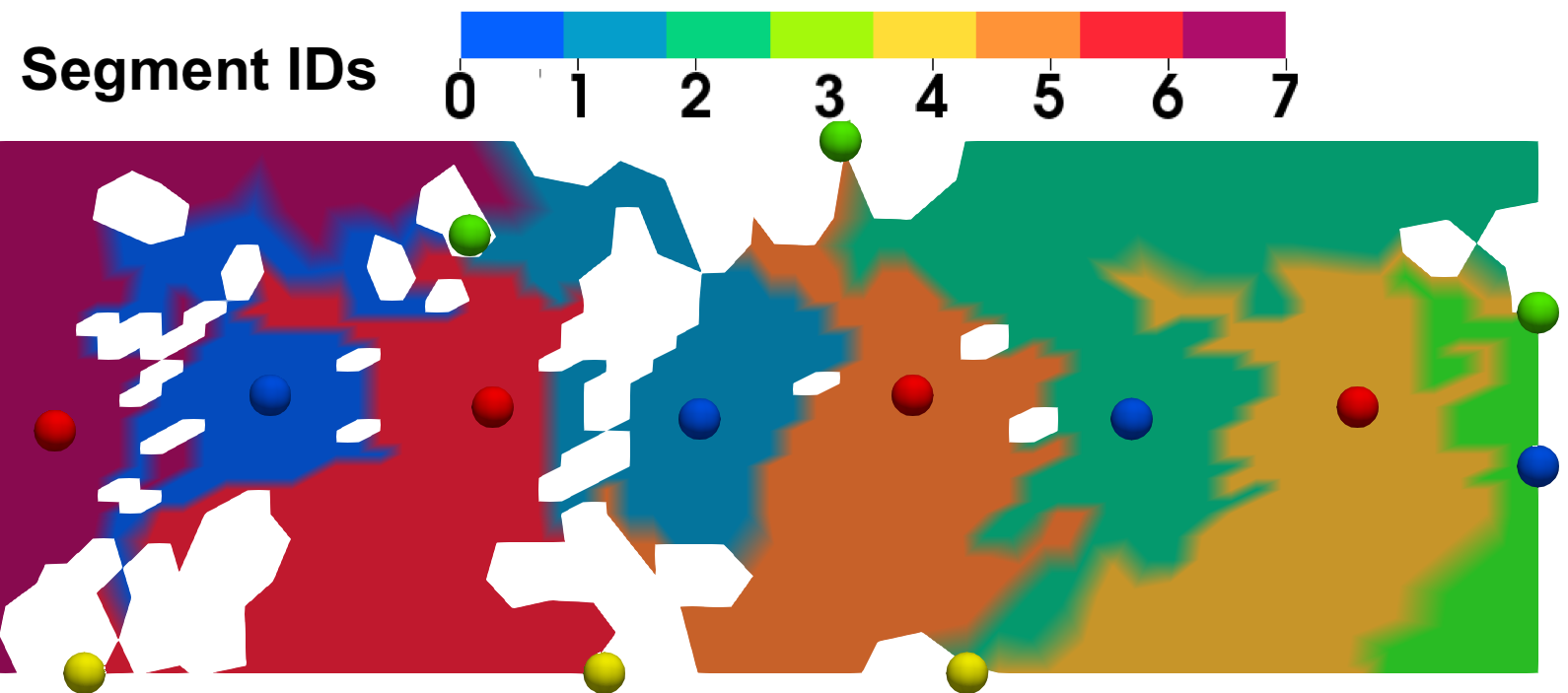}}
	
	\subcaptionbox{L-SIAC vorticity. \label{fig-2Dcyl-Seg-LSIAC}}{\includegraphics[width=0.48\linewidth,keepaspectratio]{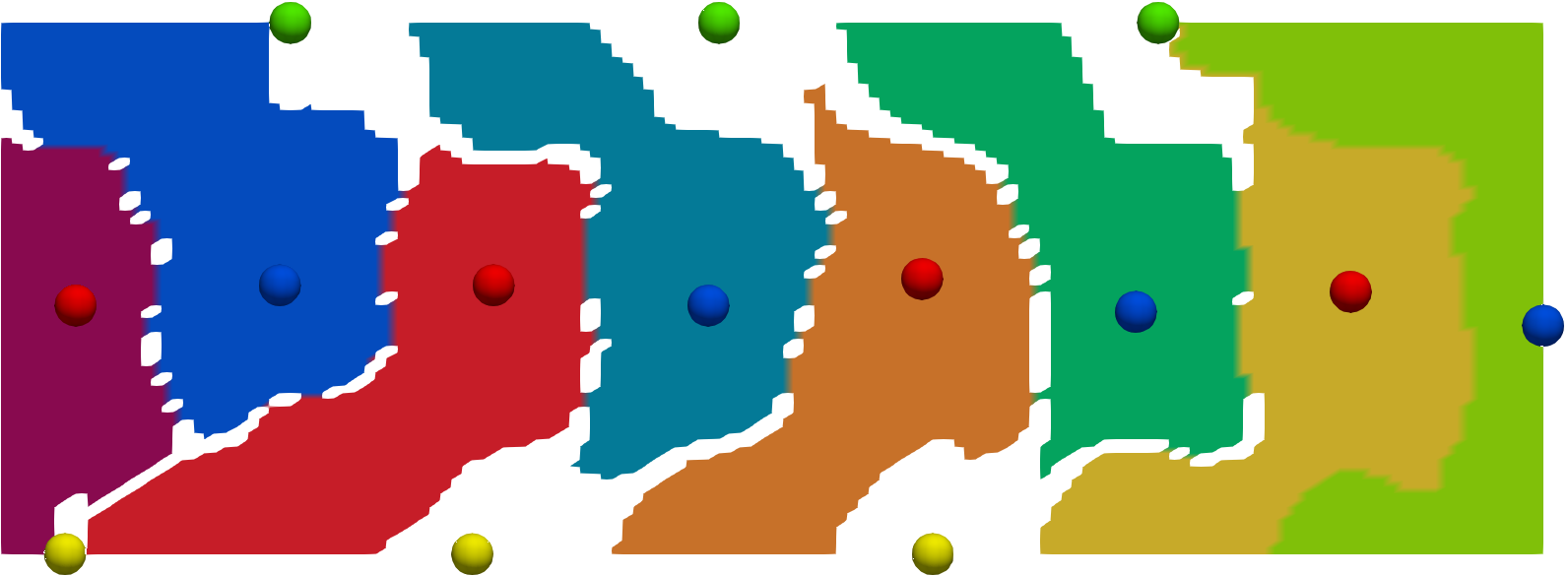}}
	
	\vspace{-0.5em}
	\caption{Segmentation of the vorticity over a flow past a cylinder described in Section \ref{subsec-flow2Dcylinder}. The segmentations are calculated using the contour tree and a persistence threshold in region $2$ of Figure \ref{fig-2Dcyl_PC_zoom}. The critical points denoted by saddle1 and saddle2 are the saddles corresponding the merges and splits, respectively. 
		\label{fig-2Dcyl-Seg-compare}}
	\vspace{-1em}
\end{figure}

The segmentation of the datasets along with the location and type of critical points are shown in Figure \ref{fig-2Dcyl-Seg-compare}, created using the contour tree and the persistence threshold in region 2 (the same ones used to create the persistence diagrams in Figure \ref{fig-2Dcyl-PD-compare}). The segmentation in all three cases captures and classif\grammar{ies} the significant part of the vortices.  Observe the critical points of the max and mins lineup at similar locations for the three datasets. The critical points\grammar{, however,} for saddle1 and saddle2 appear more uniformly distributed in the case of the L-SIAC vorticity. 
\grammar{The nonuniformly distributed critical points for saddle1 and saddle2} affects the boundaries of the segmented regions for the sampled and subdivided vorticity, which are far more irregular and exhibit resemblance to the orientations in the simulation mesh.
In the case of L-SIAC vorticity, the edges of the segmentations are smoother and \grammar{align only} with the mesh boundaries.

\begin{figure}[!ht]
	\centering
	\subcaptionbox{Sampled vorticity. \label{fig-2Dcyl-Seg-strict-Samp}}{\includegraphics[height=1.8cm,keepaspectratio]{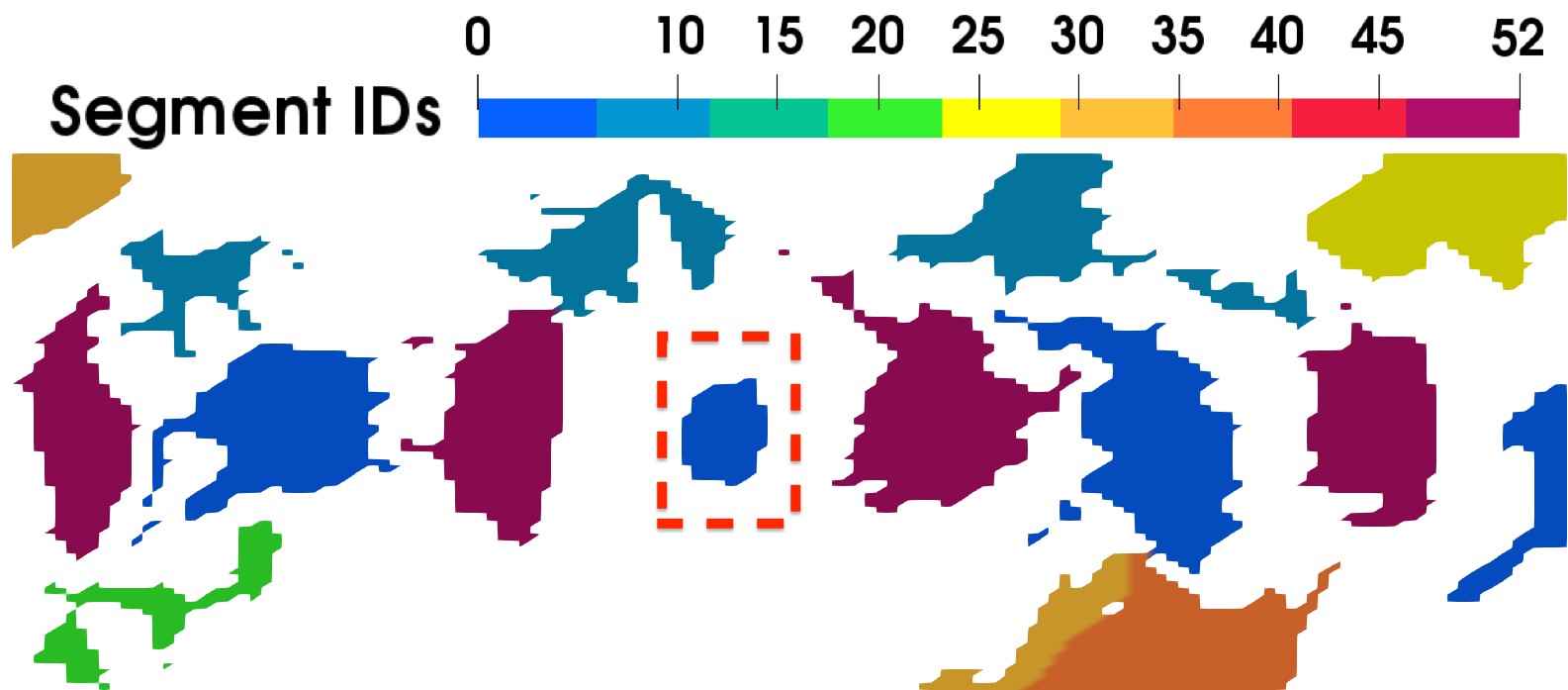}}
	\subcaptionbox{Subdivided vorticity. \label{fig-2Dcyl-Seg-strict-SubDiv}}{\includegraphics[height=1.8cm,keepaspectratio]{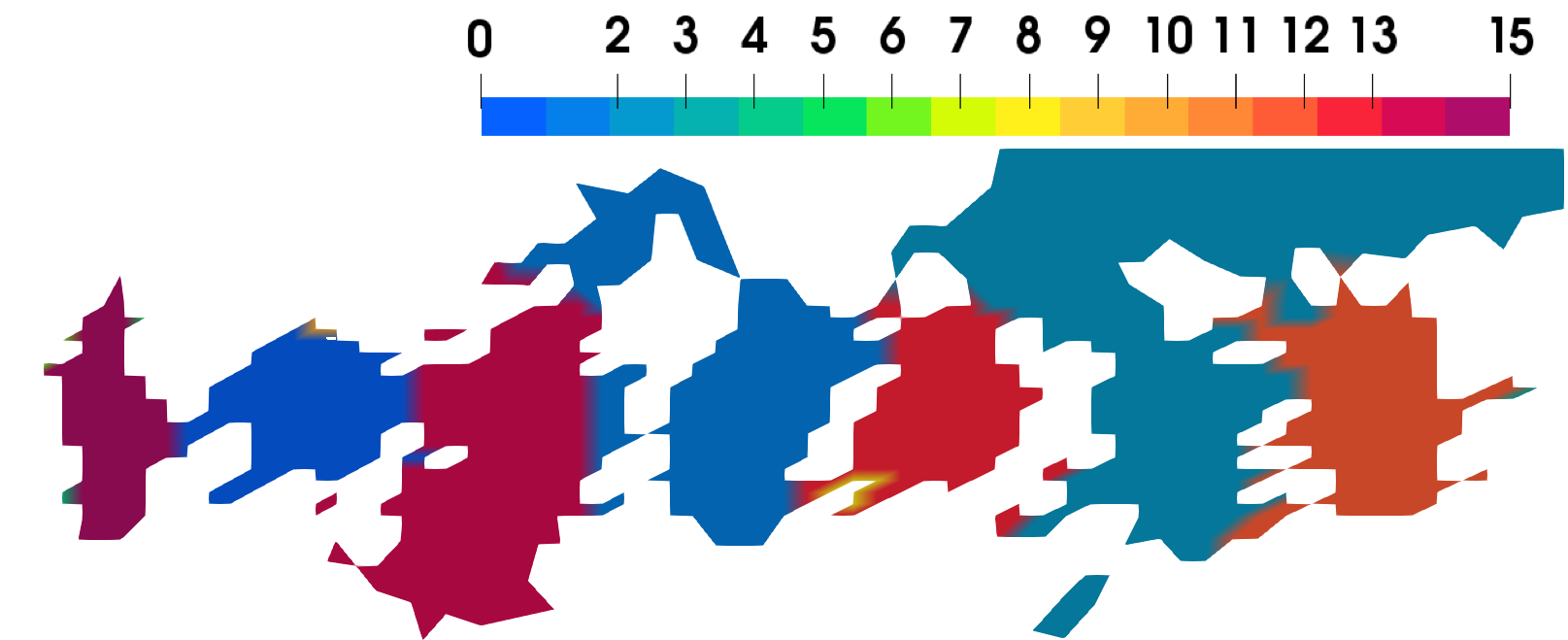}}
	
	\subcaptionbox{L-SIAC vorticity. \label{fig-2Dcyl-Seg-strict-LSIAC}}{\includegraphics[height=1.8cm,keepaspectratio]{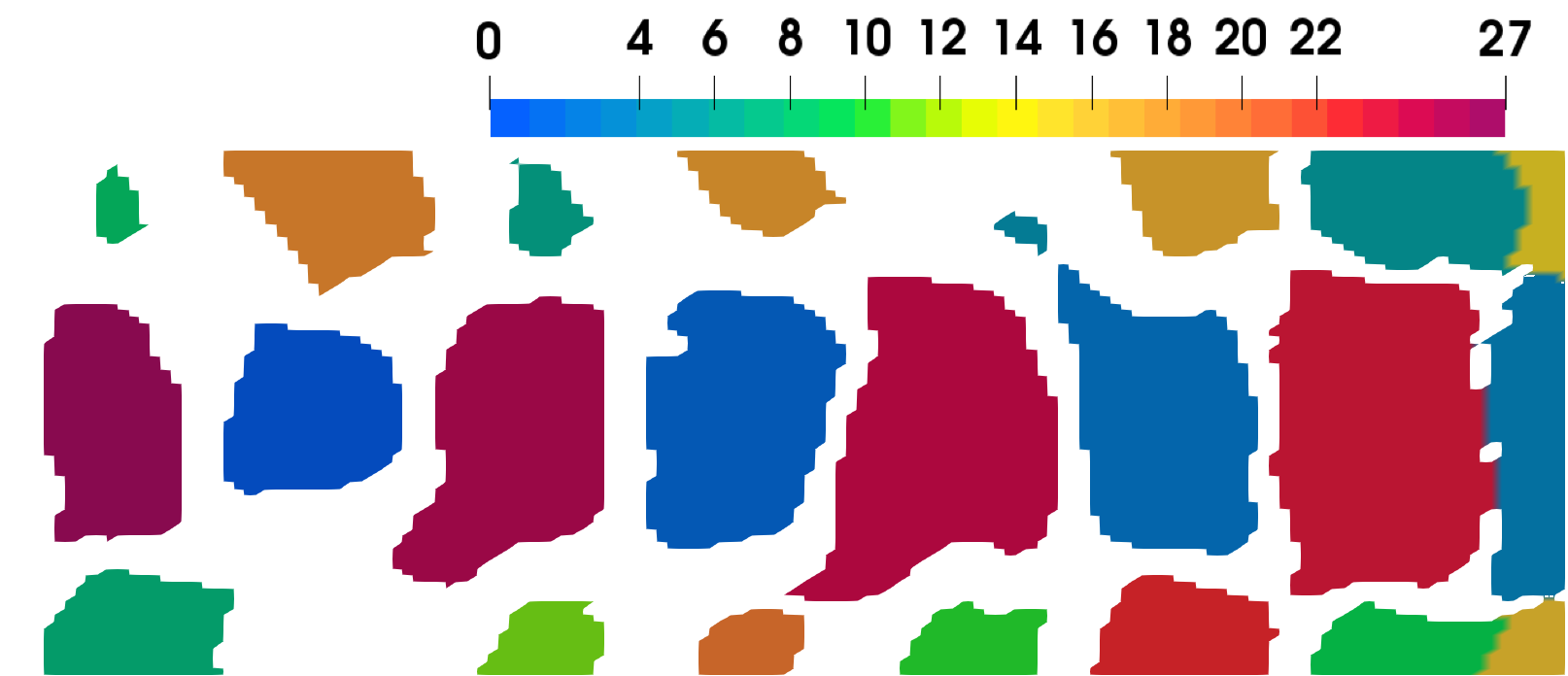}}
	\vspace{-0.5em}
	\caption{Segmentation of the vorticity for the flow past a cylinder described in Section \ref{subsec-flow2Dcylinder}. The segmentation for the sampled and the subdivided vorticity are computed using the contour tree along with a persistence threshold of $0.04$. In the case of L-SIAC vorticity, the value of persistence threshold used is $0.001$.
		\label{fig-2Dcyl-Seg-strict-compare}}
	\vspace{-1em}
\end{figure}

In an attempt to segment the vortices more stringently without collapsing features, we applied the contour tree using lower persistence thresholds to \grammar{oversegment} the domain.  We used persistence threshold of $0.04$ to the sampled and the subdivided vorticity, thus resulting in the segmentation shown in Figures \ref{fig-2Dcyl-Seg-strict-Samp} and \ref{fig-2Dcyl-Seg-strict-SubDiv}, respectively. 
In the case of the L-SIAC vorticity, we chose the persistence threshold of $0.001$, which contains all the persistence pairs (refer to Figure  \ref{fig-2Dcyl_PC_All}) and used it for segmentation (Figure \ref{fig-2Dcyl-Seg-strict-LSIAC}).  
To select the segmentations containing the vortices, we filtered them by thresholding based on their size and checking if the segment is associated with a leaf (a minima or maxima) in the contour tree. In the case of sampled vorticity (Figure \ref{fig-2Dcyl-Seg-strict-Samp}), a portion of the vortex indicated by the red dotted box was filtered out. In this case also (referring to Figure \ref{fig-2Dcyl-Seg-strict-compare}), we observed that the boundaries of segmentations for sampled and subdivided vorticity are irregular compared to the boundaries of the segments in the L-SIAC vorticity.

\subsection{Counter-Rotating Vortex}
\label{subsec:3DCounterRotVortices}

\begin{figure}[!ht]
	\begin{center}
		\includegraphics[width=0.5\linewidth]{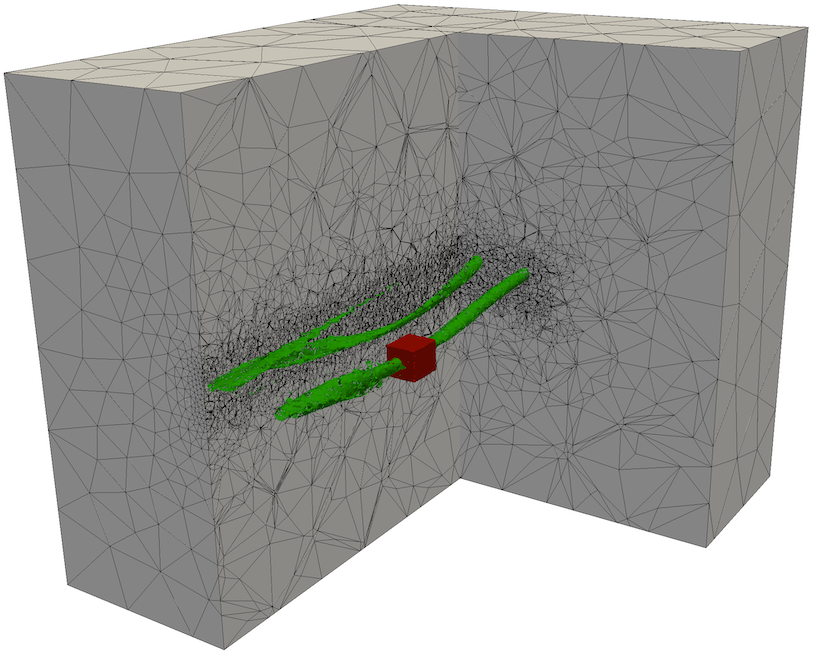}
	\end{center}
	\vspace{-0.5em}
	\caption{Input simulation of the counter-rotating vortices.  We focus on analyzing data in the red cube. The counter-rotating vortices in green are iso-surfaces of vorticity (magnitude) calculated using element-wise derivatives.}
	\label{fig:Corot_Img}
	\vspace{-1em}
\end{figure}
We also consider a 3D example for our empirical study, again using a Nektar++ simulation example. An incompressible Navier-Stokes solver with cG discretization \grammar{was} used to generate a flow scenario containing two primary vortices and their secondary counter-rotating vortices shown in Figure \ref{fig:Corot_Img}. The simulation (input) parameters were set to the following: for the advection term, the Velocity Correction Scheme~\cite{KaSh05} with SVV de-aliasing was used, and the time integration splitting scheme was set to IMEX order two. Further details on the simulations are given in \cite{Sidotthesis}.  The simulation mesh \grammar{was} adaptively refined at the location of the vortices and contains $223,837$ polynomial degree five ($P(5)$) tetrahedra. 

The simulation data from the highlighted region in Figure \ref{fig:Corot_Img} (indicated by the red cube) \grammar{was} chosen to be analyzed \grammar{topologically}. We analyzed the magnitude of vorticity as the scalar quantity to extract vortices.  We used the elementwise derivatives on the vector field quantities $(u, v, w)$ to calculate the magnitude of the vorticity vector (hereafter referred to as the vorticity). To produce the ``sampled'' vorticity, elementwise derivatives are sampled on a grid of resolution of $100\times100\times100$ and used to calculate the magnitude of vorticity.
To produce ``subdivided'' vorticity, \grammar{each tetrahedron of the input mesh, in the highlighted region was subdivided into smaller tetrahedrons($216 = 6^3$).} The vorticity \grammar{was sampled at the vertices of the subdivided mesh and 
	at the vertices having a discontinuity, the values were averaged.}
To calculate the L-SIAC vorticity, we use L-SIAC methodology proposed in Jallepalli et al.~\cite{jallepalli2017treatment}. All the required derivatives are calculated using the L-SIAC filter. The parameters used for the L-SIAC filter are B-splines of order $7$ (specifically, we use the $D^1K(11,7)$ filter, where $D^1$ represents the total derivative), the $\theta$ in direction of the derivative and the characteristic length is adapted based on the element size~\cite{jallepalli2018adaptive}. \response{The average time taken to compute vorticity at each location on a machine with a $2.5$ GHz (Intel CPU E7-8890) processor is $0.34$, $9.6e^{-4}$, and $334$ milliseconds for the sampled, subdivided, and L-SIAC vorticities, respectively.}

\begin{figure}[!ht]
	\begin{center}
		\includegraphics[width=1.0\linewidth]{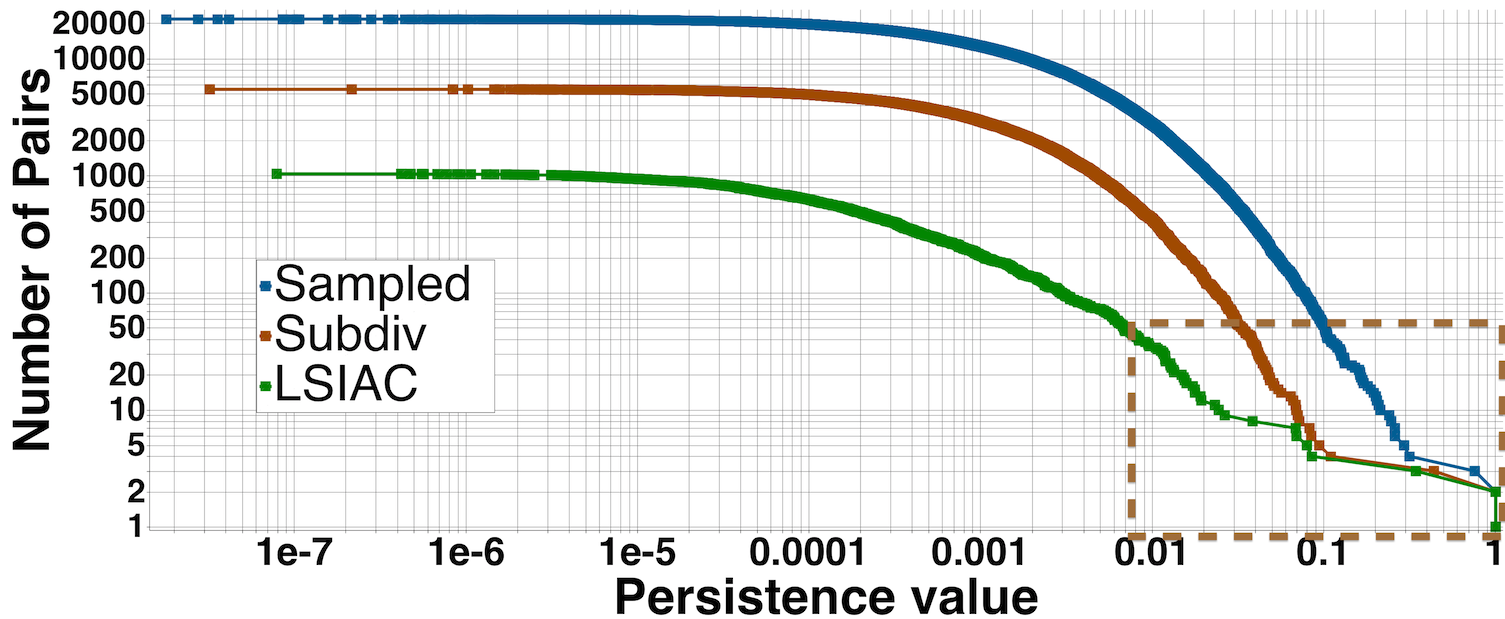}
	\end{center}
	\vspace{-0.5em}
	\caption{Persistence curves for the sampled, subdivided, and L-SIAC vorticity fields of the counter-rotating vortex.
		\label{fig-bvort_PC_All}}
	\vspace{-1em}
\end{figure}

\begin{figure}[!ht]
	\begin{center}
		\includegraphics[width=1.0\linewidth]{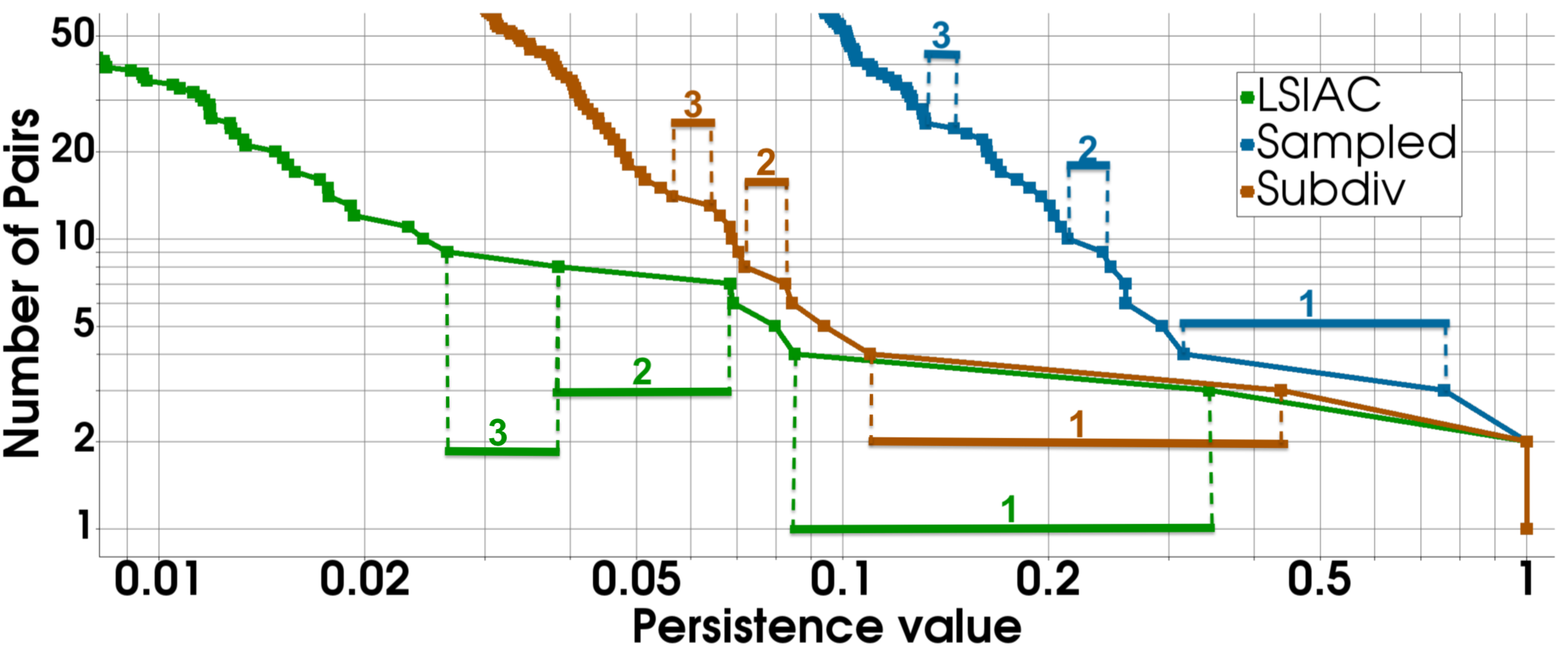}
	\end{center}		
	\vspace{-0.5em}
	\caption{Zoom in on the region associated with Figure~\ref{fig-bvort_PC_All}.  For each of datasets, we highlight three different stable regimes of critical features (numbered 1-3).
		\label{fig-bvort_PC_zoom}}
	\vspace{-1em}
\end{figure}

The datasets had vorticity ranging from $[0,84]$, $[0,88]$, and $[0,176]$ for sampled, subdivided and L-SIAC vorticity fields, respectively.  
To enable comparisons of persistence curves and diagrams, we normalized the vorticity to the same range of $[0,1]$.  
The persistence curves are shown in Figure~\ref{fig-bvort_PC_All}. We observe that the first stable regions indicated by $1$ in Figure \ref{fig-bvort_PC_zoom}, are have ranges of persistence at $[0.32, 0.76]$, $[0.11, 0.44]$, and $[0.08,0.34]$ for the sampled, subdivided, and L-SIAC vorticity. The sampled vorticity has the most extended stable region, followed by subdivided and then the L-SIAC filter.  Note that the ranges appear differently due to the log-scaled $x$-axis in Figure \ref{fig-bvort_PC_zoom}.

\begin{figure}[!ht]
	\centering
	\subcaptionbox{\grammar{Sampled vorticity at persistence threshold 0.5.} \label{fig-bvort-PD-dGSamp}}{\includegraphics[width=0.32\linewidth]{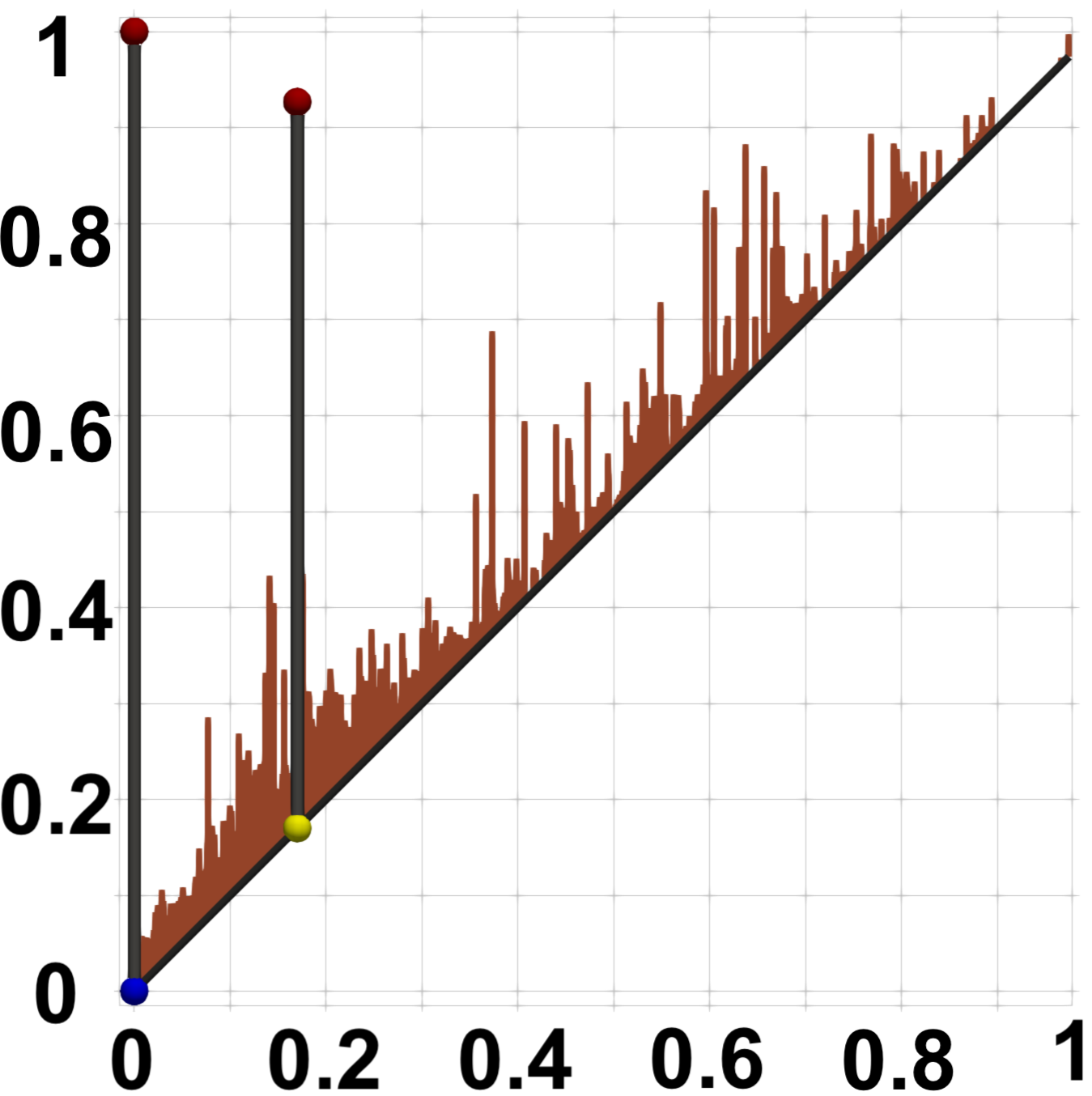}}%
	\hfill
	\subcaptionbox{\grammar{Subdivided vorticity at persistence threshold 0.2.}  \label{fig-bvort-PD-Subdiv}}{\includegraphics[width=0.32\linewidth]{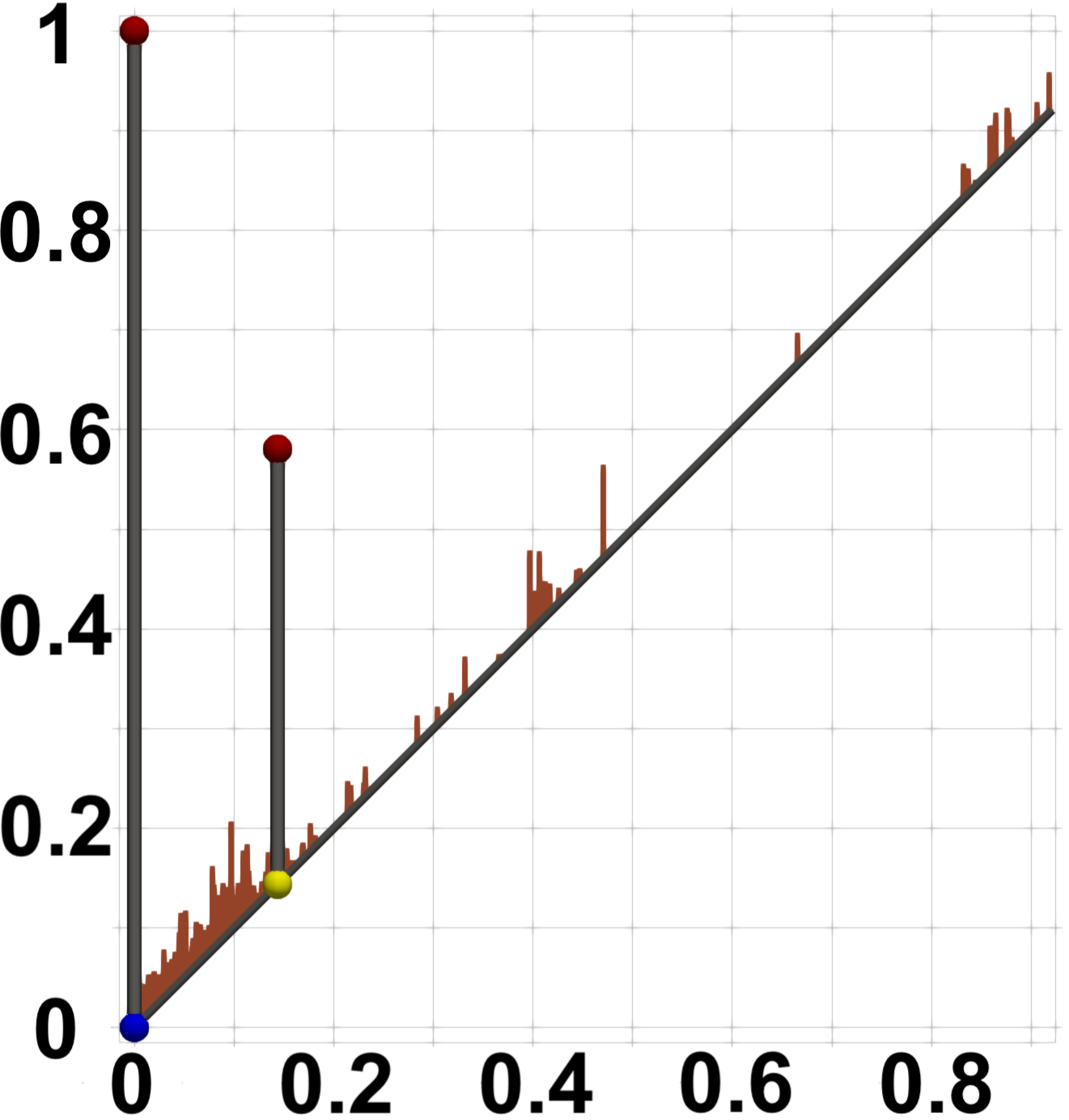}}%
	\hfill
	\subcaptionbox{\grammar{L-SIAC vorticity at persistence threshold 0.2.} \label{fig-bvort-PD-LSIAC}}{\includegraphics[width=0.32\linewidth]{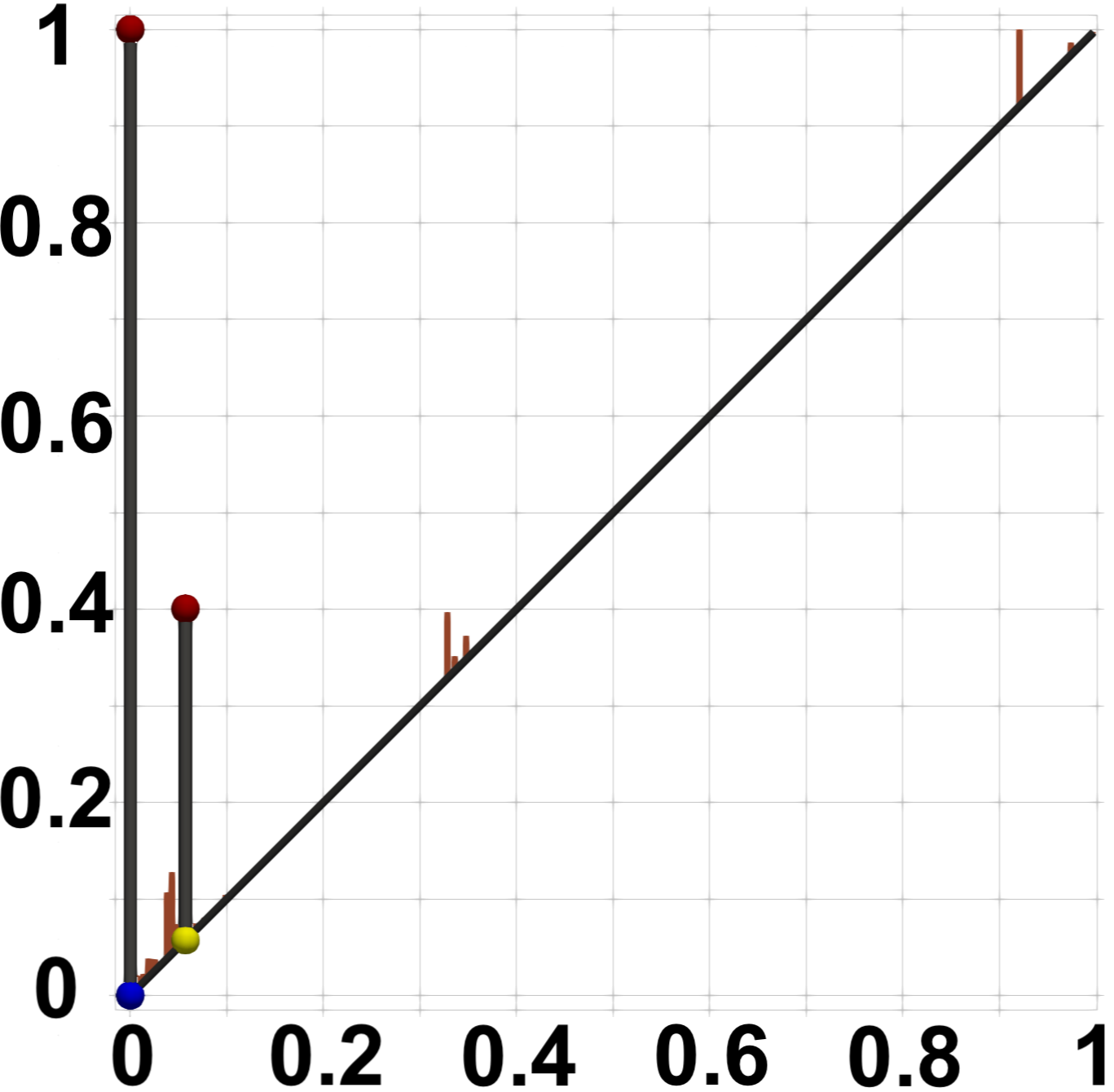}}
	\vspace{-0.5em}
	\caption{Persistence diagrams for the vorticity of the counter-rotating vortex dataset.  Orange bars highlight what was removed by topological filtering.
		\label{fig-bvort-PD-compare}}
	\vspace{-1em}
\end{figure}

Consequently, we chose different ranges of persistence values for topological simplification.  We first considered a coarse simplification into the stable region $1$.  The persistence diagrams simplified using thresholds of $0.5$ (sampled) and $0.2$ (subdivided and L-SIAC) in the respective stable regions are shown in Figures \ref{fig-bvort-PD-compare}. To highlight what was removed, the persistence pairs below the threshold are visualized as orange bars.  Unlike our other examples, we observed that sampled vorticity has a substantial number of persistence pairs far from the diagonal -- suggesting that there \grammar{was} significant topological ``noise'' in the data that makes it more difficult to separate even the most prominent vortices of interest.  By comparison, the subdivided and L-SIAC datasets appear to have more separation between the persistent features.

\begin{figure}[!ht]
	\centering
	\includegraphics[width=0.045\linewidth]{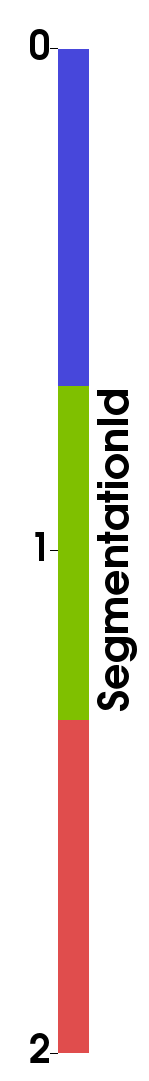}
	\subcaptionbox{Sampled vorticity.  \label{fig-bvort-Seg1-dGSamp}}{\includegraphics[width=0.31\linewidth]{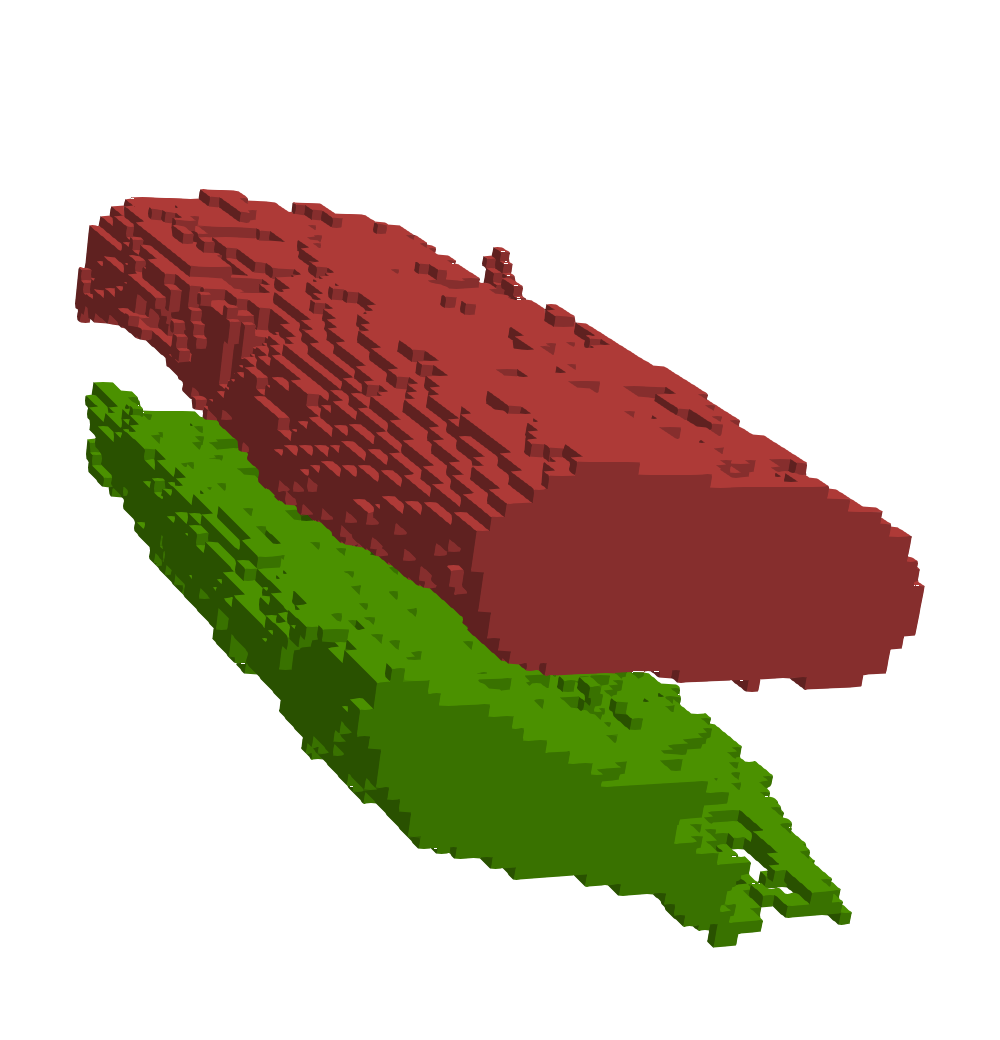}}%
	\hfill
	\subcaptionbox{Subdivided vorticity.  \label{fig-bvort-Seg1-Subdiv}}{\includegraphics[width=0.31\linewidth]{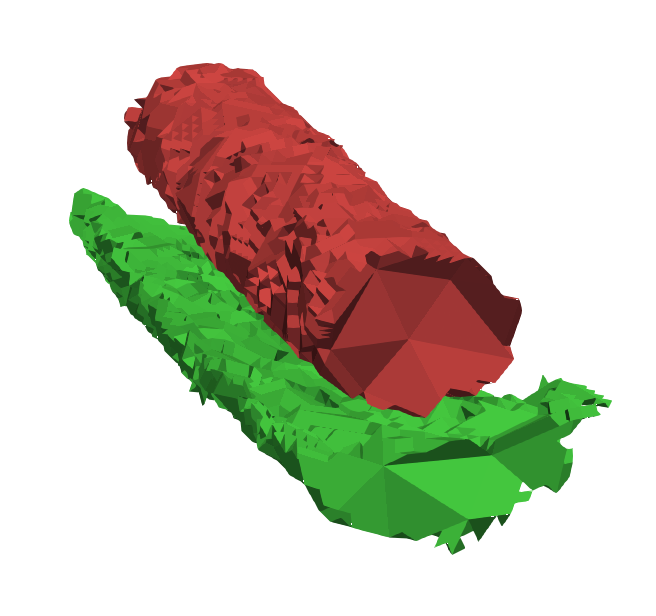}}%
	\hfill
	\subcaptionbox{L-SIAC vorticity. \label{fig-bvort-Seg1-LSIAC}}{\includegraphics[width=0.31\linewidth]{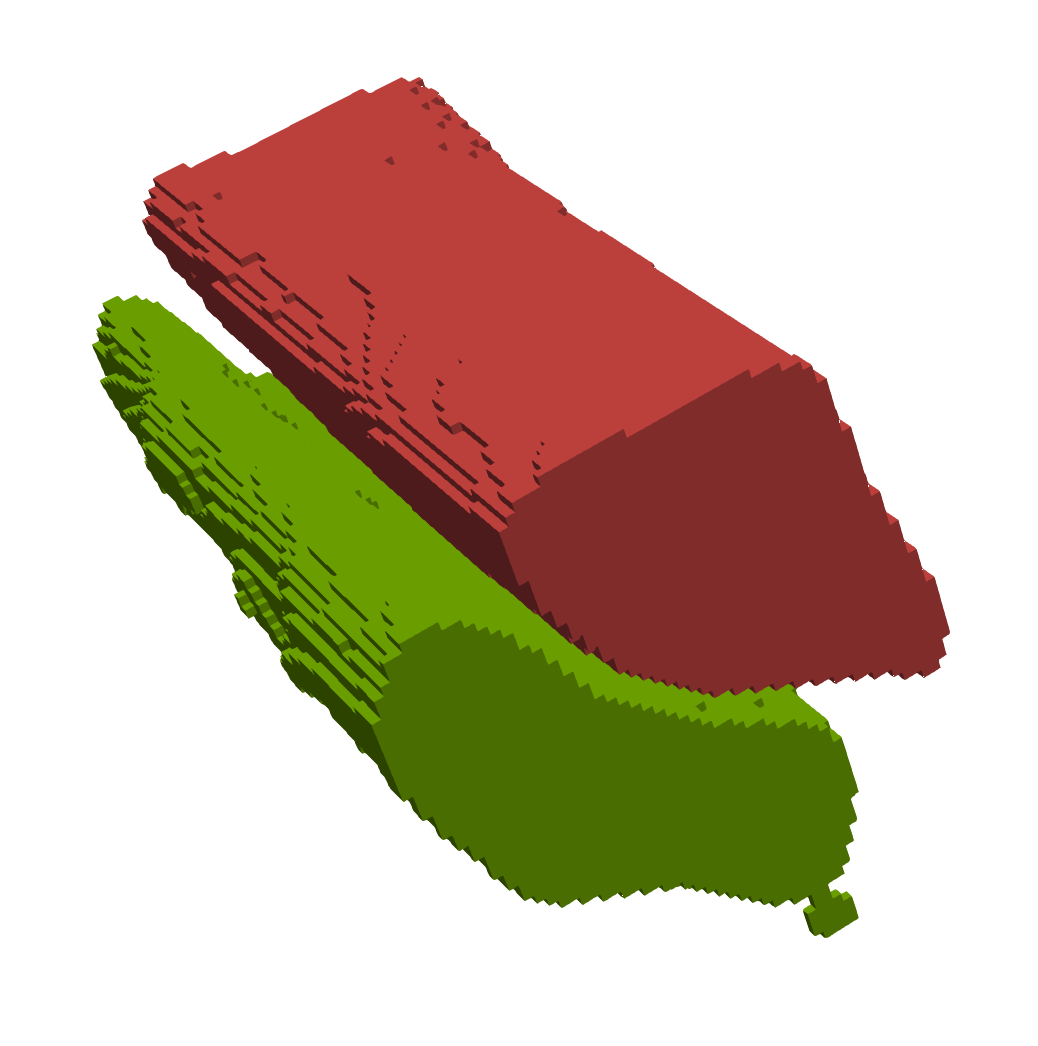}}
	\vspace{-0.5em}
	\caption{Segmentation corresponding to the topologically simplified vorticity fields as the threshold used in Figure~\ref{fig-bvort-PD-compare}.  These \grammar{figures} highlight the two main rotating vortices, as best described by each dataset.  
		\label{fig-bvort-Seg1-compare}}
	\vspace{-1em}
\end{figure}

Using the persistence threshold in the stable region $1$ ($0.5$ for sampled,  $0.2$ subdivided and L-SIAC) to simplify the dataset, we used the contour trees to segment the domain  and visualized the segmented vortex cores in Figure~\ref{fig-bvort-Seg1-compare}.  We observe\grammar{d} that all three methods had segmented significant portions of the primary and secondary vortices. The boundaries of the sampled and subdivided vorticity \grammar{were} slightly irregular compared to the L-SIAC filter.  However, at this coarse scale, they \grammar{were} not significantly worse. The subdivided mesh is expected to have irregularities due to the underlying nonuniform mesh. Thus, all the methods have identified and segmented the first and second significant features of interest in the dataset. This result demonstrates that for this dataset, topological simplification can tolerate a wide range of transformation methodologies to capture the coarsest scale of features.

\begin{figure}[!ht]
	\centering
	\subcaptionbox{\grammar{Sampled vorticity at persistence threshold $0.136$} \label{fig-bvort-PD3-dGSamp}}{\includegraphics[width=0.32\linewidth]{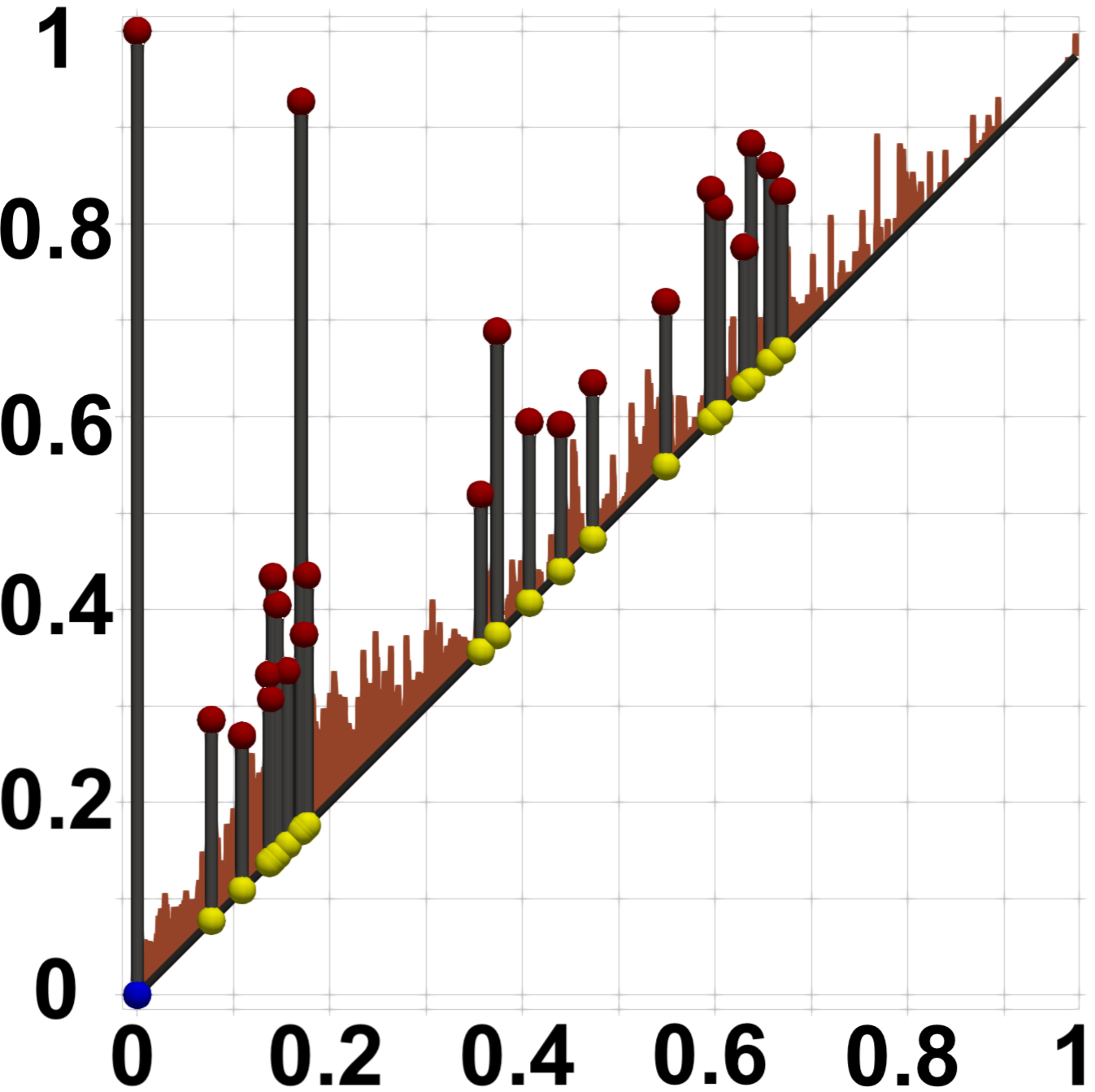}}%
	\hfill
	\subcaptionbox{\grammar{Subdivided vorticity at persistence threshold $0.06$.}  \label{fig-bvort-PD3-Subdiv}}{\includegraphics[width=0.32\linewidth]{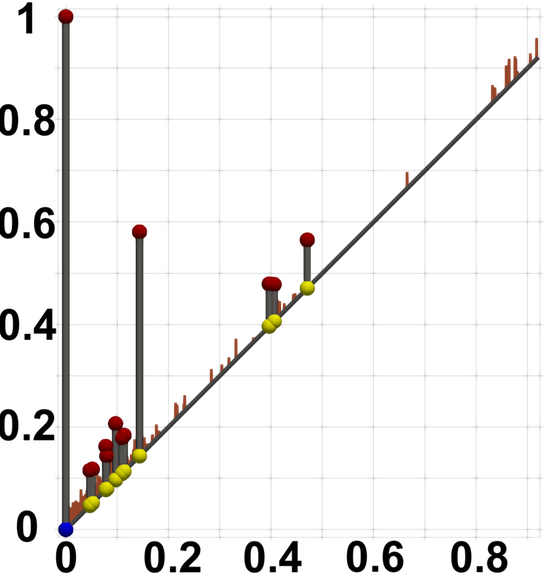}}%
	\hfill
	\subcaptionbox{\grammar{L-SIAC vorticity at persistence threshold $0.03$.} \label{fig-bvort-PD3-LSIAC}}{\includegraphics[width=0.32\linewidth]{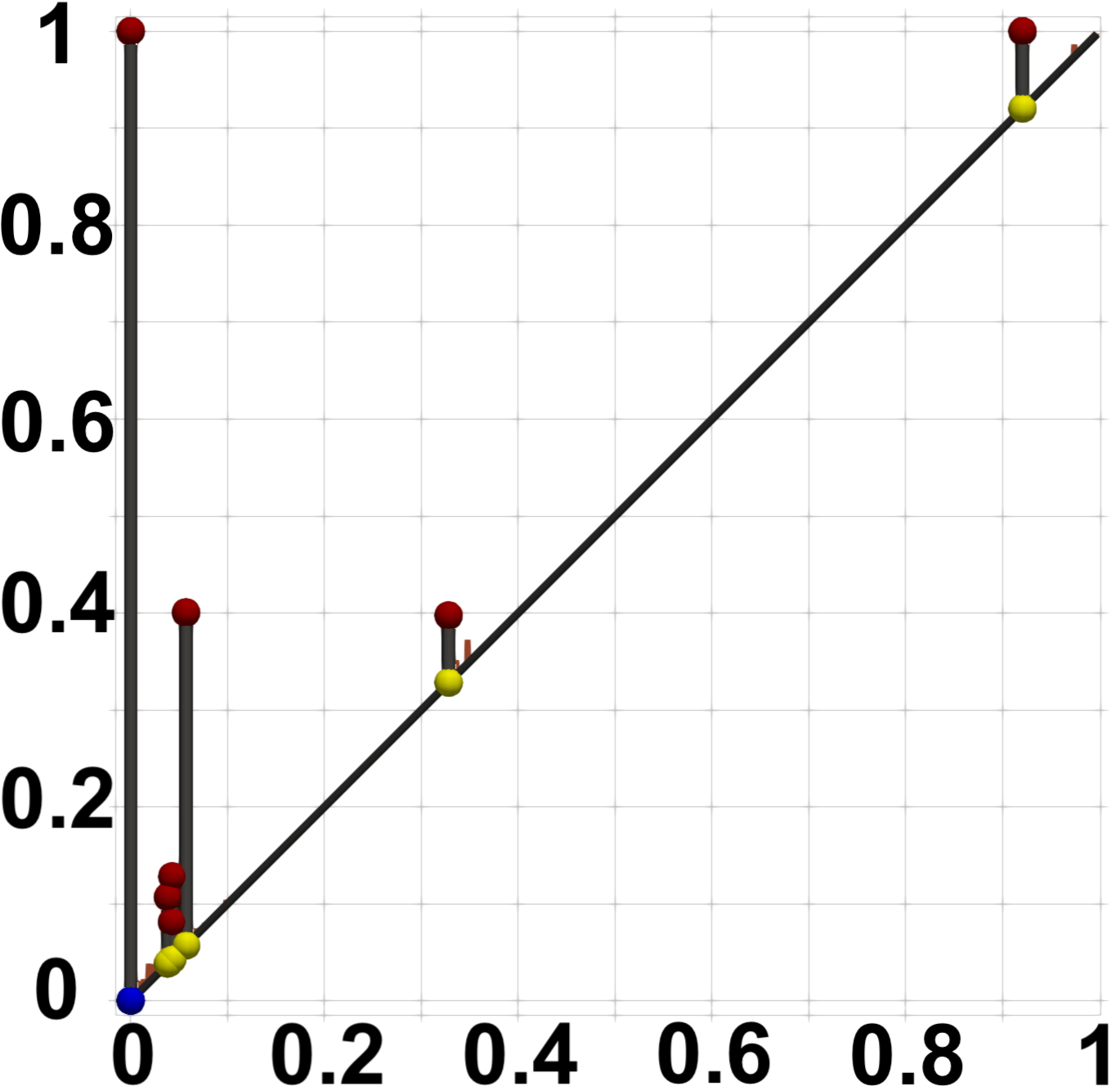}}
	\vspace{-0.5em}
	\caption{Persistence diagrams for the vorticity of the counter-rotating vortex dataset after simplifying to capture fine-scale features. 
		\label{fig-bvort-PD3-compare}}
	\vspace{-1em}
\end{figure}


To identify a second set of features from the data, we looked for smaller persistence thresholds in the datasets. In Figure \ref{fig-bvort_PC_zoom}, among the stable regions indicated by $2$ and $3$, the stable region indicated by $3$ in all the datasets yielded the best possible segmentation of the data. The persistence diagrams using the thresholds in the stable region $3$ ($0.136$ for sampled, $0.06$ for subdivided, $0.03$ for L-SIAC) are shown in Figure \ref{fig-bvort-PD3-compare}. 

\grammar{The lower threshold we selected to segment the data resulted} in numerous undesired segmentations of the data that we removed during analysis.  Specifically, we manually identified segments using their level in the contour tree, size of the segment, and the segment IDs, for segments \grammar{that} resembled vortex-like structures. 
The resulting segmentations from the datasets are shown in Figure \ref{fig-bvort-Seg2-compare}. We observe that the sampled vorticity (Figure \ref{fig-bvort-Seg2-dGSamp}) produced $4$ of $45$ segments resembling vortex-like structures, but the vortex indicated by blue was divided into smaller undesirable regions. In the case of subdivided vorticity (Figure \ref{fig-bvort-Seg2-Subdiv}), we identified $3$ of $23$ segments that resembled vortices, but the segmented \grammar{region indicated in green} has many gaps, meaning significant parts of the vortex is segmented \grammar{into} smaller segments \grammar{that} have been filtered out.  This was particularly surprising given the structure in the persistence diagram for subdivided \grammar{vorticity}, which appears to have less topological noise.   In the case of L-SIAC vorticity (Figure \ref{fig-bvort-Seg2-LSIAC}), we identified $5$ of $13$ segments resembling vortices that segment the vortices consistently, except for the segment represented by blue. A portion of this vortex \grammar{surrounds} its neighboring vortices. 

To summarize, the sampled, the subdivided, and the L-SIAC methodologies \grammar{successfully identified} the most significant features in the dataset. To identify the fine-scale features, the L-SIAC methodology performed best among the three, but we remark that it comes at a significant computational cost relative to merely sampling \cite{jallepalli2017treatment}. 
The sampled vorticity appeared to capture the shape of the vorticity features better than the subdivided vorticity.  Furthermore, it also identified an extra vortex than the subdivided vorticity.  However, the sampled data has the disadvantage that filtering to capture the best set of segments required removing a larger number of features. 

\section{Discussion}
\label{sec-conclusion}

The precise characterization of the effects of data transformations on topological analysis can manifest in subtle ways, as small changes in function values can affect the order in which features merge.  In this empirical study, we take a first step toward understanding the relationships \grammar{among} element discontinuities, sampling artifacts, and topological features.  We analyzed three methodologies that can be used to enable topological analysis of the HO-FEM data. For identifying the most significant features in HO-FEM data, we used two simple methodologies: sampling on a grid and subdividing the mesh.  While both have the advantage of simplicity, and both are successful at capturing the coarsest features, they also tend to distort the size and shape of the feature boundaries.
In particular, subdivision appears to separate features better, but sampling might do a better job of capturing the shape of the feature. Interestingly, we discovered and demonstrated the counterintuitive nature of these methodologies, since in certain ways they perform adversely with the increase in sampling resolution.

We have also shown that an extended range of features can be identified in the dataset using the L-SIAC filter.  Using the L-SIAC filter, we can improve the resolution of the features with the increase in sampling resolution. Although the L-SIAC filter may appear better suited \grammar{to} this analysis task, it is also computationally expensive as its cost of evaluation is roughly proportional to the sampling frequency.  
In future work, it would be interesting to use topological analysis on sampled or subdivided data as a guide to adaptively choose locations to apply the L-SIAC filter and use the updated results to improve the topological analysis iteratively.  In particular, at regions that are far from discontinuities, we expect all methodologies to perform better. 

\response{The current focus of this work has been on scalar fields and their topological analysis through persistence diagrams and contour trees.  A limitation of this study is its scope.  In the future, we would like to broaden this study to include gradient field topology (through Morse-Smale complexes) as well as the topology of vector and tensor data.   Nevertheless, the narrow scope we have employed has helped to isolate specific issues with both interpolants and feature extraction techniques.}
Long term, we think this study can help motivate a better set of design constraints for directly extracting topological features from higher order data without the need for data transformation. 
\response{ That said, even current tools appear capable of extracting certain coarse features if data transformation is used carefully.}

\acknowledgments{
	The authors also wish to thank Professor Spencer Sherwin (Imperial College London, UK), Mr. Alexandre Sidot, and the Nektar++ Group for the counter-rotating vortex data and helpful discussions. This material is based upon work supported by the U.S. Department of Energy, Office of Science, Office of Advanced Scientific Computing Research, under Award Number(s) DE-SC-0019039. The authors acknowledge support from ARO W911NF-15-1-0222 (Program Manager Dr. Mike Coyle).}

	\bibliographystyle{abbrv-doi}
	
	\bibliography{references,josh,ttk}
	
\end{document}